\documentclass[a4paper,onecolumn,10pt, accepted=2022-05-31]{quantumarticle}

\pdfoutput=1
\usepackage[numbers,sort&compress]{natbib}

\usepackage{amsmath}
\usepackage{amssymb}
\usepackage{graphicx}
\usepackage{float}
\usepackage[english]{babel}
\usepackage{dsfont}
\usepackage{epstopdf}
\usepackage{soul}
\usepackage{color}
\usepackage{braket}
\usepackage[colorlinks=true,
            linkcolor=magenta,
            urlcolor=blue,
            citecolor=blue]{hyperref}

\usepackage[normalem]{ulem}

\def\be{\begin{equation}}
\def\ee{\end{equation}} 
\def\bea{\begin{eqnarray}}
\def\eea{\end{eqnarray}} 
\def\ba{\begin{array}} 
\def\ea{\end{array}}
\def\pa{\partial}

\def\Om{\Omega} 
\def\om{\omega}

\def\nn{\nonumber}
\def\ket{\rangle}
\def\bra{\langle}
\def\b{\mathbf}
\def\bs{\boldsymbol}
\def\f{\frac}

\def\ra{\rightarrow}
\def\rm{\mathrm}

\newcommand{\mv}[1]{\langle #1\rangle}

\emergencystretch=\maxdimen
\begin{document}
\title{Topological phonons in arrays of ultracold dipolar particles}
\author{Marco Di Liberto} 
\email{Marco.Di-Liberto@uibk.ac.at}
\author{Andreas Kruckenhauser}
\author{Peter Zoller}
\author{Mikhail A. Baranov}
\affiliation{Institute for Quantum Optics and Quantum Information of the Austrian Academy of Sciences, 6020 Innsbruck, Austria}
\affiliation{Institute for Theoretical Physics, University of Innsbruck, 6020 Innsbruck, Austria}

\begin{abstract}

The notion of topology in physical systems is associated with the existence of a nonlocal ordering that is insensitive to a large class of perturbations. 
This brings robustness to the behaviour of the system and can serve as a ground for developing new fault-tolerant applications.
We discuss how to design and study a large variety of topology-related phenomena for phonon-like collective modes in arrays of ultracold polarized dipolar particles. 
These modes are coherently propagating vibrational excitations, corresponding to oscillations of particles around their equilibrium positions, which exist in the regime where long-range interactions dominate over single-particle motion. 
We demonstrate that such systems offer a distinct and versatile tool to investigatea wide range of topological effects in a \emph{single experimental setup} with a chosen underlying crystal structure by simply controlling the anisotropy of the interactions via the orientation of the external polarizing field. 
Our results show that arrays of dipolar particles provide a promising unifying platform to investigate topological phenomena with phononic modes. 

\end{abstract}

\maketitle

\section{Introduction}
\label{sec:intro}

Many of the intriguing phenomena in condensed matter systems have been associated and tied to the notion of topology and to the existence of topological states characterized by discrete topological indices (see, for example, Refs.~\cite{Kane2010,Chen2012, Ryu2016,Rachel2018}).
The most prominent manifestations of the underlying nontrivial topology in a many-body system are robust edge states and quantized responses to external fields, as for the celebrated Quantum Hall effect \citep{Klitzing1980,Laughlin1981}.
The family of physical systems exhibiting topological phases and effects is wide and includes topological insulators and superconductors \citep{Qi2011}, crystalline insulators \citep{Fu2011}, ultracold atoms and molecules \citep{Goldman2016a,Cooper2019}, photonic systems \citep{Ozawa2019}, electric circuits \citep{Thomale2018}, and also acoustic and mechanical systems \citep{Huber2016,Chan2019}.

Ultracold dipolar particles with long-range and anisotropic interparticle interactions provide a versatile and powerful platform for simulating many-body quantum systems, including topological ones \citep{Lahaye2009,Weimer2010,Baranov2012,Gross2017}.
Recent progress in creating systems of dipolar particles in periodic arrays, as for the case of magnetic atoms \citep{Tolra2013,Ferlaino2016}, polar molecules \citep{Ye2013} or Rydberg atoms \citep{Browaeys2016}, offers a remarkable opportunity to experimentally implement and study various topological lattice models in this setting. 
From the theory side, it has been suggested to use dipolar particles to implement topological bands \citep{Buchler2015,Buchler2018}, Hopf insulators \citep{Schuster2021}, chiral excitations \citep{Rey2014}, Weyl nodes \citep{Rey2016}, topological bound states \citep{Salerno2020}, or strongly-correlated many-body systems \citep{Manmana2013,Yao2013,Yao2018}. 
From the experimental side, the realizations of artificial gauge fields \citep{Browaeys2020} and  a symmetry-protected topological phase \citep{Browaeys2019} were recently reported.

\begin{figure}[!t]
\center \includegraphics[width=0.69\columnwidth]{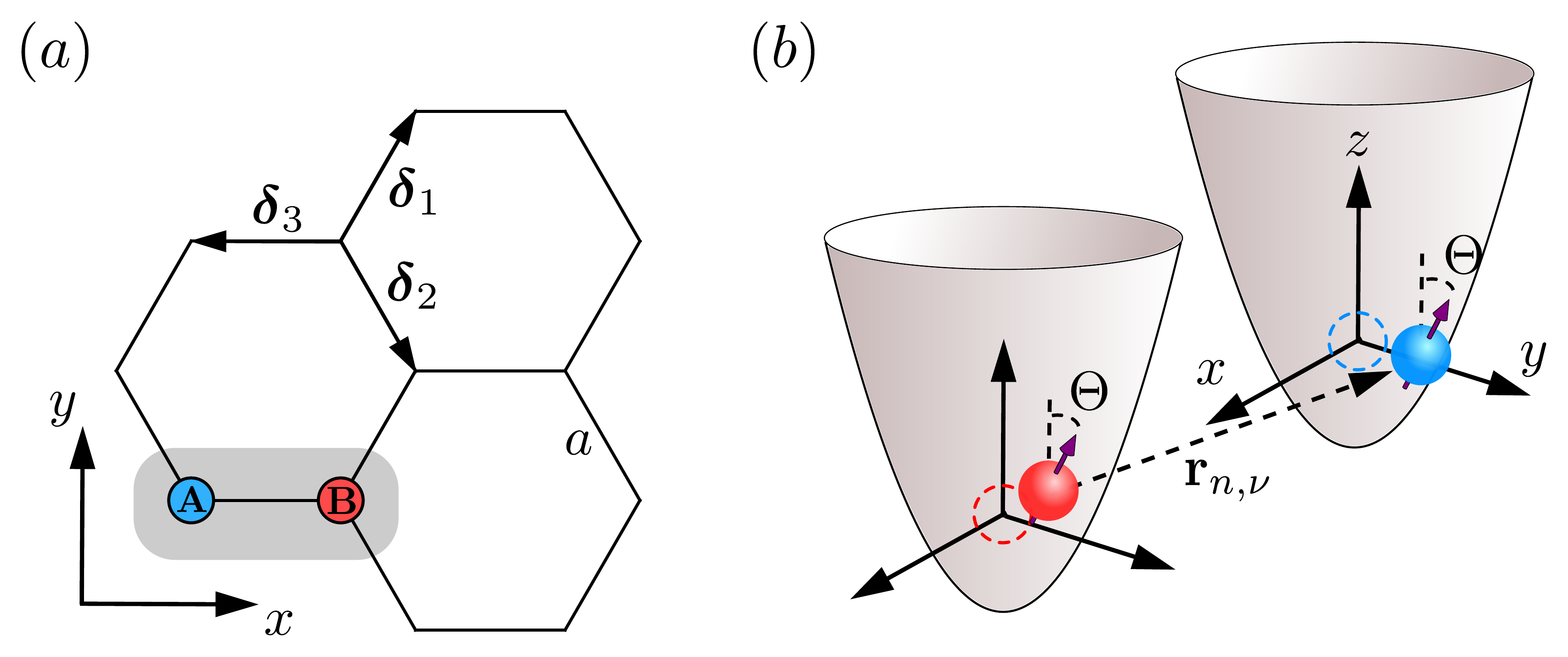}

\caption{Model. (a) Honeycomb lattice with lattice spacing $a$ and lattice vectors $\bs\delta_{1,2}=(a/2,\pm\sqrt{3}a/2)$, $\bs\delta_{3}=(-a,0)$. The unit cell containing the two inequivalent sites $A$ and $B$ is also highlighted. (b) Schematic picture of two particles interacting through a dipolar interactions $V_{\textrm{int}}$. The particles are locally trapped at the lattice sites by a deep harmonic potential and their polarization axis is aligned by a strong external field that is tilted with respect to the $\hat{z}$ axis by an angle $\Theta$. The atoms equilibrium position is indicated by dashed circles and the relative position is indicated as $\b r_{n,\nu}$ according to the convention used in the text.}
\label{fig:model} 
\end{figure}

The previously mentioned dipolar systems are characterized by very different interaction energies that are naturally to be compared with the level spacing of the confining potential. 
The interaction energy is smaller (and even much smaller) for magnetic atoms and polar molecules and much larger for Rydberg atoms. 
In this paper, we focus on the former situation which is typically described in terms of Hubbard-like models characterized by hopping particles and their interactions. 
In the presence of long-range dipole-dipole interactions, one can access physical regimes where the individual particle hopping is frozen by a deep optical lattice but excitations of individual particles can propagate via an interaction-mediated resonant exchange mechanism. 
The entire dynamics of the system in this regime is, therefore, fully controlled by the interactions. 
A common choice is to consider the \emph{internal} atomic excitations corresponding to transitions between, for example, Zeeman sublevels for magnetic atoms or rotational states for polar molecules,
which has led to a number of theoretical proposals and experiments, see for example Refs.~\cite{Yao2018, Tolra2013, Ye2013}.

Following a remarkable experimental progress in creating systems of dipolar ultracold atoms and molecules in optical lattices, we address the problem of excitation dynamics from a new perspective, and consider the propagation of \emph{vibrational} excitations of particles, which are naturally present in these systems. 
Such propagating excitations are analogous to optical phonons in crystals in the sense that they have a finite energy gap. 
In our case, these vibrational excitations correspond to transitions between $s$ and $p$ orbitals in the local harmonic oscillator basis of each well and the gap corresponds to the energy difference between these two levels. 
A large energy gap relaxes experimental requirements on temperature and technical noise, giving more precise control over the number of excitations, as compared to gapless acoustic-like modes. 
We stress that the scenario considered in our work is completely different from the one described by $p$-band Hubbard models, see Ref.~\cite{Liu2016} and references therein. In those systems, the $p$-state atoms are mobile and free to move across the lattice whereas in our setup they are localized. 
An additional advantage of freezing the atomic motion is the complete elimination of onsite particle collisions, which provide the dominant decay mechanism in $p$-band Hubbard systems \cite{Girvin2005, Muller2007, Hemmerich2011}. 

To be more specific, we consider polarized dipolar particles (magnetic atoms or polar molecules) in their lowest energy internal state, which are placed with unit filling in a deep optical lattice, thus suppressing single-particle tunneling. 
The considered dynamical degrees of freedom are therefore encoded in the collective vibrational modes - phonons - originating from the long-range dipole-dipole interactions. 
We consider the specific case of the honeycomb lattice and show how one can access a broad variety of topological effects and states by controlling the polarization direction, thus exploiting the \emph{anisotropy} of the dipole-dipole interaction. 
These include the instability of topological band degeneracies with respect to crystalline deformations, Chern and anomalous Floquet topological bands, a higher-charge monopole with a synthetic dimension, and relativistic Landau levels generated by a strain field.

We stress that, in our setup, the implementation of the topology-related effects and states mentioned above only requires the control of the dipole orientation, without changing the underlying optical lattice geometry. 
Furthermore, the orbital character of the phonons provides an additional degree of freedom that enriches the band structures.
The combination of these two factors offers definite advantages over the use of ensembles of mobile atoms in the lowest band of optical lattices, where the observation of some of these phenomena requires ad-hoc design and distinct experimental techniques. 
In contrast, our proposal provides a unifying platform, a phononic topological toolbox, to explore a large number of topological phenomena in a \emph{single experimental setup}, which have a counterpart in various classical and condensed matter phononic systems \citep{Prodan2009, Bermudez2011, Kane2014, Huber2015, Lubensky2016, Salerno2016, Li2020, Wei2021, Luo2021, Slager2022}. 

The paper is organized as follows: in Sec.~\ref{sec:results}, we introduce the model and derive the phonon dispersion relation; in Sec.~\ref{sec:control}, we analyze the topological properties of the phonon bands and how to control them; in Sec.~\ref{sec:probing}, we discuss several aspects related to the probing methods for phonons; in Sec.~\ref{sec:realization}, we analyze the experimental platform and the energy scales for different dipolar atoms and polar molecules; in Sec.~\ref{sec:conclusions}, we draw our conclusions.


\section{Phonon modes}
\label{sec:results}

\subsection{Model}
\label{sec:model}

Let us consider a system of polarized dipolar particles (either atoms or molecules) on a honeycomb lattice with one atom per site, as depicted in Fig.~\ref{fig:model}, where a deep local potential allows us to neglect single-particle tunneling processes. 
The collective dynamics of the ensemble is determined by the dipole-dipole interactions and it is governed by the Hamiltonian
\be
H = H_0 + V_{\textrm{int}} \,,
\ee
where
\be
\label{eq:H0}
H_0 = \sum_{n\in\cal A, \cal B} \f{\b p_n^2}{2m} + \f 1 2 m \om^2 \sum_{n\in\cal A, \cal B} (\b r_n - \b R_n)^2\,,
\ee
describes the single-particle Hamiltonian, with $\b r_n \!\equiv\! \b r(\b R_n)$ and $\b p_n \!\equiv\! \b p(\b R_n)$ the coordinate and momentum of the particle at the site $n$ of position $\b R_n$, $m$ is the particle mass, and $\om$ the local harmonic oscillator frequency of the potential minima, which we take as isotropic for simplicity of notation. 
We assume the low-temperature condition $k_B T \ll \hbar \om$ such that we neglect the thermal occupation of excited states in each harmonic well. 
The nearest-neighbor dipole-dipole interaction takes the form  
\be
\label{eq:H1}
V_{\textrm{int}} = V_{\textrm{dd}} \sum_{n\in \mathcal B, \nu} \f{|\b r_{n,\nu}|^2 - 3 (\hat{\b d} \cdot \b r_{n,\nu})^2}{|\b r_{n,\nu}|^5}\,,
\ee
where we defined the inter-particle distance $\b r_{n,\nu} = \b r(\b R_n) - \b r(\b R_n+\bs \delta_\nu)$, $V_{\textrm{dd}}$ is the interaction strength, $\bs \delta_\nu$ are the lattice vectors shown in Fig.~\ref{fig:model}(a) and the anisotropy of dipolar interactions is described by the orientation of the dipole moment $\hat{\b d} = (\sin \Theta \cos \Phi, \sin \Theta \sin \Phi, \cos \Theta)$, where the angles are chosen as in Fig.~\ref{fig:model}(b). In particular, for $\Theta \!=\! 0$ the dipoles are oriented orthogonal to the honeycomb lattice plane and the interactions are therefore isotropic.

\subsection{Dispersion relation}
\label{sec:dispersion1}

In order to find the collective vibrational modes, we define the small displacements $\b u^A(\b R_n + \bs \delta_\nu)$, $\b u^B(\b R_n)$,  $n\in \cal B$, with respect to the equilibrium position 
\begin{align}
\b r(\b R_n ) &= \b R_n + \b u^B(\b R_n)\,,\nn\\
\b r( \b R_n + \bs\delta_\nu) &= \b R_n + \bs\delta_\nu + \b u^A(\b R_n + \bs \delta_\nu) \,,
\end{align}
and we obtain the dynamical matrix
\be
\label{eq:dyn_mat}
D_{ij}^{\alpha\beta}(\b R', \b R'')= m\om_i^2 \delta_{ij}\delta_{\alpha\beta} + \f{\pa^2 V_{\textrm{int}}}{\pa u_i^\alpha(\b R') \pa u_j^\beta(\b R'')} \,,
\ee
with $i,j=x,y,z$ the orthogonal directions of oscillation. Notice that we have allowed for different harmonic frequencies $\om_i$ along the three spatial dimensions at each lattice site. The Hamiltonian can therefore be recast in momentum space as 
\be
\label{eq:ham_ksp}
H = \sum_{i,\alpha,\b k} \f{p_{i,\b k}^\alpha \,p_{i,-\b k}^\alpha}{2m} + \f 1 2 \sum_{i,j,\alpha,\beta, \b k} u_{i,\b k}^\alpha D_{ij}^{\alpha\beta} (\b k ) u_{j,-\b k}^{\beta}\,,
\ee
where we only kept quadratic terms in $u_{i,\b k}^\alpha$ and the explicit form of $D_{ij}^{\alpha\beta}(\b k )$ is given in Appendix \ref{sec:dynmat}. The phonon spectrum $\hbar \om_{\b k}$ and the corresponding eigenmodes are then obtained by solving the eigenvalue problem \cite{Bruus}
\be
\b D(\b k) \cdot \b u_{\b k} = m \om_{\b k}^2\, \b u_{\b k}\,.
\ee
Let us now consider a large isotropic trapping frequency $\hbar \om \gg V_{\text{dd}} \ell^2/a^5$ where $\ell = (\hbar/m\om)^{1/2}$ is the harmonic oscillator length at each site. In this case, we can write the eigenfrequencies as 
\be
\label{eq:highfreq}
\om_{\b k} \approx \om + \f{\epsilon_{\b k}}{2m\om}\,,
\ee
where $\epsilon_{\b k}$ are the eigenvalues of $\b D^0 (\b k) \equiv \b D(\b k)|_{\om=0}$. In the rest of this work, we will show the corresponding phonon bands of dipolar atoms on the honeycomb lattice obtained by diagonalizing $\b D^0 (\b k)$ and we will use the quantity $E_{\text{int}} = V_{\text{dd}} \ell^2/2a^5$ as energy unit, where the factor two in the denominator is taken as in Eq.~\eqref{eq:highfreq}.

The phonon dispersion obtained from $\b D^0 (\b k)$ displays six bands in total. The in-plane modes $u_{x,\b k}$ and $u_{y,\b k}$ are however decoupled from the out-of-plane modes $u_{z,\b k}$ when $\Theta = 0$, as it can be directly seen by inspecting the explicit form of the dynamical matrix. However, in the rest of this work we will consider the in-plane and out-of-plane modes as independent also when $\Theta \neq 0$ by assuming $|\om_z - \om_{x,y}| \gg \Delta E_B/\hbar$, where $\Delta E_B$ is the typical bandwidth of the phonon modes. 

\section{Controlling topological effects}
\label{sec:control}

\subsection{Band crossing points for $\Theta=0$}
\label{sec:bandcrosspt}

\begin{figure}[!t]
\center
\includegraphics[width=0.69\columnwidth]{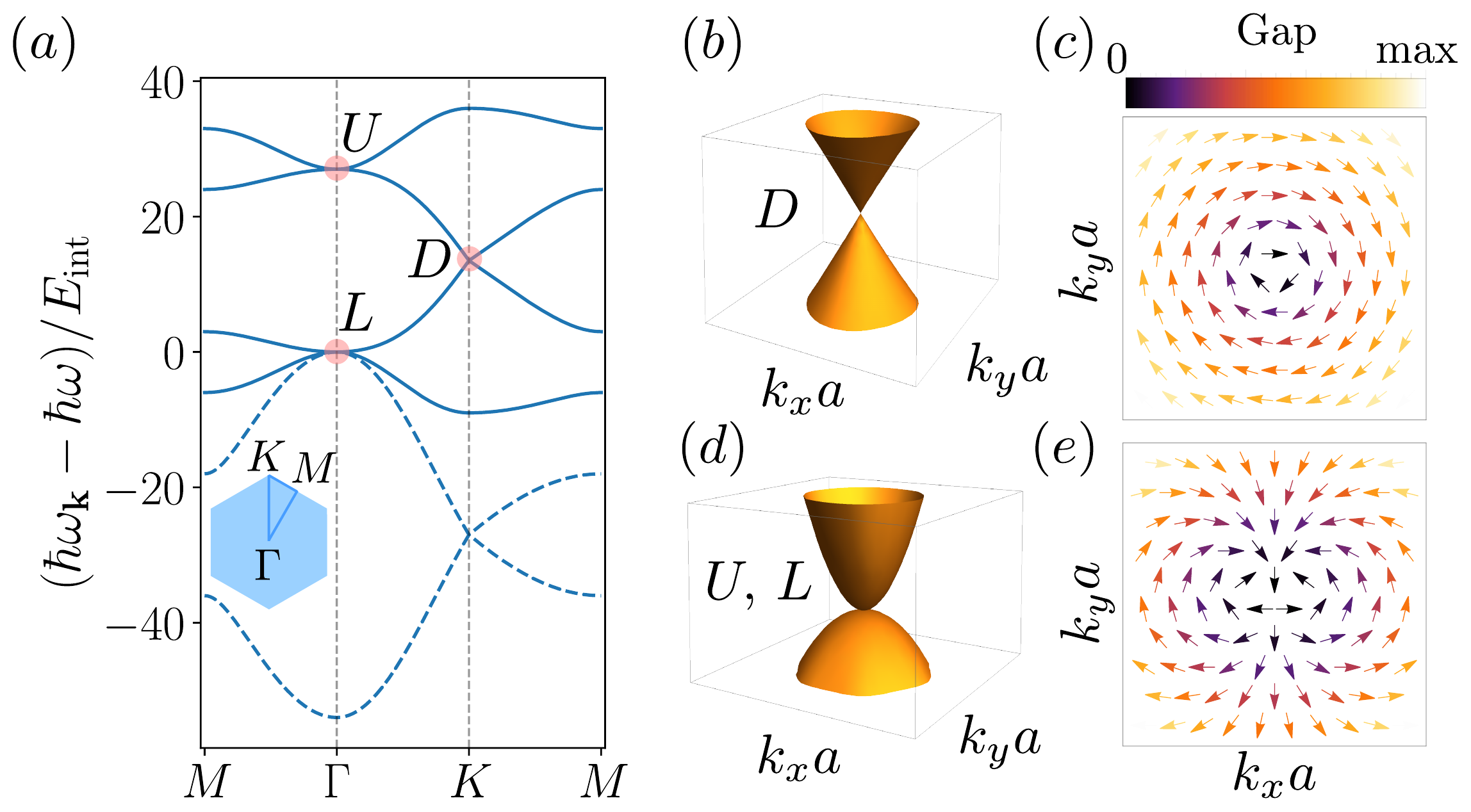}

\caption{Phonon dispersion for $\Theta=0$. (a) Dispersion of the in-plane modes $u_{x, \b k}$, $u_{y, \b k}$ (solid line) and out-of-plane modes $u_{z, \b k}$ (dashed line) for $\Theta=0$, namely with the polarization orthogonal to the honeycomb lattice plane. (b) Dirac dispersion at $\b K=(0,4\pi/3\sqrt 3 a)$ for the in-plane modes and (c) corresponding winding of the $\b d_{\text{2D}}$ vector displaying a winding number $w=1$. (d) Upper quadratic band-touching point at $\bs \Gamma = (0,0)$ and (e) corresponding winding of the $\b d_{\text{2D}}$ vector displaying a winding number $w=2$. The different letters $U$, $L$, $D$ indicate the band crossing points analyzed in the main text.
}
\label{fig:band}
\end{figure}

The full dispersion relation is shown in Fig.~\ref{fig:band}(a) in the regime of isotropic interactions, namely for $\Theta=0$ and $\om_x\!=\!\om_y\!=\!\om_z\!=\!\om$, which displays several band crossing points \cite{Schnyder2018, Feng2021}. The in-plane modes display two types of band crossings. There are two quadratic band touching points (QBTPs) at $\bs \Gamma$ at different energies, whereas there are two Dirac points with a relativistic linear dispersion at $\b K$ and $\b K'=-\b K$. The dispersion of the in-plane phonon modes shares many similarities with the single-particle tight-binding $p$-band spectrum on the honeycomb lattice ~\cite{Wu2008}, which has been observed with polaritons \cite{Amo2014} and electronic lattices \cite{Gardenier2020}, except for the fact that here the upper and lower bands are dispersive instead of flat. The out-of-plane modes, which resemble the $p_z$-orbital modes of graphene \cite{CastroNeto2009}, also display Dirac cones at $\b K$ and $\b K'$. Notice the existence of a triple point degeneracy at the $\bs \Gamma$ point near zero energy for $\Theta = 0$, which originates from the translational symmetry of interactions. However, this band degeneracy has no interesting topological properties as the in-plane modes are decoupled from the out-of-plane modes. Triple band degeneracies at the $\bs \Gamma$ point are common in phononic systems and they can be nevertheless characterized by unconventional topological indices \cite{Park2021}. 

The topological properties of two-band crossings in the vicinity of a momentum $\b k_0$ are universally described by an effective continuum model for the dynamical matrix that reads
\be
\b D^0_{\text{eff}} (\b k_0, \b q) = d_0(\b q) \mathcal I_{2\times2} + \sum_{i=x,y,z} d_i(\b q) \sigma_i \,,
\ee
where $\b q = \b k - \b k_0$ with $|\b q|a \ll 1$, $\sigma_i$ are the Pauli matrices and $\mathcal I_{2\times2}$ is the identity matrix. The derivation of the specific form of the $\b D^0_{\text{eff}} (\b k_0, \b q)$ for the different cases discussed in this work is based on L\"odwin's perturbative method, which we outline in Appendix \ref{sec:efftheo}. 

Let us focus for the moment on the in-plane modes $u_{x,\b k}$ and $u_{y,\b k}$. At the $\b K$ point, the leading term of the effective dynamical matrix is linear in momentum (see Fig.~\ref{fig:band}(b)) and we can therefore neglect quadratic terms originating from higher-order perturbation theory contributions. The resulting effective model reads
\be
\label{eq:Dirac}
\b D^0_{\text{eff},D} (\b K, \b q) = v_F^{} (q_y \sigma_x - q_x \sigma_y ) \,,
\ee
with $v_F^{} = 45 V_{\text{dd}}/4a^4$, which describes a massless Dirac particle in two dimensions. We have omitted a diagonal term $d_0 = 27 V_{\text{dd}}/2a^5$ that shifts the energies but does not change the topological properties carried by the eigenvectors. As a result of the effective theory description \eqref{eq:Dirac}, the phonon modes experience a Berry phase $\pi$ when adiabatically encircling the $\b K$ point, corresponding to a winding $w = 1$ of the vector $\b d_{\text{2D}}=(d_x,d_y)$, see Fig.~\ref{fig:band}(c). Unless time-reversal or inversion symmetries are broken, the theory remains gapless and the winding number or the Berry phase is a well defined topological invariant \cite{Bernevig2013}. Similarly to what happens in graphene, the winding at $\b K'=-\b K$ is opposite to the one at $\b K$.

In order to derive the effective models describing the QBTPs at the $\b \Gamma$ point, it is necessary to perform second order perturbation theory, as they contribute with quadratic terms in momentum, which are the leading terms of a QBTP. 
The dispersion near the upper QBTP, see Fig.~\ref{fig:band}(d), namely near $\epsilon_{\bs \Gamma} = 27 V_{\text{dd}}/a^5$, therefore reads
\be
\label{eq:QBTP}
\b D_{\textrm{eff},U}^0(\bs \Gamma, \b q) = 
\begin{pmatrix}
t_2 (q_x^2-q_y^2) & t_1 q_x q_y   \\
t_1 q_x q_y & -t_2 (q_x^2-q_y^2)
\end{pmatrix}\,,
\ee
with $t_1 = - 45 V_{\text{dd}}/8 a^3$ and $t_2 = t_1/2$. As for the Dirac point, we did not include an overall diagonal contribution that reads $d_0 = 3(144 + 7 |\b q|^2 a^2 )V_{\text{dd}}/16a^5$. A similar expression can be obtained for the lower QBTP around $\epsilon_{\bs \Gamma} = 0$. The effective theory is gapless and the winding of $\b d_{\text{2D}}$ around the $\bs \Gamma$ point is $w \!=\! 2$, as shown in Fig.~\ref{fig:band}(e), unless time-reversal or the discrete $C_3$ rotational symmetries are broken \cite{Sun2009}. We will explore how to break the symmetry protection of the degeneracy points in the next sections.

\subsection{Tilting the polarization axis: $C_3$ crystal symmetry breaking}
\label{sec:tilt}

\begin{figure}[t]
\center
\includegraphics[width=0.69\columnwidth]{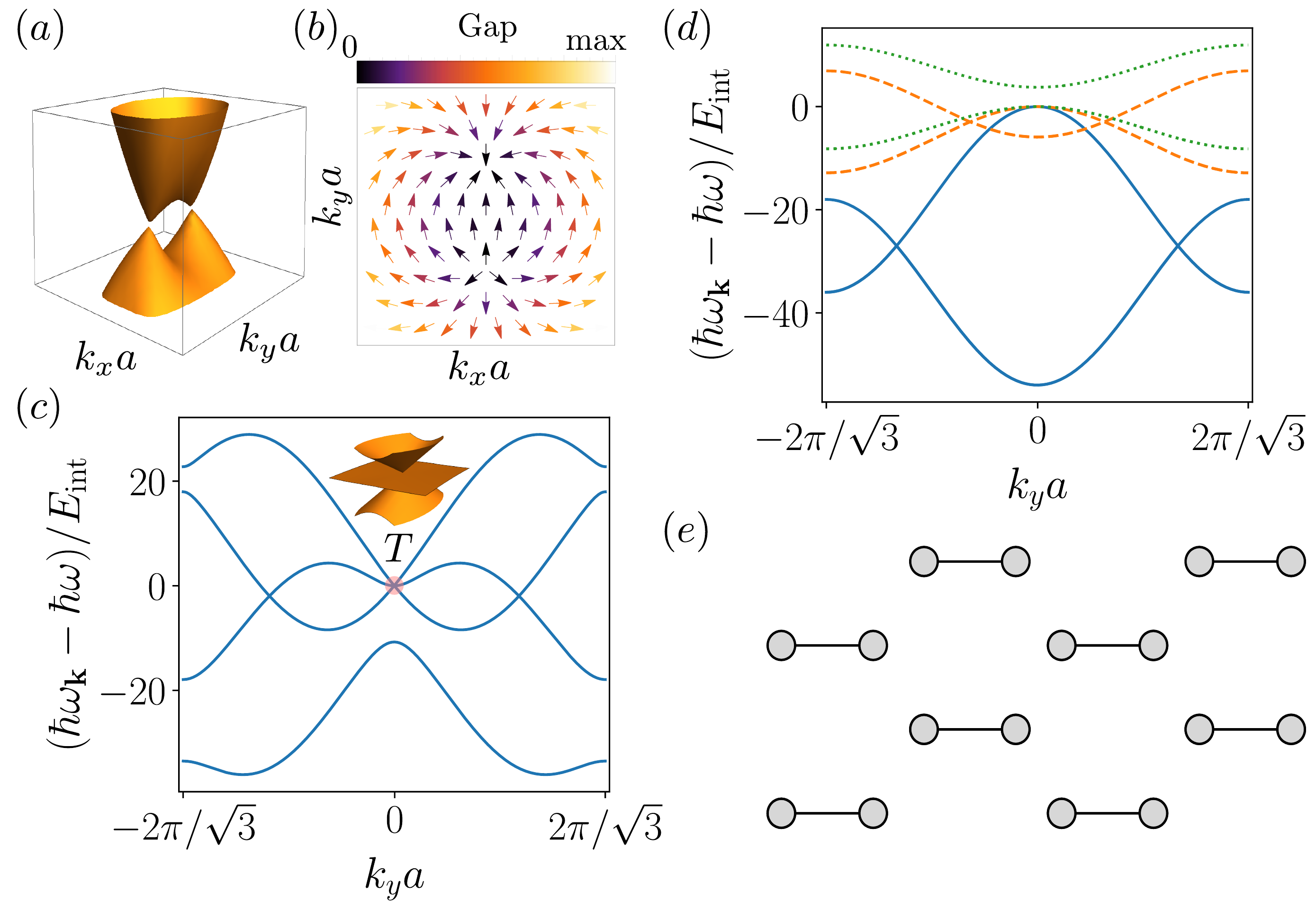}

\caption{Effects of interactions anisotropy. (a) Splitting of the upper QBTP shown in Fig.~\ref{fig:band}(d) for $\Theta \!=\! 0.03 \pi$ into two Dirac cones and (b) the corresponding winding of the $\b d_{\text{2D}}$ vector. (c) Dispersion of the in-plane modes for $k_x = 0$ at the critical angle $\Theta_c=\arccos(-3/5)/2$. Inset shows the triple band-crossing point near the $\bs{\Gamma}$ point. (d) Dispersion of out-of-plane modes for $k_x = 0$ as a function of $\Theta$: (solid line) $\Theta=0$, (dashed line) $\Theta = 0.28\pi$ and (dotted line) $\Theta = 0.32\pi$ showing that Dirac cones merge and then gap out. (e) Disconnected dimers regime for $\Theta=\Theta_{2c}$ for the out-of-plane modes.
}
\label{fig:tilt}
\end{figure}

The anisotropy of dipolar interactions provides a knob to manipulate the phonon band structure by breaking certain crystal symmetries that are present when the polarizing field is orthogonal to the lattice plane. In particular, let us consider a tilt of the polarization axis by an angle $\Theta$, while keeping $\Phi=0$, which breaks the discrete $C_3$ rotational symmetry. For small $\Theta\ll 1$, the dynamical matrix for the upper QBTP, which we discussed in the previous section, has a correction
\be
\Delta \b D_{\textrm{eff},U}^0 = \Theta^2 \left( \Delta d_0 \, \mathcal I_{2\times2} + \Delta d_z \sigma_z\right)\,,
\ee
where $\Delta d_0 = -81V_{\text{dd}}/2a^5$ and $\Delta d_z \!=\! 27 V_{\text{dd}}/4 a^5$. As shown in Fig.~\ref{fig:tilt}(a), the QBTP then splits into two Dirac cones as a consequence of the momentum independent term $\Delta d_z \sigma_z$ \cite{Montambaux2018}. The winding of $\b d_{\text{2D}}$ for $\Theta \neq 0$ is shown in Fig.~\ref{fig:tilt}(b), which clearly displays a winding $w=1$ around each Dirac cone such that the winding $w=2$ of the QBTP is conserved.

The lower QBTP at $\epsilon_{\bs \Gamma} = 0$ behaves differently. A direct calculation shows that a small angle $\Theta$ does not affect the effective theory and the touching point is therefore protected for a small anisotropy of the dipolar interaction.   
However, for larger values of $\Theta$, the bands are more severely affected. Among the four eigenvalues at $\bs \Gamma$, two of them are always zero, $\epsilon_{\bs \Gamma}^{1,2} = 0$. The other two eigenvalues read $\epsilon_{\bs \Gamma}^3 = 27 V_{\text{dd}}(3+5\cos2\Theta)/8a^5$ and $\epsilon_{\bs \Gamma}^4 = 27 V_{\text{dd}} (1+7\cos2\Theta)/8a^5$. This means that there are two tilt angles that provide a triple degeneracy at $\epsilon_{\bs \Gamma} = 0$. In particular, we focus on the angle $\Theta_c = \arccos(-3/5)/2 \approx 0.352 \pi$. The spectrum is shown in Fig.~\ref{fig:tilt}(c) and it is captured by the effective theory
\be
\label{eq:spin1}
\b D_{\textrm{eff},T}^0(\bs \Gamma, \b q) = 
\begin{pmatrix}
0 & -i v_x q_x & i v_y q_y  \\
i v_x q_x & 0 & -i v_y q_y  \\
-i v_y  q_y & i v_y q_y & 0
\end{pmatrix}\,,
\ee
where $v_x \!=\! 27 V_{\text{dd}}/2a^4$ and $v_y \!=\! v_x/\sqrt 2$, which describes a two-dimensional (pseudo)spin-1 relativistic particle. Indeed, the model above can be rewritten as $\b D_{\textrm{eff},T}^0(\bs \Gamma, \b q)$ $=$ $\sum_{i=x,y,z}d_i(\b q) S_i = \b{d}_i(\b q)\cdot\b S$, where $\b S = (S_x, S_y, S_z) = (\lambda_2, \lambda_5, \lambda_7)$ is a spin-1 representation of the SU(2) algebra satisfying $[S_i, S_j] = i\, \epsilon_{ijk}S_k$, $\lambda_i$ are Gell-Mann matrices, whereas $d_x=v_x q_x$, $d_y=-v_y q_y$ and $d_z = v_y q_y$. A discussion of the topological characterization of a general spin-1 model in 2D has been carried out in Ref.~\cite{Varma2012}, where it is applied to the Lieb lattice case. For the time-reversal invariant regime that we are considering here, a topological invariant associated to each band is the Berry phase around the degeneracy point. A direct calculation, however, shows that the Berry phase of each band is trivially zero. In Sec.~\ref{sec:monopole}, we will however show that this triple band-crossing point can be exploited to construct a higher-dimensional higher-charge monopole by using the angle $\Phi$ and the harmonic trapping as additional degrees of freedom.

We conclude this section by studying the effect of the polarization tilt on the out-of-plane phonons $u_{z,\b k}$. These modes are described by a graphene-like Hamiltonian that reads
\be
\label{eq:offplane}
\b D^0_z(\b k) = \begin{pmatrix}
- V(0) & V(\b k) \\ 
V(\b k)^\dag & - V(0)
\end{pmatrix}\,,
\ee
where $V(\b k) = -J_1 e^{i \b k \cdot \bs \delta_1} - J_2 e^{i \b k \cdot \bs \delta_2} - J_3 e^{i \b k \cdot \bs \delta_3}$ and 
\begin{align}
\label{eq:J_dipol}
J_{1,2} &= -\f{3 V_{\text{dd}}}{8 a^5} \left(11+ 13 \cos 2\Theta \right)\,, \nn \\
J_{3} &= -\f{3 V_{\text{dd}}}{2 a^5} \left(-1+ 7 \cos 2\Theta \right) \,.
\end{align}
The corresponding dispersion relation is shown in Fig.~\ref{fig:tilt}(d) for $\Theta = 0$ (solid line). Applying a tilt to the polarization along one of the symmetry axis of the lattice, namely changing $\Theta$ while keeping $\Phi=0$, creates an anisotropy among the couplings, $J_3 \neq J_{1,2}$. This effect shifts the Dirac cones position (and the overall energy of the band) until the Dirac cones merge at the $\bs \Gamma$ point for $\Theta=\Theta_{c1} = -\arccos(-1/3)/2 \approx 0.304\pi$. For $\Theta > \Theta_{1c}$, the spectrum gaps out since the two Dirac cones have opposite winding \cite{Montambaux2009,Montambaux2018}, as shown in Fig.~\ref{fig:tilt}(d). Such a merging transition has been observed with ultracold atoms \cite{Tarruell2012}, photonic \cite{Bellec2013, Rechtsman2013,Amo2019} and electronic systems \cite{Su2015}.


Furthermore, at a larger value of the tilt angle $\Theta_{c2} = \arccos(-11/13)/2\approx 0.410\pi$, we instead find a complete suppression of the couplings, $J_1(\Theta_{c2}) = J_2(\Theta_{c2}) = 0$, thus leaving only disconnected dimers coupled by $J_3(\Theta_{c2})\neq 0$ (see Fig.~\ref{fig:tilt}(e)). This regime opens the opportunity to realize an anomalous Floquet topological system \cite{Kitagawa2010}, by tilting the dipolar axis along the three crystal symmetry axes in a clockwise or counterclockwise sequence: at each time step, the $u_{z,\b k}$ modes are only coupled as dimers, like in Fig.~\ref{fig:tilt}(d), and the bulk dynamics is fully localized. This out-of-equilibrium phase has vanishing Chern number in the bulk bands but nevertheless topologically protected edge modes. The ideal protocol where at each time step only disconnected dimers are present and the corresponding Floquet quasienergy bands are flat, as proposed by Kitagawa et al.~\cite{Kitagawa2010}, has been realized with optical waveguides \cite{Mukherjee2017,Szameit2017}. 
The anomalous Floquet regime has also been accessed with ultracold atoms in optical lattices by appropriately deforming the relative position of the optical lattice potential minima \cite{Wintersperger2020}. However, the ideal flat band case remains a more challenging scenario to be realized with ultracold atoms \cite{Quelle2017}.
The degree of control of the nearest-neighbor couplings for the phonon modes of dipolar atoms therefore paves the way to achieve the flat-band anomalous Floquet regime, where the effect of interactions can be enhanced and thus play a nontrivial role.

\subsection{High-frequency rotation of the polarization axis: time-reversal symmetry breaking}
\label{sec:rotation}

Besides breaking crystalline symmetries, the control over the tilt angle can also be exploited to break time-reversal symmetry by applying a fast rotation of the polarization axis with respect to the $\hat z$ axis. As we shall show below, this method can be used to mimic the effect of a synthetic magnetic field on the phonon modes, as realized for neutral atoms by circularly shaking the lattice \cite{Cooper2019}. 

\begin{figure}[!t]
\center
\includegraphics[width=0.69\columnwidth]{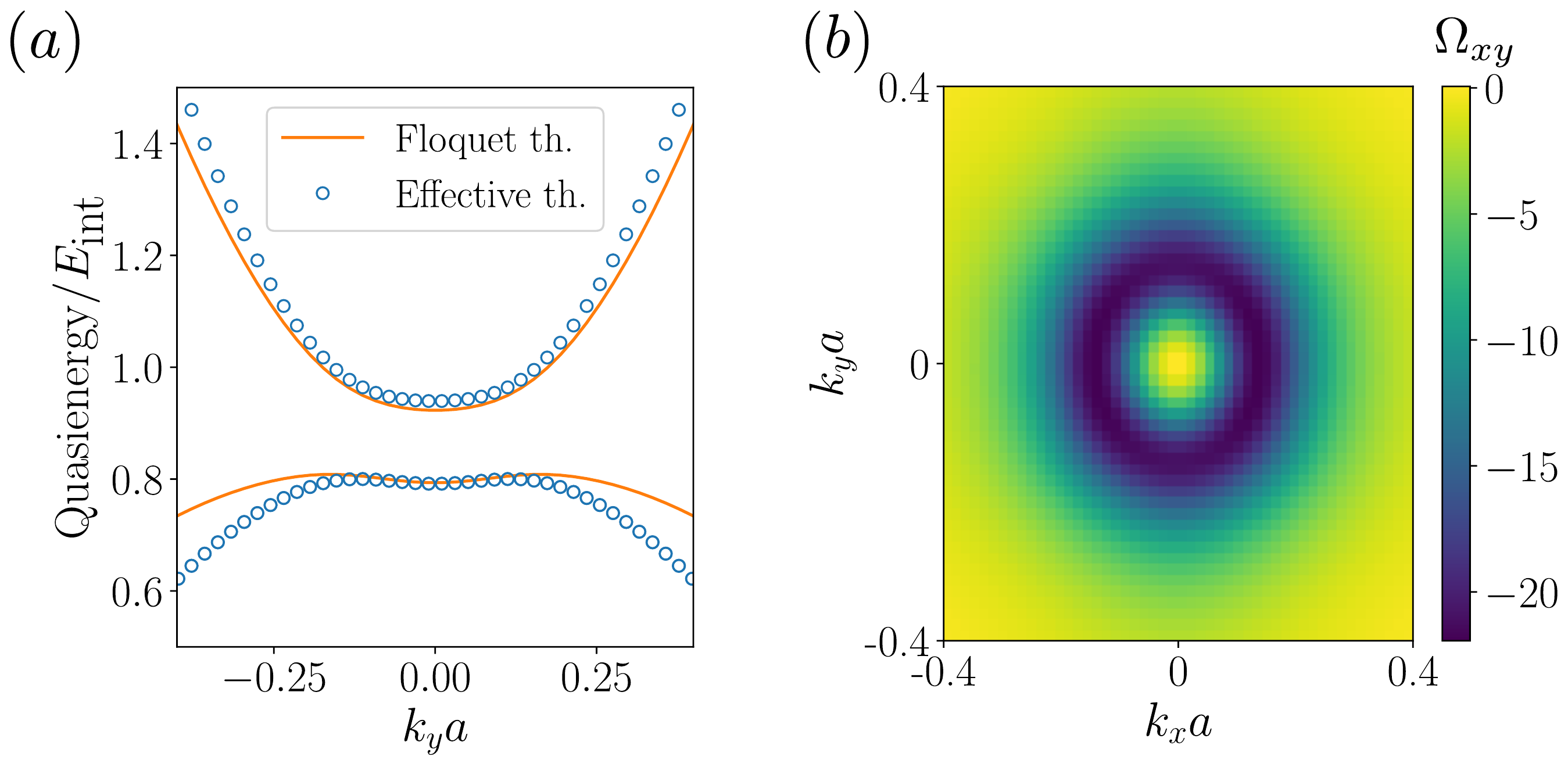}

\caption{Time-reversal symmetry breaking for the QBTP. (a) Quasienergy spectrum near the upper QBTP for $\Theta = 0.127\pi$ and $\Omega = 4 \, E_{\text{int}}/\hbar$ computed by diagonalizing the full time-evolution operator (solid line) and the effective  Floquet theory obtained in perturbation theory (empty circles). (b) Berry curvature of the lowest band with the typical annular shape of a QBTP. 
}
\label{fig:floquet}
\end{figure}

Let us consider a polarization axis tilted by a fixed small angle $\Theta\ll 1$ and apply a time-dependence on the angle $\Phi = \Omega t$, where $\Omega$ is the rotation frequency, which generates a fast precession of the polarization axis. This method has already been used to control the dipole-dipole interaction strength and sign \cite{Pfau2002,Lev2018}, under the adiabatic assumption that the rotation frequency is small as compared, for example, to the Zeeman splitting in the case of dipolar magnetic atoms.  We assume the condition $\hbar \Omega \gg \Lambda$ to be valid, with $\Lambda$ playing the role of an energy cut-off for the time-dependent theory that allows us to apply a high-frequency Floquet expansion. Let us focus on the upper QBTP at the $\bs \Gamma$ point, since the lower QBTP is not affected by small tilting angles, as discussed above. By performing a high-frequency expansion, we obtain the effective Floquet (time-independent) dynamical matrix, see Appendix \ref{sec:floquetheo} for further details, 
\be
\label{eq:floq}
\b D_{\textrm{eff},U}^F \approx \b D_{\textrm{eff},U}^{(0)} + \Delta \b D_{\textrm{eff},U}^F \,, 
\ee
where $\b D_{\textrm{eff},U}^{(0)} $ is the lowest order term when $\Om\ra\infty$ and $\Delta \b D_{\textrm{eff},U}^F$ is the leading correction in $1/\Omega$. The first term, $\b D_{\textrm{eff},U}^{(0)} $, has the same form as Eq.~\eqref{eq:QBTP} with parameters: $t_1 \!=\! - 45 V_{\text{dd}}/8 a^5$, $t_2\!=\!t_1/2$ and $d_0 \!=\! 3 \left[ 72(2\!-\!3\Theta^2\!+\!\Theta^4) + 7 |\b q|^2a^2 \right] V_{\text{dd}} / 16 a^5$. We have kept terms up to $\Theta^4$ as they will appear in the next-to-leading term and assumed that they are of the same magnitude as the terms $q^2 a^2$ that characterize the QBTP. Therefore, we have neglected terms $q^2 a^2 \Theta^2$, which are perturbatively negligible under the previous assumption.
Time-reversal symmetry is broken if $\Delta \b D_{\textrm{eff},U}^F$ contains a term proportional to $\sigma_y$ that is even in $\b q$. Indeed, a direct calculation yields
\be
\Delta \b D_{\textrm{eff},U}^F = \sigma_y \f{729\, \Theta^4}{32} \f{V_{\text{dd}}^2}{a^{10}} \f{(2m\om)^{-1}}{2\hbar\Om}\,,
\ee
which confirms that the rotation of the dipolar axis has the effect of breaking time-reversal symmetry and thus opens a gap at the $\bs \Gamma$ point. This result is validated numerically in Fig.~\ref{fig:floquet}(a), where we compare the effective Floquet theory derived above with the numerical quasienergy spectrum obtained by the full time-dependent dynamical matrix $\b D^0(\b k, t)$. The corresponding Berry curvature $\Omega_{xy}(\b k)$ calculated from the lower band eigenvectors is shown in Fig.~\ref{fig:floquet}(b) and has the typical annular shape expected when gapping a QBTP. After integrating over momentum, one finds a Chern number $\nu = \int \text d^2\b k\, \Omega_{xy}(\b k)/2\pi = -1$, where the integration area has been extended to a full 2D plane in momentum space because the Berry curvature is localized in the vicinity of the $\bs\Gamma$ point. A similar analysis can be performed for the $\b K$ and $\b K'$ points, yielding a theory of the form \eqref{eq:Dirac} for $\b D^F_{\text{eff}} (\b K, \b q) $. The perturbative correction from Floquet theory is in this case
\be
 \Delta \b D^F_{\text{eff}} (\b K, \b q) = \sigma_z \f{99225\,  \Theta^4}{64} \f{V_{\text{dd}}^2}{a^{10}} \f{(2m\om)^{-1}}{2\hbar\Om} \,,
\ee
which also breaks time-reversal symmetry for the Dirac model and opens a gap. A similar analysis can be carried out at $\b K'$ and the resulting effective theory has the same form, so the Berry curvature of the two valleys adds up to $\nu = -1$ for the upper band. Notice that the upper band contribution at the Dirac points and the lowest band contribution at the $\bs \Gamma$ point actually belong to the same band (see Fig.~\ref{fig:band}). Therefore, the total Chern number is $\nu = -2$. 

Let us now summarize the Chern numbers of the resulting bands. In total, we have three set of bands: an upper band with $\nu = +1$, with Berry curvature around the $\bs \Gamma$ point; a middle band with $\nu = -2$ with Berry curvature around the $\bs \Gamma$ and the $\bs K$, $\bs K'$ points; a lower set of two bands with $\nu = +1$ with Berry curvature around the $\bs K$, $\bs K'$ points. Recall that the lower QBTP is not affected by the dipolar axis rotation and therefore the two lowest bands remain degenerate at $\bs \Gamma$.


\begin{figure}[!]
\center
\includegraphics[width=0.69\columnwidth]{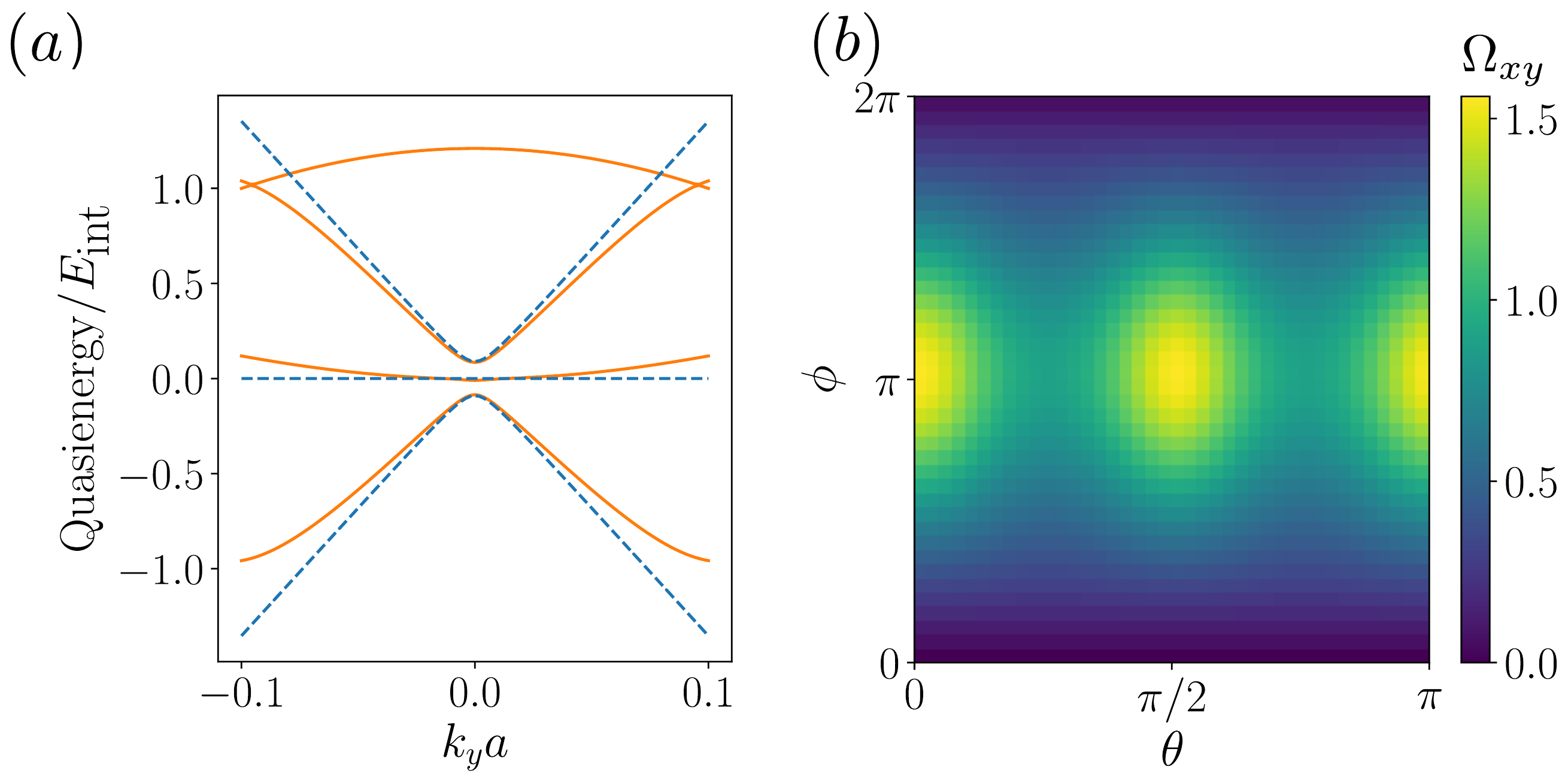}

\caption{Higher-charge monopole using a synthetic dimension. (a) Exact quasienergy spectrum near the triple point for $\Theta = \Theta_c = \arccos(-3/5)/2 $, $\Phi_0 = 0.5$, $\alpha = \pi/2$, $W_0=0.5\, E_{\text{int}}a^{-2}$,  $\Omega = 3 \, E_{\text{int}}/\hbar$ (solid line) and effective perturbative Floquet theory (dashed lines). (b) Berry curvature of the lowest band on a sphere enclosing the singularity point $q_1=q_2=q_3=0$. The corresponding monopole charge and Chern number is $\nu = 2$.
}
\label{fig:floquettriple}
\end{figure}

\subsection{Modulation of the harmonic frequency: Higher-charge monopole using a synthetic dimension}
\label{sec:monopole}

In Sec.~\ref{sec:tilt}, we found that the in-plane phonon dispersion relation displays a peculiar behaviour at a critical angle $\Theta_c$ where it describes relativistic excitations of pseudo-spin $S\!=\!1$. As anticipated, the band topological properties are trivial, as a direct calculation of the Berry phase of each band shows. However, the model \eqref{eq:spin1} provides the starting point to engineer a topological monopole in 3D, as we discuss here below. In order to obtain this result, we will exploit the concept of synthetic dimension \cite{Celi2014, Price2019}. 

Let us consider the correction to the dynamical matrix near the triple point obtained when changing the angle $\Phi$, with $\Phi \ll \pi$, and the trapping frequency along the $x$ direction $\om_x \ra \om + \delta \om_x$, with $\delta\om_x\ll \om$, while keeping $\om_y = \om$ fixed. We obtain 
\begin{align}
\label{eq:spin1dtheta}
\Delta \b D_{\textrm{eff},T}^0 &= 
\begin{pmatrix}
0 & 0 & v\, \Phi  \\
0 & 0 & v\, \Phi  \\
v\, \Phi  & v\, \Phi & 2 m\, \om\,\delta \om_x
\end{pmatrix} = \f{W}{3}\mathcal I - \f{W}{\sqrt 3} \lambda_8 +v\, \Phi (\lambda_4 + \lambda_6)\,,
\end{align}
where in the second equality we wrote $\Delta \b D_{\textrm{eff},T}^0 $ in terms of Gell-Mann matrices and we defined $v=27 V_{\textrm{dd}}/5\sqrt 2 a^5$ and $W=2 m\, \om\,\delta \om_x$. By noticing that the following commutation relation holds, $i [\lambda_8, \lambda_4+\lambda_6] = -\sqrt 3 (\lambda_5 + \lambda_7)$, one realizes that perturbative Floquet theory will preserve the SU(2) algebraic structure of the time-independent model $\b D_{\textrm{eff},T}^0$, Eq.~\eqref{eq:spin1}, which is described by $(\lambda_2, \lambda_5, \lambda_7)$, as discussed in Sec.~\ref{sec:tilt}. Another crucial observation is that the commutator calculated above yields $\lambda_5$ and $\lambda_7$ with same sign, whereas they appear with opposite sign in $\b D_{\textrm{eff},T}^0$ (see Eq.~\eqref{eq:spin1}). 

Let us then apply a time modulation of the form $\Phi = \Phi_0 \cos (\Omega t+\alpha)$ and $W = W_0 \cos \Omega t$. We thus find an effective Floquet theory
\be
\b D_{\textrm{eff},T}^F \approx \b D_{\textrm{eff},T}^0 - \f{(2m\om)^{-1} W_0\Phi_0 v}{2\hbar \Om} \sin\alpha\, (\lambda_5+\lambda_7)\,.
\ee
We can now define $q_1 = v_x q_x$, $q_2 = v_y q_y$ and $q_3 = (2m\om)^{-1} \sin \alpha \, W_0 \Phi_0 v/2\hbar\Om$ and obtain 
\be
\label{eq:spin1-3D}
\b D_{\textrm{eff},T}^F = 
\begin{pmatrix}
0 & -i q_1 &  i (q_2+q_3)  \\
i q_1 & 0 & -i (q_2-q_3)  \\
-i (q_2+q_3) & i (q_2-q_3) & 0
\end{pmatrix}\,,
\ee
which corresponds to the Hamiltonian of a relativistic particle of pseudo-spin $S\!=\!1$ in the three-dimensional parameter space $(q_1,q_2,q_3)$ \footnote{One can actually redefine $q_2 + q_3 \ra - q_2$ and $q_2 - q_3 \ra q_3$ to obtain a more symmetric structure of the matrix that corresponds to the model introduced in \cite{Bradlyn2016}, but here we have decided to keep the current parametrization to avoid further manipulation of the model since each parameter $q_i$ can be controlled by a single experimental parameter.}, as introduced in Ref.~\cite{Bradlyn2016}. Similar types of dispersions \cite{Chamon2010,Fulga2017,Zhu2017,Yang2018,Zhang2018} can describe nontrivial modes beyond the conventional massless Dirac/Weyl representation \cite{Armitage2018}. In particular, here we have a monopole in momentum space with charge $\nu\!=\!2$, which provides a Berry curvature flux on an arbitrary surface enclosing the triple band degeneracy at $q_1 = q_2 = q_3=0$. This can be easily verified by going to polar coordinates, $q_1 = r \sin\theta \cos\phi$, $q_2 = r \sin\theta\sin\phi$, $q_3=r\cos\theta$. Let us then calculate the Berry curvature on the sphere enclosing the singularity for the lowest band, which in polar coordinates reads \cite{Niu2010}
\be
\label{eq:bcurv}
\Omega_{\theta\phi} = i \sum_{n > 0} \f{ \bra u_0 | \pa_\theta \b D_{\textrm{eff},T}^F | u_n\ket \bra u_n | \pa_\phi \b D_{\textrm{eff},T}^F | u_0\ket  - (\theta \leftrightarrow \phi ) }{ (\epsilon_n-\epsilon_0)^2}\,,
\ee
where we have indicated the eigenvalues of $\b D_{\textrm{eff},T}^F$ as $\epsilon_{0,2} = \mp \sqrt{q_1^2 + (q_2+q_3)^2+(q_2-q_3)^2}$ and $\epsilon_1=0$, whereas $|u_m\ket$ indicate the corresponding eigenvectors. The Chern number, namely the charge that is the source of Berry flux, can then be straightforwardly obtained as $\nu=(1/2\pi)\int_{S^2} \mathrm d\theta\mathrm d \phi \,\Omega_{\theta\phi} = 2$, whereas for the upper band one finds $\nu=-2$ and a vanishing Chern number for the middle band \cite{Bradlyn2016}. 

We have verified these results by first comparing the exact Floquet phonon quasienergies with the effective model \eqref{eq:spin1-3D}, which is shown in Fig.~\ref{fig:floquettriple}(a). Furthermore, we have numerically computed the Berry curvature $\Omega_{\theta\phi}$ from the Floquet eigenstates of the full time-dependent problem and the result is shown in Fig.~\ref{fig:floquettriple}(b). Numerical integration gives $\nu=2$, in agreement with the result from the effective theory \eqref{eq:spin1-3D}. A few experiments have realized synthetic monopoles in dimensions $d\!>\!2$ with ultracold atoms, for example a Weyl point in 3D \cite{Pan2021} or a Yang monopole in 5D \cite{Spielman2018}, by exploiting internal atomic levels as parameter space. However, to the best of our knowledge no realization of a 3D monopole of higher charge ($\nu=2$), as the one discussed here by combining real ($q_1$, $q_2$) and synthetic ($q_3$) dimensions, has been achieved so far.

\subsection{Gradient of the polarizing external field: strain and relativistic Landau levels}
\label{sec:strain}

In solid state systems as graphene, the possibility to stretch a material along certain directions by applying strain offers the opportunity to observe interesting physical phenomena. The application of strain has the effect of deforming the crystal structure and it therefore changes the lattice spacings, which become space dependent. As a result, under specific strain configurations the effective theory near the Dirac cones resembles the one of a two-dimensional relativistic particle in a magnetic field, which displays gapped relativistic Landau levels \cite{Guinea2010}. Similar physics has been observed, for instance, in optical systems \cite{Rechtsman2013a, Jamadi2020} where the lattice spacing can be engineered at will in the fabrication process. The simulation of strain with ultracold atoms instead has not been realized so far, although theoretical proposals are already available \cite{Pekker2015,Jamotte2022}. The purpose of this section is to show how the physics of strain can be explored with dipolar particles. In particular, we show how the gradient of the external polarizing (electric or magnetic) field can be used to simulate a strain field for the out-of-plane modes $u_{z,\b k}$ and observe the corresponding (pseudo)-Landau levels. 

For simplicity, we exploit the angle dependence of the couplings in Eq.~\eqref{eq:J_dipol} by assuming that the orientation of the polarizing external field is space dependent as $\Theta = \Theta_0 +\tau\, x / a$, where $\Theta_0 =\pi/4$ and $\tau$ is a dimensionless parameter quantifying the angle gradient. In this case, the couplings take the approximate form
\begin{align}
\label{eq:Jlinear}
J_{1,2} &= -\left(\f{33}{8} - \f{39\tau}{4a} x \right)\f{V_\textrm{dd}}{a^5}\,, \nn \\
J_{3} &= -\left(-\f{3}{2} - \f{21\tau}{a} x \right)\f{V_\textrm{dd}}{a^5}\,.
\end{align}
The space independent anisotropic contribution provides a shift of the Dirac cones, as we have already shown in Fig.~\ref{fig:tilt}(d), which are now located at $\b K_s = (0, \pm2\arccos(2/11)/\sqrt{3}a)$. The space dependent part instead generates the strain field, which translates into a gauge potential near the Dirac points. Indeed, by expanding near $\b K_s$, we obtain the low-energy theory
\begin{align}
\label{eq:dirac_LL}
\b D^0_{\text{eff},z}(\b K_s, \b q) =&\, -v_F^x (q_x+\mathcal A_x) \sigma_y - v_F^y(q_y + \mathcal A_y) \sigma_x - ( 27 V_{\text{dd}}/4a^5 - \mathcal E_x x) \,\mathcal I_{2\times2}  \,,
\end{align}
where $\b A = (0, 270\, \tau x/11a^2)$ is an emergent vector potential in the Landau gauge corresponding to a pseudo-magnetic field $\mathcal B=\pa_x \mathcal A_y - \pa_y \mathcal A_x = 270\tau/11a^2$, $\mathcal E_x=81\tau V_{\text{dd}}/2a^6$ is an emergent (pseudo)electric field, $v_F^x= 9 V_{\text{dd}}/4a^4$ and $v_F^y=9\sqrt{39}V_{\text{dd}}/8a^4$ are anisotropic velocities. Note that in deriving the low-energy theory we have neglected the spatial dependences of the Fermi velocities \cite{Salerno2015}.    
We have therefore found that near the Dirac cone at $\b K_s$ the model describes a particle in the presence of a (pseudo)magnetic and a (pseudo)electric field. Near the other cone at $-\b K_s$, the opposite (pseudo)magnetic field will appear, as time reversal is not broken \cite{Goerbig2011}. 

\begin{figure}[!t]
\center
\includegraphics[width=0.69\columnwidth]{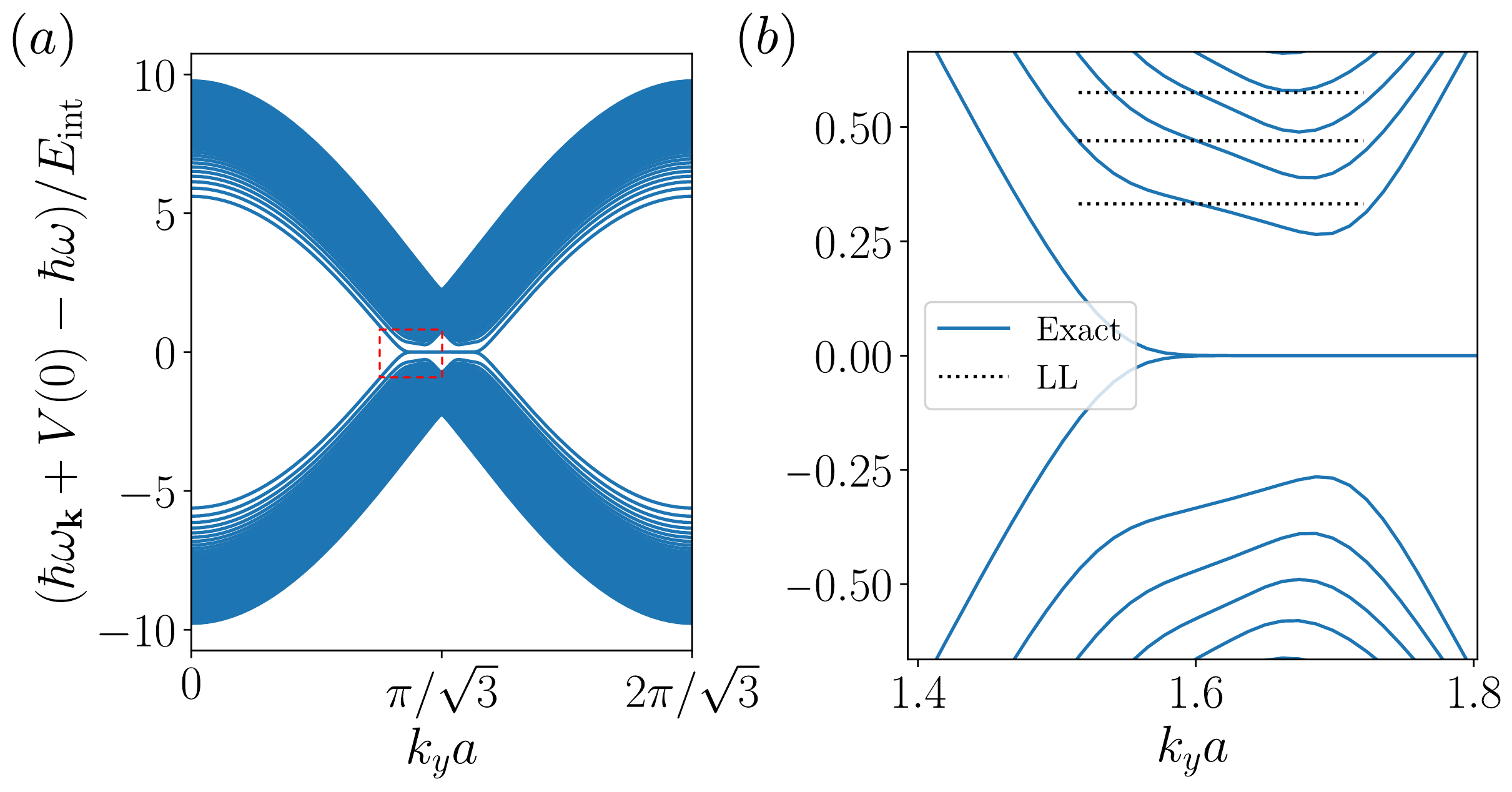}

\caption{Simulating relativistic Landau levels from strain. (a) Spectrum of out-of-plane modes with $\Theta_0=\pi/4$ and $\tau=0.001$ on a lattice that is infinite in the $y$ direction and has $N_x = 50$ unit cells in the $x$ direction. We have subtracted the diagonal term $V(0)$. (b) Zoom of the spectrum shown in panel (a) near one of the Dirac points showing (solid line) the exact result, where the couplings are parametrized as in Eq.~\eqref{eq:J_dipol}, and (dashed line) the Landau levels $\epsilon^{\textrm{LL}}_n$ from Eq.~\eqref{eq:enLL} for $n=1,2,3$.
}
\label{fig:strain}
\end{figure}

The (pseudo)electric field $\mathcal E_x$ is quite strong. Indeed, by direct inspection one obtains the value $\mathcal E_x /v_F\mathcal B >1$, where $v_F=\sqrt{v^x_F v^y_F}$ is the Fermi velocity geometric mean. As a result, the drift velocity $v_D = \mathcal E/\mathcal B$ is larger than the mean Fermi velocity $v_F$, thus destroying the Landau level structure of the spectrum originating from the (pseudo)magnetic field. An overview of the different regimes is provided in Ref.~\cite{Goerbig2011}. The most convenient strategy to observe Landau levels physics in our setup is therefore to compensate for $\mathcal E_x$, \emph{e.\,g.} by applying an external linear force of opposite sign to the phonons. We discuss in the next section how such a force can be engineered.
When $\mathcal E_x = 0$, the spectrum of the low-energy theory \eqref{eq:dirac_LL} takes the form of relativistic Landau levels that reads
\be
\label{eq:enLL}
\epsilon^{\text{LL}}_n = \pm \sqrt{v^x_F v^y_F \mathcal B |n|}\,,
\ee
where we have omitted an overall energy shift $ -27 V_{\text{dd}}/4a^5$. 
In Fig.~\ref{fig:strain}(a), we show the exact spectrum for a system described by the hopping amplitudes \eqref{eq:Jlinear} with $\tau = 0.001$ on a lattice that is infinite in the $y$ direction and has $N_x=50$ unit cells in the $x$ direction. In order to compare with the effective Landau level picture discussed above, we subtracted the space-dependent contribution $V(0) = -(J_1+J_2+J_3)$ (see Eq.~\eqref{eq:offplane}), which corresponds to taking $\mathcal E_x = 0$ in the low-energy theory.
In Fig.~\ref{fig:strain}(b), we compare the dispersion with the analytical result \eqref{eq:enLL}. From the inspection of the spectrum, we see that the energy levels display a momentum dependence, which originates from a spatial dependence of the Fermi velocities that we dropped in the derivation of the effective low-energy theory \cite{Salerno2015}. 


The previous results are obtained by considering a small gradient of the magnetic field around the tilt angle $\Theta_0 = \pi/4$, which was chosen to minimizes the corrections beyond linear order in Eqs.~\eqref{eq:Jlinear}. However, large tilt angles reduce the overall bandwidth and bring the Dirac points close to each other near the gapping transition, as shown in Fig.~\ref{fig:tilt}(d). To circumvent these possible limitations, it could therefore be convenient to reduce the tilt angle in the experimental setting in order to increase the overall bandwidth and the Landau levels energy spacing. In that case, however, the corrections to Eqs.~\eqref{eq:Jlinear} will be of quadratic order in $x$ instead of cubic order, which occurs for the case $\Theta_0 = \pi/4$ discussed here. The strong anisotropy, may also affect the wavepacket dynamics, as we shall see in the next section. A distinct possibility to reach a more ideal regime where to observe Landau level physics could be to start from a highly anisotropic honeycomb lattice from the beginning with $|\bs\delta_3|>|\bs\delta_{1,2}|$ and restore the $C_3$ crystal symmetry through the polarization tilt. By following this strategy, one could therefore optimize the bandwidth, suppress unwanted strong anisotropic effects and have a sufficiently homogeneous pseudo(magnetic) field.


\begin{figure}[!t]
\center
\includegraphics[width=0.69\columnwidth]{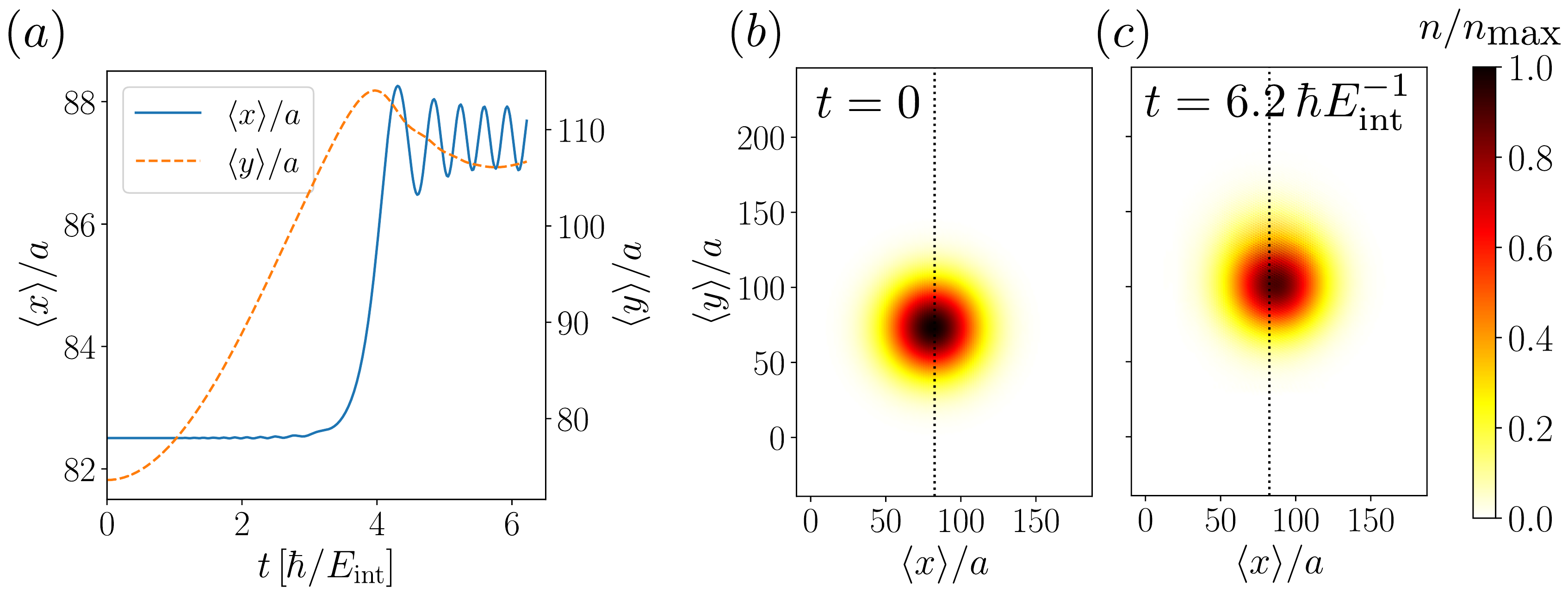}

\caption{Hall deflection of a phonon wavepacket. (a) Semiclassical wavepacket dynamics for the out-of-plane modes with $\b F = (0,F_y)$, $F_y = 0.63 \,E_{\text{int}}/a$ and width $\sigma=30a$. An onsite energy difference $\Delta = 4.5 \, E_{\text{int}}/a$ has been applied to open the gap at $\b K$, thus generating a non-vanishing Berry curvature that is responsible for the Hall drift along $x$. The small oscillations are caused by non-adiabatic processes originating from transfer to the upper band. (b) Initially prepared gaussian wavepacket, where the dashed vertical line indicates the initial $\mv x$ position. (c) Final state of the wavepacket showing a horizontal displacement with respect to the initial $\mv x$ position. In panels $(b)$ and $(c)$, the normalized wavepacket density $n$ is shown.
}
\label{fig:wpacket}
\end{figure}

\section{Probing methods}
\label{sec:probing}

In this Section, we discuss several aspects related to the control of a phonon wavepacket (see the end of the Section for its preparation) in order to explore the  topological properties of the phonon bands, as the Berry curvature or the Berry phase. A strategy that is well established with ultracold atoms requires a tilted potential, which plays the role of an electric field for charged particles, to investigate the transport properties of wavepackets in individual or multiple bands. As a result, the wavepacket satisfies semiclassical equations, which for a non-degenerate band read
\be
\label{eq:semicl}
\dot{\mathbf{r}}_c = \nabla_{\mathbf k} \epsilon(\mathbf k_c) - \mathbf F \times \mathbf\Omega_{xy}(\mathbf k_c)\,, \quad \dot{\mathbf{k}}_c = \mathbf F \,,
\ee
where $\b r_c$ and $\b k_c$ are the center-of-mass position and momentum of the wavepacket, $\b F$ is the applied force, $\epsilon(\mathbf k_c)$ the band dispersion and $\mathbf\Omega_{xy}(\mathbf k_c)$ the Berry curvature \cite{Cooper2019}. The force $\b F$, which moves an atomic wavepacket in real space, is typically obtained by accelerating the lattice or by applying a magnetic field gradient. However, these techniques do not produce a force on a phonon wavepacket and a different approach, which we detail below, must be followed.

In order to correctly reproduce the effects of an applied force on the phonon modes, we find it useful to recast the formalism that we developed in Sec.~\ref{sec:dispersion1} into a tight-binding model in real space, as described in Ref.~\cite{Bermudez2011} for the phonon modes of trapped ion arrays. As outlined in Appendix \ref{sec:tightb}, the Hamiltonian reads

\begin{align}
\label{eq:tbph}
H  &= \sum_n \hbar \tilde \om_{i,n} \, a^\dag_{i,n} a_{i,n}^{}  \!+\!  \sum_{\substack{ i,j \\ \mv{n,n'} } } \f{\hbar D_{ij}^{nn'}}{2m\sqrt{\tilde\om_{i,n} \tilde\om_{j,n'}}} (a_{i,n}^\dag a_{j,n'}^{} \!+\!\text{h.c.} ) \nn \\
& \equiv \sum_{i,n} \mu_{i,n}^{} \, a^\dag_{i,n} a_{i,n}^{}  +  \sum_{\substack{ i,j \\ \mv{n,n'} } } J_{ij}^{nn'}(a_{i,n}^\dag a_{j,n'}^{} +\text{h.c.} ) \,,
\end{align}
where $\tilde \omega^2_{i,n}  = \omega^2_{i,n} + D_{ii}^{nn}/m$ and we neglected phonon non-conserving terms like $a^{}_{i,n}a^{}_{j,n'}$ or $a^{\dag}_{i,n}a^\dag_{j,n'}$ under the rotating wave approximation, which is valid when $\hbar \om_{i,n} \gg V_{\text{dd}} \ell^2/a^5$. Notice that we have recovered the result in Eq.~\eqref{eq:highfreq} for the spectrum of the model and at the same time we have recast the Hamiltonian into an effective tight-binding description with $\mu_{i,n}$ the onsite energies and $J_{ij}^{nn'}$ the hopping amplitudes. The first term in Eq.~\eqref{eq:tbph}, namely $\mu_{i,n}$, clearly shows that a space-dependent harmonic frequency, \emph{e.\,g.} $\om_{i,n} = \om_{i,0} + \Delta \om \, X_n/a$ corresponds to an onsite potential $\mu_{i,n}$ for the phonon modes with a nonvanishing space gradient in the $x$ direction, which can therefore be used to generate the force $\b F$ acting on a phonon wavepacket. This approach would naturally also lead to space-dependent couplings, as evident from the second term of Eq.~\eqref{eq:tbph}. However, adiabaticity requires that the magnitude of the force must be sufficiently small to prevent Landau-Zener transitions when the phonon wavepacket approaches a band gap. This would reflect into a slow variation of the couplings $J_{ij}^{nn'}$, which needs to be minimized in order to avoid detrimental effects on the topological gap. 

In Fig.~\ref{fig:wpacket}, we show an example of a gaussian wavepacket dynamics on the honeycomb lattice for the $u_{z,\b k}$ modes. We assume a uniform and isotropic case, with $J_1\!=\!J_2\!=\!J_3\!=\!J$ and we break inversion symmetry by changing the relative onsite energy between the $A$ and $B$ sites by imposing an offset $\Delta$, which can be obtained by making the harmonic frequencies on the two sublattices different, \emph{i.e.} $\om^A\neq \om^B$. We prepare a gaussian wavepacket at $\bs \Gamma$ and apply a linear potential corresponding to a force $\b F = (0, F_y)$. The offset $\Delta$ opens a gap at the $\b K$ and $\b K'$ points with a nonvanishing (but opposite at the two valleys) Berry curvature. As a result, the wavepacket experiences a Hall deflection after crossing the $\b K$ point under the applied $\b F$ force, as expected from Eq.~\eqref{eq:semicl}. This strategy can be applied to control a phonon wavepacket in order to probe the Berry curvature generated by the driving schemes discussed before or to measure the Berry phase around the band degeneracy points using interferometric methods \cite{Duca2015}.

\begin{figure}[!t]
\center
\includegraphics[width=0.69\columnwidth]{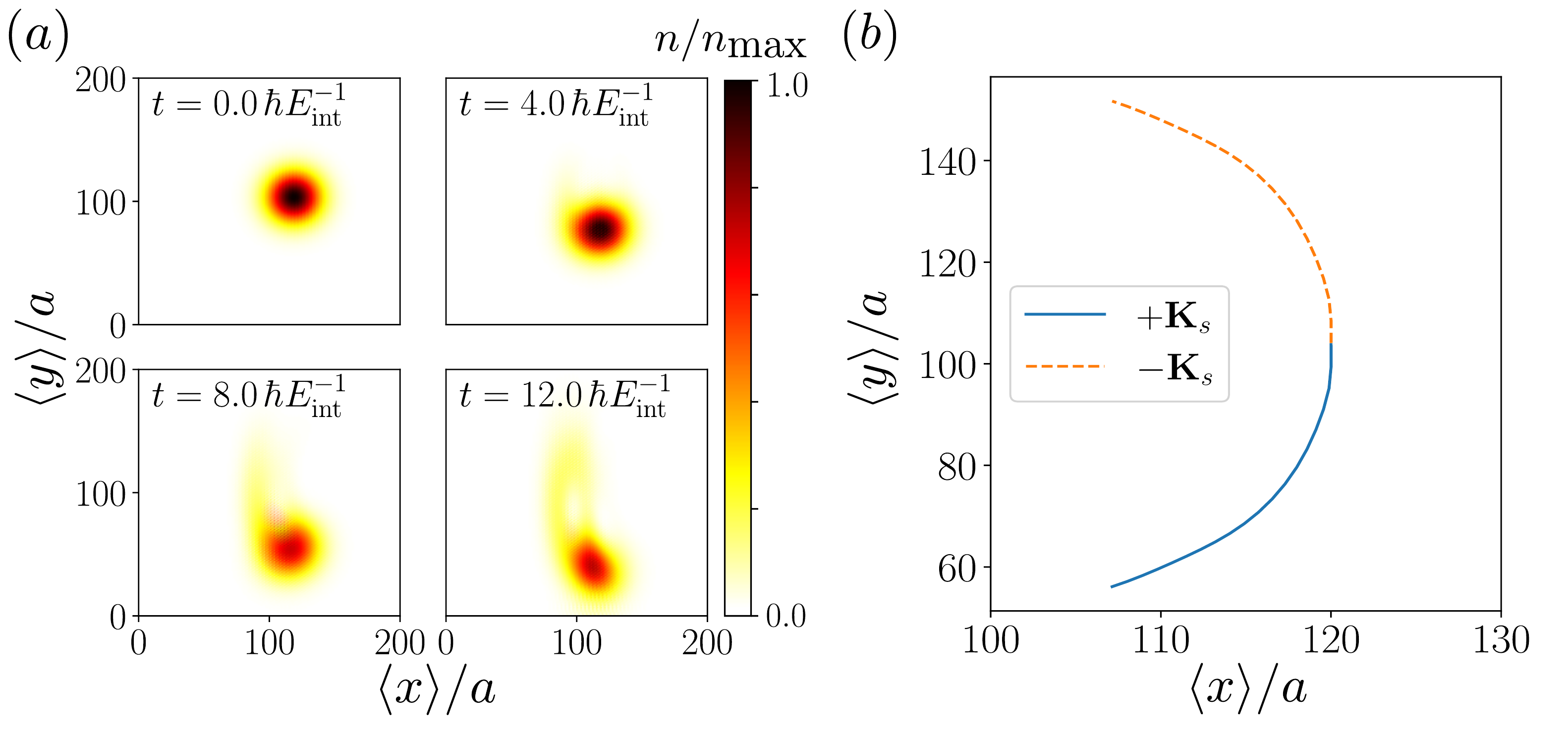}

\caption{Cyclotron orbit motion in the presence of strain. (a) A gaussian wavepacket of out-of-plane modes, width $\sigma=20a$, is prepared at the $\bs\Gamma$ and rapidly accelerated to the $\textbf{K}_s$ point, where it is left to evolve in the absence of any force. The polarization tilt parameters are chosen as $\Theta_0 = \pi/4$ and $\tau = 0.001$. The large difference in the $x$ and $y$ motion reflects the strong anisotropy of the Dirac velocity. (b) Center of mass motion up to $t=12 \,\hbar E_{\text{int}}^{-1}$ after preparing the wavepacket at the two valleys $\pm\textbf{K}_s$. The trajectory shows opposite chiral motion, as expected from the underlying time-reversal symmetry of the model.
}
\label{fig:orbit}
\end{figure}

For the anisotropic and nonuniform case with space dependent couplings $J_\nu(x)$, as in Eq.~\eqref{eq:Jlinear}, we saw in Sec.~\ref{sec:strain} that a strong pseudo-electric field contribution $\mathcal E_x$ must be compensated in order to access the Landau levels regime. In this case, after noticing that the onsite energies read $\tilde \om  = \om \!-\!J_T(x) /2m\om$, where $J_T(x)\equiv(J_1(x)\!+\!J_2(x)\!+\!J_3(x))$, and therefore $\mathcal E_x \propto \pa_x J_T(x)$, the strong electric field can be suppressed by an appropriate space-dependent trapping frequency $\om(x) = \om_0 + \Delta \om \,x/a$ by taking $\Delta \om / a  = \pa_x J_T/2m\om_0$ such that $\tilde\om$ is approximately a constant. If $\tilde\om$ has no space space dependence or a residual small one, the Landau level structure is preserved and the related physics can therefore be observed. We prepared a wavepacket with momentum $\b K_s$ that we let evolve in the presence of strain. We show in Fig.~\ref{fig:orbit}(a) the expected dynamics, where the wavepacket performs a chiral motion corresponding to cyclotron orbits, originating from the nonvanishing pseudo-magnetic field $\mathcal B$ induced by strain. In Fig.~\ref{fig:orbit}(b), we verify that the motion displays opposite chirality at $-\b{K}_s$, as expected from time-reversal symmetry, namely the fact that the pseudo-magnetic field $\mathcal B$ changes sign from one valley to the other. The strong anisotropy reflects into substantially different Fermi velocities in the two spatial directions, which cause a larger displacement of the wavepacket along the $y$ direction as compared to the $x$ direction. 

In order to resolve the phonons occupation and monitor the position of the phonon wavepacket, as required to observe the Hall deflection and more generally transport properties, we may take advantage of single-site microscopy. A phonon excitation locally corresponds to a superposition of the atomic (or molecular) wavefunction including the lowest $s$-orbital of each site and the higher orbitals, dominantly the $p$-orbitals. A measurement of the phonon wavepacket density could then proceed in two steps: in the first step, a $\pi$ microwave pulse resonantly couples the $s$ orbitals to a different atomic state that is not trapped by the laser confinement. As a result, the ground state occupation can be removed, thus projecting the wavefunction onto the excited orbitals, which map the phonon modes occupation. In a second step, single site microscopy is then employed to image the position over each lattice site, thus resolving the phonon wavepacket real-space distribution.

While the discussion on the probing methods has mainly focused on the transport properties of a phonon wavepacket to extract the topological properties of the bands, as Berry phases or Berry curvatures, a distinct, complementary and obvious possibility is to perform spectroscopic measurements of the phonon bands. In this case, one can access the phonon dispersion and directly map out the band singularities, their splitting or the gap opening events. Raman \cite{Muller2007} or Bragg \cite{Sengstock2010} beams can therefore be exploited to perform band spectroscopy of the phonon dispersion, while being at the same time employed for the state preparation of a phonon wavepacket. 


\section{Experimental realizations}
\label{sec:realization}

\label{sec:realization} Let us now discuss potential experimental realizations of our phonon system and present the corresponding estimates of the relevant energy scales. The most natural candidates for dipolar particles are magnetic atoms (and also diatomic molecules made of magnetic atoms) and polar molecules placed in a constant magnetic or electric field, respectively. 

More precisely, we consider magnetic atoms or diatomic molecules (Er, Dy atoms or $\mathrm{Er_{2}}$, $\mathrm{Dy_{2}}$ diatomic molecules, for example) in their lowest energy state in a magnetic field of the order of $0.1\div1$ Gauss which corresponds to energy gaps of the order of $1\div 50\, h\,\mathrm{MHz}$ between Zeemann levels. In the case of polar molecules (KRb, RbCs, NaK, NaRb, for example), we consider them in their electronic and vibrational ground state in an external constant electric field $\mathbf{E}$  which induces an electric dipole moment.
For our estimates, we take the achievable value of $12\,\mathrm{kV}/\mathrm{cm}$. In Table \ref{tab:molparameters}, we present the values of masses, magnetic/electric dipole moments, as well as the corresponding values of the dipole-dipole interaction parameter $V_{\mathrm{dd}}=\mu_{0}\mu^{2}/(4\pi)$ or $V_{\mathrm{dd}}=d^{2}/(4\pi\epsilon_{0})$ for two magnetic ($\mu$) or electric ($d$) dipole moments, respectively, which correspond to their interaction at the distance $1\mu m$, for several magnetic/electric dipolar particles.

\begin{table}[t]
  \centering
  \begin{tabular}{l | c | c | c | c }
{} & ${m}/{\mathrm{u}}$ & \multicolumn{2}{|c|}{$\mu/\mu_0$} & $V_\mathrm{dd}/(h\,\mathrm{kHz}~\mu \mathrm{m}^3)$ \\ 
\hline
Er& $164$ & \multicolumn{2}{|c|}{$7$} & $0.00064$\\
Dy& $161$ & \multicolumn{2}{|c|}{$10$} & $0.0013$\\
Er\textsubscript{2} & $328$ & \multicolumn{2}{|c|}{$\lesssim 14$} & $0.0025$\\
Dy\textsubscript{2} & $322$ &\multicolumn{2}{|c|}{$\lesssim 20$} & $0.0052$  \\
{} & {} &\multicolumn{2}{|c|}{} & {} \\
{} & {} & ${\mathcal{D}}/{\mathrm{D}}$ & $d/{\mathrm{D}}$ & {}\\ 
\hline
KRb  & $127$ & $0.57$  & $0.34$ & $0.017$\\ 
RbCs & $220$ & $1.17$  & 0.95 & $0.136$ \\ 
NaK & $63$ & $2.72$  & $1.90$ & $0.547$\\ 
NaRb & $110$ & $3.2$   & 2.50 & $0.916$\\ 
LiRb & $91$ & $3.99$ &  2.35 & $0.836$ 
\end{tabular}
  \caption{Parameters for a variety of magnetic atoms/molecules and polar molecules.  For all polar molecular species the offset electric field is chosen to be $|\mathbf{E}| = 12\,\mathrm{kV}/\mathrm{cm}$. Parameters for different magnetic atoms can be found in Ref.~\cite{Frisch2014} and for polar molecules in Ref.~\cite{Aymar2005}.}
  \label{tab:molparameters}
\end{table}

To get an estimate of the phonon bandwidth, we consider
a 1D situation where the dipolar particles are placed in a deep optical
lattice with a lattice spacing $a=\lambda/2$ and a depth $V_{0}E_{R}$,
where $E_{R}$ is the recoil energy $E_{R}=\pi^{2}\hbar^{2}/(2ma^{2})$
and $V_{0}\gg1$. In this case, the hopping amplitude $t$ between different
sites of the lattice is exponentially small. For example, $t/E_{R}=(4/\sqrt{\pi})V_{0}^{3/4}\exp(-2\sqrt{V_{0}})$
\cite{Bloch2008} for the nearest-neighbor tunneling of particles between
the lowest Wannier states of each site, such that the individual particles tunneling dynamics is frozen. When considering one particle per site, the spatial
degrees of freedom reduce to local oscillations/vibrations
which, in the harmonic approximations, can be characterized by the
oscillator frequency $\omega\!=\!\sqrt{2V_{0}}E_{R}/\hbar$ and by the
oscillator length $\ell\!=\!\sqrt{\hbar/(m\omega)}\!=\!a\pi^{-1}(2/V_{0})^{1/4}$.
To be more precise, the harmonic approximation, resulting in an equidistant
spectrum, is only valid for a few lowest energy states in a very deep
optical lattices. For experimentally realistic situations, the level
spacing decreases with increasing the energy, and $\hbar\omega$ gives
the difference between the ground and the first excited states ($s$
and $p$ orbitals in the 3D language). By using the expression $12V_\mathrm{dd}/a^{5}$ for the off-diagonal element of the dynamical matrix [see~Eq.~\eqref{eq:formfact} adapted to the 1D case discussed here], the phonon bandwidth $\Delta E_{B}$ is given by $\Delta E_{B}$ $=$ $24(\ell/a)^{2}V_\mathrm{dd}/a^{3}=24(\sqrt{2/V_{0}}/\pi^{2})V_\mathrm{dd}/a^{3}$, such that $\Delta E_{B}/(\hbar\omega)=(24/\pi^{2})(V_{0}E_{R})^{-1}V_\mathrm{dd}/a^{3}\sim a^{-1}$.

\begin{figure}[!t]
\center
\includegraphics[width=0.69\columnwidth]{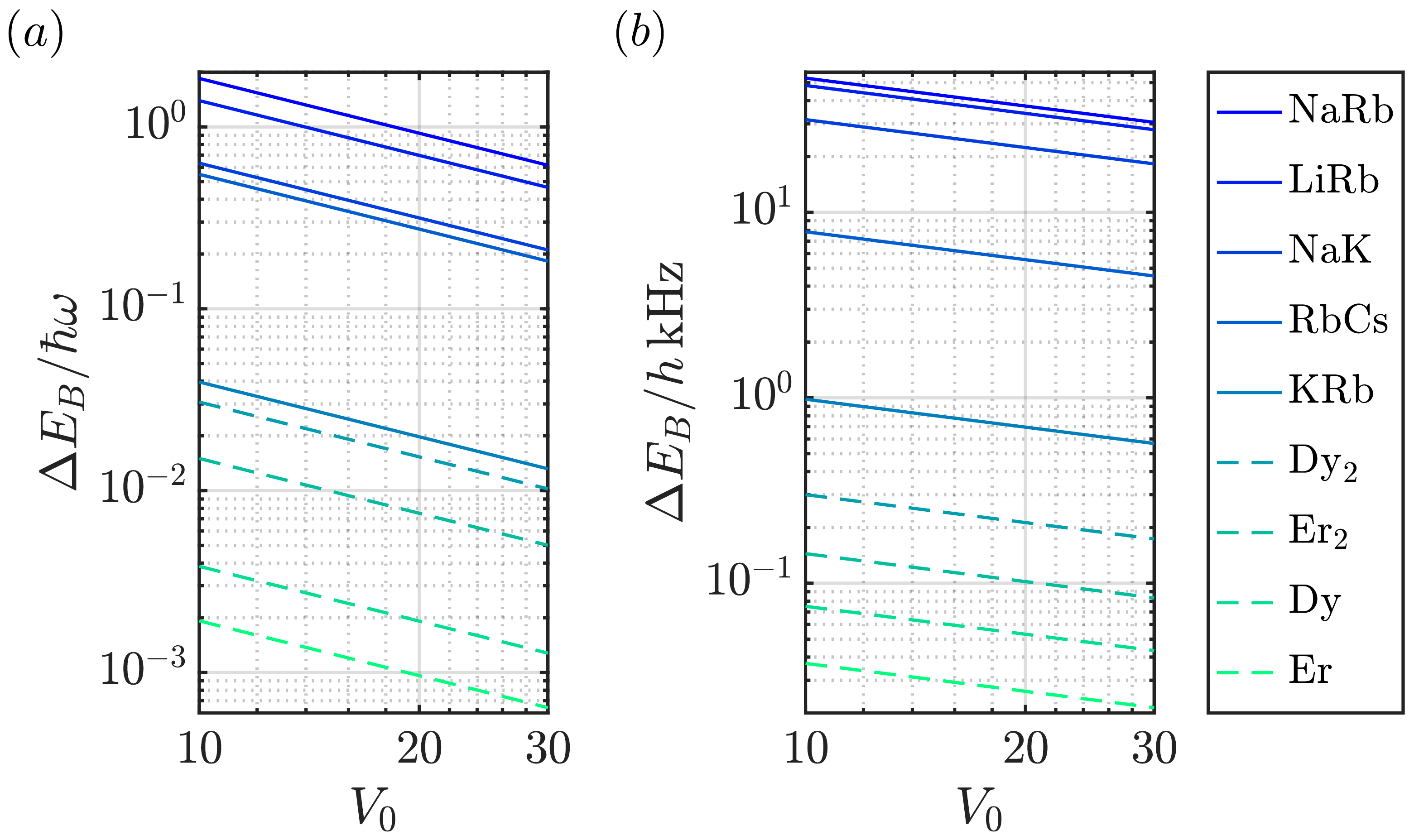}

\caption{Phonon bandwidth estimate for dipolar atoms and polar molecules. (a) Phonon bandwidth in units of the harmonic trapping frequency as a function of the lattice depth $V_0$ for a lattice spacing $a = 0.266\,\mu\mathrm{m}$. (b) Phonon bandwidth in energy units as a function of the lattice depth for the same lattice spacing used in panel (a). 
}
\label{fig:scaling}
\end{figure}

In Fig.~\ref{fig:scaling}, we present the values of $\Delta E_{B}$ for a fixed
lattice spacing $a=0.266\mathrm{\mu m}$ and different lattice depths
$V_{0}.$ We see that the values of $\Delta E_{B}$ range from $10\div300\,h\,\mathrm{Hz}$
for magnetic systems to $1\div50\,h\,\mathrm{kHz}$ for molecular ones.
We mention also that $\Delta E_{B}$ can
reach values of more than $100\,h\,\mathrm{kHz}$ for polar molecules
with larger dipole moment (NaCs and LiCs have a dipole moment $d=3.8\,\mathrm{D}$ and $d=3.67\,\mathrm{D}$, respectively, with an electric field as in Table~\ref{tab:molparameters}, for example) but this situation
is not covered by our approximations as the phonon bandwidth becomes larger than the harmonic frequency.

Finally, we mention another possible experimental implementation of
optical phonons which is based on Rydberg atoms in tweezer arrays.
This case, however, requires more careful analysis because of much
smaller values of the parameter $\ell/a$, as compared to optical lattices.
As a result, phonon bands of similar widths would require much
stronger values of the interparticle interaction and, therefore, the creation
of stable structures can be problematic. Similar arguments also apply to the
case of molecules with large dipole moments.


\section{Discussion and perspective}
\label{sec:conclusions}

In this work, we have shown that arrays of ultracold dipolar particles offer a versatile platform to explore the physics of topological states of matter by studying the vibrational phonon modes in a deep optical lattice. 
The current experimental progress provides all the necessary ingredients to pursue this approach.
We have shown how the sole static and driven control of the dipole-dipole interactions anisotropy can be instrumental to access a variety of topological effects. 
While the results of this work have been derived for the honeycomb lattice case, our methods are quite general. They can be applied to other two-dimensional lattices, \emph{e.\,g.} the Lieb or the Kagome, as well as to one- and three-dimensional lattices, where new topological phenomena can be accessed. 
Our analysis has been simplified under the condition that the dynamics of in-plane modes is independent of the out-of-plane modes one, and viceversa. 
However, relaxing this condition in three dimensions can offer an additional nontrivial set of possibilities that is left to a future investigation \cite{Park2021}.

An interesting perspective to be investigated is the fate of the edge modes that are expected in the gaps of the topological phonon bands. 
This discussion requires a comprehensive analysis of the boundary details. 
The region of the system where the density drops to zero provides a natural boundary for the existence of edge modes. 
If the external confinement is a box potential, this corresponds to a hard boundary and the edge modes will therefore be present. 
Less clear is the existence of edge modes in a different scenario, namely in the presence of a confining harmonic potential. 
This case corresponds to a soft boundary because the particle density smoothly decreases, thus breaking the condition of unit filling with unclear consequences for the dynamics of phonon excitations in this region. 
Note also that our setup offers a distinct and rather simple opportunity to create and study topological modes localized at defects in the bulk of the system. Such defects with arbitrary shape can indeed be created by removing dipolar particles with single-site addressing techniques.

In our analysis, we assumed a low phonon density, which allowed us to neglect phonon-phonon interactions. These will provide nontrivial scattering processes for phonons as they will couple the different directions of oscillations, namely their orbital degrees of freedom.
Collisions therefore play an important role to establish intra- and inter-band scattering events that can lead, for example, to the ``thermalization" of the phonon system. 
In this sense, a comprehensive analysis of the regime with a large phonon density offers the opportunity to explore remarkable orbital-like many-body phenomena \cite{Hemmerich2011}. 

We mention also that the strategies considered in this paper, which exploit the anisotropy of the dipole-dipole interaction to manipulate the band structure of the phonon excitations, can also be applied to other types of excitations whose transport is governed by a resonant dipole-exchange mechanism.
Besides the case of spin-like excitations with magnetic atoms and polar molecules, an interesting possibility is offered by the high-energy orbital excitations in arrays of Rydberg atoms \cite{Weimer2010, Buchler2018, Browaeys2019, Browaeys2020}. 


\section{Acknowledgements}

We acknowledge discussions with F. Ferlaino, N. Goldman, M. Mark, M. Norcia, G. Palumbo, G. Salerno, J. Yu.
We also acknowledge support by the European Union program Horizon 2020 under Grant No. 817482 (PASQuanS), by the QuantERA grant MAQS via the Austrian Science Fund FWF No I4391-N, by the Simons Collaboration on Ultra-Quantum Matter, which is a grant from the Simons Foundation (651440, PZ), and by a joint-project grant from the FWF (Grant No. I 4426, RSF/Russia 2019).



\appendix

\section{Dynamical matrix}
\label{sec:dynmat}

The phonon spectrum is obtained by diagonalizing the dynamical matrix, which for the system considered in this work reads
\begin{align}
D_{ij}^{\alpha \alpha}(\b k) &= m\om_i^2 \delta_{ij} + \sum_\nu f_{ij}^\nu \,,\nn\\
D_{ij}^{AB}(\b k) &= - \sum_\nu f_{ij}^\nu\, e^{i\b k \cdot \bs \delta_\nu}\,, \nn\\
D_{ij}^{BA}(\b k) &= D_{ij}^{AB}(-\b k) \,,
\end{align}
with 
\begin{align}
\label{eq:formfact}
\left(\f{V_{\text{dd}}}{a^5}\right)^{-1} f_{ij}^\nu =& \,3\, \delta_{ij} \left[ -1 + 5a^{-2} (\hat{\b d} \cdot \bs{\delta}_\nu )^2 \right] - 6\, \hat d_i \hat d_j  \nn \\
& + 30\, a^{-2}  (\hat{\b d}\cdot\bs{\delta}_\nu)\left[ \hat d_i \, \delta_{\nu}^j + \hat d_j \, \delta_{\nu}^i \right] \nn\\ 
& + 15\, \delta_\nu^i \delta_\nu^j \left[ 1 - 7a^{-2}  (\hat{\b d} \cdot \bs{\delta}_\nu )^2 \right]\,.
\end{align}

\section{Effective continuum theory}
\label{sec:efftheo}

Near a band touching point occurring at $\b k_0$ and for small momenta $|\b q|a \equiv |\b k - \b k_0|a \ll 1$, the effective two-band theory can generally be cast in the form
\be
\b D^0_{\text{eff}} (\b k_0, \b q) = d_0(\b q) \mathcal I_{2\times2} + \sum_{i=x,y,z} d_i(\b q) \sigma_i \,,
\ee
where $\sigma_i$ are the Pauli matrices and $\mathcal I_{2\times2}$ is the identity matrix. In order to obtain the form of the effective theory, we use L\"odwin's method \cite{Lim2020}. We therefore split the modes into a low-energy sector, which lies in energy window near the degeneracy point, and into a high-energy sector, which is treated in perturbation theory. We therefore construct the low-energy projector $P=\sum_{\rho,\sigma}' |\epsilon_{\b k_0}^\rho \ket \bra \epsilon_{\b k_0}^\sigma |$ and the high-energy projector $Q=\sum_{\rho,\sigma}'' |\epsilon_{\b k_0}^\rho \ket \bra \epsilon_{\b k_0}^\sigma |$, where $|\epsilon_{\b k_0}^\rho \ket$ represent the eigenvectors at $\b k_0$ and the symbols $\sum'$ and $\sum ''$ indicate that the summation is restricted to the low-energy and high-energy manifolds, respectively. We obtain the effective low-energy theory
\be
\label{eq:lodwin}
\b D^0_{\textrm{eff}} = P\b D^0(\b k_0 \!+\! \b q) P + P \b{\tilde{D}} Q (\epsilon_{\b k_0}'-Q \b{\tilde{D}} Q)^{-1} (Q \b{\tilde{D}} P)\,,
\ee
where $\b{\tilde{D}}(\b q) \!=\! \b D^0(\b k_0 \!+\! \b q) - \b D^0(\b k_0)$ and $\epsilon_{\b k_0}'$ is the characteristic energy of the low-energy sector.

\section{Floquet perturbation theory}
\label{sec:floquetheo}

Let us consider a theory for the phonon modes subject to a periodic drive described by a dynamical matrix $\b D^0(t)$. We can therefore decompose the matrix as follows 
\be
\b D^0(t) = \sum_l e^{il\Omega t} \b D^{(l)}\,,
\ee
where $\b D^{(l)}$ denotes the $l$-th Fourier component and $\Omega$ the driving frequency. This problem can be described through Floquet theory \cite{Goldman2014b} by constructing a time-independent Floquet dynamical matrix $\b D^F$ that governs the stroboscopic time evolution, which reads
\be
\f{\b D^F}{2m\om} = \f{i \hbar}{T} \log \b U(T)\,,
\ee
where $\b U(t) = \mathcal T \exp(-i \hbar^{-1} \int_0^t \text d t \, \b D(t)/ 2m\om)$ is the time-evolution operator and $T=2\pi/\Om$.

By performing a high-frequency expansion, the Floquet effective (time-independent) dynamical matrix can be computed as
\be
\b D^F \approx \b D^{(0)} + \f{(2m\om)^{-1}}{\hbar\Omega} \sum_{l>0} \f 1 l [\b D^{(+l)},\b D^{(-l)}]\,,
\ee
where we have indicated only the leading order in $1/\Omega$.


\section{Tight-binding description}
\label{sec:tightb}

The phonon modes description that we have used throughout the manuscript is based on the construction of a dynamical matrix describing the small vibrations of the ensemble. However, under the circumstances described here below, the description can be recast in a tight-binding model picture. The Hamiltonian in momentum space reads 
\be
H = \sum_{i,\alpha,\b k} \f{p_{i,\b k}^\alpha \,p_{i,-\b k}^\alpha}{2m} + \f 1 2 \sum_{i,j,\alpha,\beta, \b k} u_{i,\b k}^\alpha D_{ij}^{\alpha\beta} (\b k ) u_{j,-\b k}^{\beta}\,,
\ee
which can be cast in real space as
\begin{align}
H &= \sum_{i,n} \f{p_{i,n}^2}{2m} + \f 1 2 \sum_{i,j,n,n'} u_{i,n} D_{i j}^{n n'} u_{j,n'}\, \\
&= \sum_{i,n} \f{p_{i,n}^2}{2m} + \f 1 2 m  \sum_{i,n}  \tilde \om^2_{i,n} u_{i,n}^2 + \f 1 2 \sum_{\substack{ i,j \\ \mv{n,n'} } } u_{i,n} D_{i j}^{n n'} u_{j,n'}\,, \nn
\end{align}
where $n,n'$ indicate the lattice sites (irrespective of the sublattice degree of freedom), $i,j=x,y,z$ and $\tilde \omega^2_{i,n}  = \omega_{i,n}^2 + D_{ii}^{nn}/m$. We now introduce the local harmonic oscillator creation and annihilation operators as follows
\begin{align}
u_{i,n} &= \sqrt{\f{\hbar}{2\,m\,\tilde\om_{i,n}}}  (a_{i,n}^{} + a_{i,n}^\dag)\,, \\
p_{i,n} &= i \sqrt{\f{m\,\hbar \,\tilde \om_{i,n}}{2}} (a_{i,n}^\dag - a_{i,n}^{})\,,
\end{align}
and rewrite the Hamiltonian as
\begin{align}
\label{eq:tbph2}
H  &= \sum_n \hbar \tilde \om_{i,n} \, a^\dag_{i,n} a_{i,n}^{}  \!+\!  \sum_{\substack{ i,j \\ \mv{n,n'} } } \f{\hbar D_{ij}^{nn'}}{2m\sqrt{\tilde\om_{i,n} \tilde\om_{j,n'}}} (a_{i,n}^\dag a_{j,n'}^{} \!+\!\text{h.c.} ) \nn \\
& \equiv \sum_{i,n} \mu_{i,n}^{} \, a^\dag_{i,n} a_{i,n}^{}  +  \sum_{\substack{ i,j \\ \mv{n,n'} } } J_{ij}^{nn'}(a_{i,n}^\dag a_{j,n'}^{} +\text{h.c.} ) \,,
\end{align}
where we neglected phonon non-conserving terms like $a^{}_{i,n}a^{}_{j,n'}$ or $a^{\dag}_{i,n}a^\dag_{j,n'}$ under the rotating wave approximation (RWA), which is valid when $\hbar \om_{i,n} \gg V_{\text{dd}} \ell^2/a^5$. As a result, the RWA allows us to reinterpret the phonon dynamics in terms of particles hopping on a lattice. This picture become extremely convenient in order, for instance, to understand how to manipulate a phonon wavepacket, as discussed in the main text.

\vspace{1cm}

\bibliographystyle{quantum}
\bibliography{biblio}

\begin{thebibliography}{99}%
\makeatletter
\providecommand \@ifxundefined [1]{%
 \@ifx{#1\undefined}
}%
\providecommand \@ifnum [1]{%
 \ifnum #1\expandafter \@firstoftwo
 \else \expandafter \@secondoftwo
 \fi
}%
\providecommand \@ifx [1]{%
 \ifx #1\expandafter \@firstoftwo
 \else \expandafter \@secondoftwo
 \fi
}%
\providecommand \natexlab [1]{#1}%
\providecommand \enquote  [1]{``#1''}%
\providecommand \bibnamefont  [1]{#1}%
\providecommand \bibfnamefont [1]{#1}%
\providecommand \citenamefont [1]{#1}%
\providecommand \href@noop [0]{\@secondoftwo}%
\providecommand \href [0]{\begingroup \@sanitize@url \@href}%
\providecommand \@href[1]{\@@startlink{#1}\@@href}%
\providecommand \@@href[1]{\endgroup#1\@@endlink}%
\providecommand \@sanitize@url [0]{\catcode `\\12\catcode `\$12\catcode
  `\&12\catcode `\#12\catcode `\^12\catcode `\_12\catcode `\%12\relax}%
\providecommand \@@startlink[1]{}%
\providecommand \@@endlink[0]{}%
\providecommand \url  [0]{\begingroup\@sanitize@url \@url }%
\providecommand \@url [1]{\endgroup\@href {#1}{\urlprefix }}%
\providecommand \urlprefix  [0]{URL }%
\providecommand \Eprint [0]{\href }%
\providecommand \doibase [0]{http://dx.doi.org/}%
\providecommand \selectlanguage [0]{\@gobble}%
\providecommand \bibinfo  [0]{\@secondoftwo}%
\providecommand \bibfield  [0]{\@secondoftwo}%
\providecommand \translation [1]{[#1]}%
\providecommand \BibitemOpen [0]{}%
\providecommand \bibitemStop [0]{}%
\providecommand \bibitemNoStop [0]{.\EOS\space}%
\providecommand \EOS [0]{\spacefactor3000\relax}%
\providecommand \BibitemShut  [1]{\csname bibitem#1\endcsname}%
\let\auto@bib@innerbib\@empty
\bibitem [{\citenamefont {Hasan}\ and\ \citenamefont {Kane}(2010)}]{Kane2010}%
  \BibitemOpen
  \bibfield  {author} {\bibinfo {author} {\bibfnamefont {M.~Z.}\ \bibnamefont
  {Hasan}}\ and\ \bibinfo {author} {\bibfnamefont {C.~L.}\ \bibnamefont
  {Kane}},\ }\href {\doibase 10.1103/RevModPhys.82.3045} {\bibfield  {journal}
  {\bibinfo  {journal} {Rev. Mod. Phys.}\ }\textbf {\bibinfo {volume} {82}},\
  \bibinfo {pages} {3045} (\bibinfo {year} {2010})}\BibitemShut {NoStop}%
\bibitem [{\citenamefont {Chen}\ \emph {et~al.}(2012)\citenamefont {Chen},
  \citenamefont {Gu}, \citenamefont {Liu},\ and\ \citenamefont
  {Wen}}]{Chen2012}%
  \BibitemOpen
  \bibfield  {author} {\bibinfo {author} {\bibfnamefont {X.}~\bibnamefont
  {Chen}}, \bibinfo {author} {\bibfnamefont {Z.-C.}\ \bibnamefont {Gu}},
  \bibinfo {author} {\bibfnamefont {Z.-X.}\ \bibnamefont {Liu}}, \ and\
  \bibinfo {author} {\bibfnamefont {X.-G.}\ \bibnamefont {Wen}},\ }\href
  {\doibase 10.1126/science.1227224} {\bibfield  {journal} {\bibinfo  {journal}
  {Science}\ }\textbf {\bibinfo {volume} {338}},\ \bibinfo {pages} {1604}
  (\bibinfo {year} {2012})}\BibitemShut {NoStop}%
\bibitem [{\citenamefont {Chiu}\ \emph {et~al.}(2016)\citenamefont {Chiu},
  \citenamefont {Teo}, \citenamefont {Schnyder},\ and\ \citenamefont
  {Ryu}}]{Ryu2016}%
  \BibitemOpen
  \bibfield  {author} {\bibinfo {author} {\bibfnamefont {C.-K.}\ \bibnamefont
  {Chiu}}, \bibinfo {author} {\bibfnamefont {J.~C.~Y.}\ \bibnamefont {Teo}},
  \bibinfo {author} {\bibfnamefont {A.~P.}\ \bibnamefont {Schnyder}}, \ and\
  \bibinfo {author} {\bibfnamefont {S.}~\bibnamefont {Ryu}},\ }\href {\doibase
  10.1103/RevModPhys.88.035005} {\bibfield  {journal} {\bibinfo  {journal}
  {Rev. Mod. Phys.}\ }\textbf {\bibinfo {volume} {88}},\ \bibinfo {pages}
  {035005} (\bibinfo {year} {2016})}\BibitemShut {NoStop}%
\bibitem [{\citenamefont {Rachel}(2018)}]{Rachel2018}%
  \BibitemOpen
  \bibfield  {author} {\bibinfo {author} {\bibfnamefont {S.}~\bibnamefont
  {Rachel}},\ }\href {\doibase 10.1088/1361-6633/aad6a6} {\bibfield  {journal}
  {\bibinfo  {journal} {Reports on Progress in Physics}\ }\textbf {\bibinfo
  {volume} {81}},\ \bibinfo {pages} {116501} (\bibinfo {year}
  {2018})}\BibitemShut {NoStop}%
\bibitem [{\citenamefont {Klitzing}\ \emph {et~al.}(1980)\citenamefont
  {Klitzing}, \citenamefont {Dorda},\ and\ \citenamefont
  {Pepper}}]{Klitzing1980}%
  \BibitemOpen
  \bibfield  {author} {\bibinfo {author} {\bibfnamefont {K.~v.}\ \bibnamefont
  {Klitzing}}, \bibinfo {author} {\bibfnamefont {G.}~\bibnamefont {Dorda}}, \
  and\ \bibinfo {author} {\bibfnamefont {M.}~\bibnamefont {Pepper}},\ }\href
  {\doibase 10.1103/PhysRevLett.45.494} {\bibfield  {journal} {\bibinfo
  {journal} {Phys. Rev. Lett.}\ }\textbf {\bibinfo {volume} {45}},\ \bibinfo
  {pages} {494} (\bibinfo {year} {1980})}\BibitemShut {NoStop}%
\bibitem [{\citenamefont {Laughlin}(1981)}]{Laughlin1981}%
  \BibitemOpen
  \bibfield  {author} {\bibinfo {author} {\bibfnamefont {R.~B.}\ \bibnamefont
  {Laughlin}},\ }\href {\doibase 10.1103/PhysRevB.23.5632} {\bibfield
  {journal} {\bibinfo  {journal} {Physical Review B}\ }\textbf {\bibinfo
  {volume} {23}},\ \bibinfo {pages} {5632} (\bibinfo {year}
  {1981})}\BibitemShut {NoStop}%
\bibitem [{\citenamefont {Qi}\ and\ \citenamefont {Zhang}(2011)}]{Qi2011}%
  \BibitemOpen
  \bibfield  {author} {\bibinfo {author} {\bibfnamefont {X.-L.}\ \bibnamefont
  {Qi}}\ and\ \bibinfo {author} {\bibfnamefont {S.-C.}\ \bibnamefont {Zhang}},\
  }\href {\doibase 10.1103/RevModPhys.83.1057} {\bibfield  {journal} {\bibinfo
  {journal} {Rev. Mod. Phys.}\ }\textbf {\bibinfo {volume} {83}},\ \bibinfo
  {pages} {1057} (\bibinfo {year} {2011})}\BibitemShut {NoStop}%
\bibitem [{\citenamefont {Fu}(2011)}]{Fu2011}%
  \BibitemOpen
  \bibfield  {author} {\bibinfo {author} {\bibfnamefont {L.}~\bibnamefont
  {Fu}},\ }\href {\doibase 10.1103/PhysRevLett.106.106802} {\bibfield
  {journal} {\bibinfo  {journal} {Phys. Rev. Lett.}\ }\textbf {\bibinfo
  {volume} {106}},\ \bibinfo {pages} {106802} (\bibinfo {year}
  {2011})}\BibitemShut {NoStop}%
\bibitem [{\citenamefont {Goldman}\ \emph {et~al.}(2016)\citenamefont
  {Goldman}, \citenamefont {Budich},\ and\ \citenamefont
  {Zoller}}]{Goldman2016a}%
  \BibitemOpen
  \bibfield  {author} {\bibinfo {author} {\bibfnamefont {N.}~\bibnamefont
  {Goldman}}, \bibinfo {author} {\bibfnamefont {J.~C.}\ \bibnamefont {Budich}},
  \ and\ \bibinfo {author} {\bibfnamefont {P.}~\bibnamefont {Zoller}},\ }\href
  {https://doi.org/10.1038/nphys3803} {\bibfield  {journal} {\bibinfo
  {journal} {Nature Physics}\ }\textbf {\bibinfo {volume} {12}},\ \bibinfo
  {pages} {639} (\bibinfo {year} {2016})}\BibitemShut {NoStop}%
\bibitem [{\citenamefont {Cooper}\ \emph {et~al.}(2019)\citenamefont {Cooper},
  \citenamefont {Dalibard},\ and\ \citenamefont {Spielman}}]{Cooper2019}%
  \BibitemOpen
  \bibfield  {author} {\bibinfo {author} {\bibfnamefont {N.~R.}\ \bibnamefont
  {Cooper}}, \bibinfo {author} {\bibfnamefont {J.}~\bibnamefont {Dalibard}}, \
  and\ \bibinfo {author} {\bibfnamefont {I.~B.}\ \bibnamefont {Spielman}},\
  }\href {\doibase 10.1103/RevModPhys.91.015005} {\bibfield  {journal}
  {\bibinfo  {journal} {Rev. Mod. Phys.}\ }\textbf {\bibinfo {volume} {91}},\
  \bibinfo {pages} {015005} (\bibinfo {year} {2019})}\BibitemShut {NoStop}%
\bibitem [{\citenamefont {Ozawa}\ \emph {et~al.}(2019)\citenamefont {Ozawa},
  \citenamefont {Price}, \citenamefont {Amo}, \citenamefont {Goldman},
  \citenamefont {Hafezi}, \citenamefont {Lu}, \citenamefont {Rechtsman},
  \citenamefont {Schuster}, \citenamefont {Simon}, \citenamefont {Zilberberg},\
  and\ \citenamefont {Carusotto}}]{Ozawa2019}%
  \BibitemOpen
  \bibfield  {author} {\bibinfo {author} {\bibfnamefont {T.}~\bibnamefont
  {Ozawa}}, \bibinfo {author} {\bibfnamefont {H.~M.}\ \bibnamefont {Price}},
  \bibinfo {author} {\bibfnamefont {A.}~\bibnamefont {Amo}}, \bibinfo {author}
  {\bibfnamefont {N.}~\bibnamefont {Goldman}}, \bibinfo {author} {\bibfnamefont
  {M.}~\bibnamefont {Hafezi}}, \bibinfo {author} {\bibfnamefont
  {L.}~\bibnamefont {Lu}}, \bibinfo {author} {\bibfnamefont {M.~C.}\
  \bibnamefont {Rechtsman}}, \bibinfo {author} {\bibfnamefont {D.}~\bibnamefont
  {Schuster}}, \bibinfo {author} {\bibfnamefont {J.}~\bibnamefont {Simon}},
  \bibinfo {author} {\bibfnamefont {O.}~\bibnamefont {Zilberberg}}, \ and\
  \bibinfo {author} {\bibfnamefont {I.}~\bibnamefont {Carusotto}},\ }\href
  {\doibase 10.1103/RevModPhys.91.015006} {\bibfield  {journal} {\bibinfo
  {journal} {Rev. Mod. Phys.}\ }\textbf {\bibinfo {volume} {91}},\ \bibinfo
  {pages} {015006} (\bibinfo {year} {2019})}\BibitemShut {NoStop}%
\bibitem [{\citenamefont {Lee}\ \emph {et~al.}(2018)\citenamefont {Lee},
  \citenamefont {Imhof}, \citenamefont {Berger}, \citenamefont {Bayer},
  \citenamefont {Brehm}, \citenamefont {Molenkamp}, \citenamefont {Kiessling},\
  and\ \citenamefont {Thomale}}]{Thomale2018}%
  \BibitemOpen
  \bibfield  {author} {\bibinfo {author} {\bibfnamefont {C.~H.}\ \bibnamefont
  {Lee}}, \bibinfo {author} {\bibfnamefont {S.}~\bibnamefont {Imhof}}, \bibinfo
  {author} {\bibfnamefont {C.}~\bibnamefont {Berger}}, \bibinfo {author}
  {\bibfnamefont {F.}~\bibnamefont {Bayer}}, \bibinfo {author} {\bibfnamefont
  {J.}~\bibnamefont {Brehm}}, \bibinfo {author} {\bibfnamefont {L.~W.}\
  \bibnamefont {Molenkamp}}, \bibinfo {author} {\bibfnamefont {T.}~\bibnamefont
  {Kiessling}}, \ and\ \bibinfo {author} {\bibfnamefont {R.}~\bibnamefont
  {Thomale}},\ }\href {https://doi.org/10.1038/s42005-018-0035-2} {\bibfield
  {journal} {\bibinfo  {journal} {Communications Physics}\ }\textbf {\bibinfo
  {volume} {1}},\ \bibinfo {pages} {39} (\bibinfo {year} {2018})}\BibitemShut
  {NoStop}%
\bibitem [{\citenamefont {Huber}(2016)}]{Huber2016}%
  \BibitemOpen
  \bibfield  {author} {\bibinfo {author} {\bibfnamefont {S.~D.}\ \bibnamefont
  {Huber}},\ }\href {https://doi.org/10.1038/nphys3801} {\bibfield  {journal}
  {\bibinfo  {journal} {Nature Physics}\ }\textbf {\bibinfo {volume} {12}},\
  \bibinfo {pages} {621} (\bibinfo {year} {2016})}\BibitemShut {NoStop}%
\bibitem [{\citenamefont {Ma}\ \emph {et~al.}(2019)\citenamefont {Ma},
  \citenamefont {Xiao},\ and\ \citenamefont {Chan}}]{Chan2019}%
  \BibitemOpen
  \bibfield  {author} {\bibinfo {author} {\bibfnamefont {G.}~\bibnamefont
  {Ma}}, \bibinfo {author} {\bibfnamefont {M.}~\bibnamefont {Xiao}}, \ and\
  \bibinfo {author} {\bibfnamefont {C.~T.}\ \bibnamefont {Chan}},\ }\href
  {\doibase 10.1038/s42254-019-0030-x} {\bibfield  {journal} {\bibinfo
  {journal} {Nature Reviews Physics}\ }\textbf {\bibinfo {volume} {1}},\
  \bibinfo {pages} {281} (\bibinfo {year} {2019})}\BibitemShut {NoStop}%
\bibitem [{\citenamefont {Lahaye}\ \emph {et~al.}(2009)\citenamefont {Lahaye},
  \citenamefont {Menotti}, \citenamefont {Santos}, \citenamefont {Lewenstein},\
  and\ \citenamefont {Pfau}}]{Lahaye2009}%
  \BibitemOpen
  \bibfield  {author} {\bibinfo {author} {\bibfnamefont {T.}~\bibnamefont
  {Lahaye}}, \bibinfo {author} {\bibfnamefont {C.}~\bibnamefont {Menotti}},
  \bibinfo {author} {\bibfnamefont {L.}~\bibnamefont {Santos}}, \bibinfo
  {author} {\bibfnamefont {M.}~\bibnamefont {Lewenstein}}, \ and\ \bibinfo
  {author} {\bibfnamefont {T.}~\bibnamefont {Pfau}},\ }\href {\doibase
  10.1088/0034-4885/72/12/126401} {\bibfield  {journal} {\bibinfo  {journal}
  {Reports on Progress in Physics}\ }\textbf {\bibinfo {volume} {72}},\
  \bibinfo {pages} {126401} (\bibinfo {year} {2009})}\BibitemShut {NoStop}%
\bibitem [{\citenamefont {Weimer}\ \emph {et~al.}(2010)\citenamefont {Weimer},
  \citenamefont {M{\"u}ller}, \citenamefont {Lesanovsky}, \citenamefont
  {Zoller},\ and\ \citenamefont {B{\"u}chler}}]{Weimer2010}%
  \BibitemOpen
  \bibfield  {author} {\bibinfo {author} {\bibfnamefont {H.}~\bibnamefont
  {Weimer}}, \bibinfo {author} {\bibfnamefont {M.}~\bibnamefont {M{\"u}ller}},
  \bibinfo {author} {\bibfnamefont {I.}~\bibnamefont {Lesanovsky}}, \bibinfo
  {author} {\bibfnamefont {P.}~\bibnamefont {Zoller}}, \ and\ \bibinfo {author}
  {\bibfnamefont {H.~P.}\ \bibnamefont {B{\"u}chler}},\ }\href {\doibase
  10.1038/nphys1614} {\bibfield  {journal} {\bibinfo  {journal} {Nature
  Physics}\ }\textbf {\bibinfo {volume} {6}},\ \bibinfo {pages} {382} (\bibinfo
  {year} {2010})}\BibitemShut {NoStop}%
\bibitem [{\citenamefont {Baranov}\ \emph {et~al.}(2012)\citenamefont
  {Baranov}, \citenamefont {Dalmonte}, \citenamefont {Pupillo},\ and\
  \citenamefont {Zoller}}]{Baranov2012}%
  \BibitemOpen
  \bibfield  {author} {\bibinfo {author} {\bibfnamefont {M.~A.}\ \bibnamefont
  {Baranov}}, \bibinfo {author} {\bibfnamefont {M.}~\bibnamefont {Dalmonte}},
  \bibinfo {author} {\bibfnamefont {G.}~\bibnamefont {Pupillo}}, \ and\
  \bibinfo {author} {\bibfnamefont {P.}~\bibnamefont {Zoller}},\ }\href
  {\doibase 10.1021/cr2003568} {\bibfield  {journal} {\bibinfo  {journal}
  {Chemical Reviews}\ }\textbf {\bibinfo {volume} {112}},\ \bibinfo {pages}
  {5012} (\bibinfo {year} {2012})}\BibitemShut {NoStop}%
\bibitem [{\citenamefont {Gross}\ and\ \citenamefont
  {Bloch}(2017)}]{Gross2017}%
  \BibitemOpen
  \bibfield  {author} {\bibinfo {author} {\bibfnamefont {C.}~\bibnamefont
  {Gross}}\ and\ \bibinfo {author} {\bibfnamefont {I.}~\bibnamefont {Bloch}},\
  }\href {\doibase 10.1126/science.aal3837} {\bibfield  {journal} {\bibinfo
  {journal} {Science}\ }\textbf {\bibinfo {volume} {357}},\ \bibinfo {pages}
  {995} (\bibinfo {year} {2017})}\BibitemShut {NoStop}%
\bibitem [{\citenamefont {de~Paz}\ \emph {et~al.}(2013)\citenamefont {de~Paz},
  \citenamefont {Sharma}, \citenamefont {Chotia}, \citenamefont {Mar\'echal},
  \citenamefont {Huckans}, \citenamefont {Pedri}, \citenamefont {Santos},
  \citenamefont {Gorceix}, \citenamefont {Vernac},\ and\ \citenamefont
  {Laburthe-Tolra}}]{Tolra2013}%
  \BibitemOpen
  \bibfield  {author} {\bibinfo {author} {\bibfnamefont {A.}~\bibnamefont
  {de~Paz}}, \bibinfo {author} {\bibfnamefont {A.}~\bibnamefont {Sharma}},
  \bibinfo {author} {\bibfnamefont {A.}~\bibnamefont {Chotia}}, \bibinfo
  {author} {\bibfnamefont {E.}~\bibnamefont {Mar\'echal}}, \bibinfo {author}
  {\bibfnamefont {J.~H.}\ \bibnamefont {Huckans}}, \bibinfo {author}
  {\bibfnamefont {P.}~\bibnamefont {Pedri}}, \bibinfo {author} {\bibfnamefont
  {L.}~\bibnamefont {Santos}}, \bibinfo {author} {\bibfnamefont
  {O.}~\bibnamefont {Gorceix}}, \bibinfo {author} {\bibfnamefont
  {L.}~\bibnamefont {Vernac}}, \ and\ \bibinfo {author} {\bibfnamefont
  {B.}~\bibnamefont {Laburthe-Tolra}},\ }\href {\doibase
  10.1103/PhysRevLett.111.185305} {\bibfield  {journal} {\bibinfo  {journal}
  {Phys. Rev. Lett.}\ }\textbf {\bibinfo {volume} {111}},\ \bibinfo {pages}
  {185305} (\bibinfo {year} {2013})}\BibitemShut {NoStop}%
\bibitem [{\citenamefont {Baier}\ \emph {et~al.}(2016)\citenamefont {Baier},
  \citenamefont {Mark}, \citenamefont {Petter}, \citenamefont {Aikawa},
  \citenamefont {Chomaz}, \citenamefont {Cai}, \citenamefont {Baranov},
  \citenamefont {Zoller},\ and\ \citenamefont {Ferlaino}}]{Ferlaino2016}%
  \BibitemOpen
  \bibfield  {author} {\bibinfo {author} {\bibfnamefont {S.}~\bibnamefont
  {Baier}}, \bibinfo {author} {\bibfnamefont {M.~J.}\ \bibnamefont {Mark}},
  \bibinfo {author} {\bibfnamefont {D.}~\bibnamefont {Petter}}, \bibinfo
  {author} {\bibfnamefont {K.}~\bibnamefont {Aikawa}}, \bibinfo {author}
  {\bibfnamefont {L.}~\bibnamefont {Chomaz}}, \bibinfo {author} {\bibfnamefont
  {Z.}~\bibnamefont {Cai}}, \bibinfo {author} {\bibfnamefont {M.}~\bibnamefont
  {Baranov}}, \bibinfo {author} {\bibfnamefont {P.}~\bibnamefont {Zoller}}, \
  and\ \bibinfo {author} {\bibfnamefont {F.}~\bibnamefont {Ferlaino}},\ }\href
  {\doibase 10.1126/science.aac9812} {\bibfield  {journal} {\bibinfo  {journal}
  {Science}\ }\textbf {\bibinfo {volume} {352}},\ \bibinfo {pages} {201}
  (\bibinfo {year} {2016})}\BibitemShut {NoStop}%
\bibitem [{\citenamefont {Yan}\ \emph {et~al.}(2013)\citenamefont {Yan},
  \citenamefont {Moses}, \citenamefont {Gadway}, \citenamefont {Covey},
  \citenamefont {Hazzard}, \citenamefont {Rey}, \citenamefont {Jin},\ and\
  \citenamefont {Ye}}]{Ye2013}%
  \BibitemOpen
  \bibfield  {author} {\bibinfo {author} {\bibfnamefont {B.}~\bibnamefont
  {Yan}}, \bibinfo {author} {\bibfnamefont {S.~A.}\ \bibnamefont {Moses}},
  \bibinfo {author} {\bibfnamefont {B.}~\bibnamefont {Gadway}}, \bibinfo
  {author} {\bibfnamefont {J.~P.}\ \bibnamefont {Covey}}, \bibinfo {author}
  {\bibfnamefont {K.~R.~A.}\ \bibnamefont {Hazzard}}, \bibinfo {author}
  {\bibfnamefont {A.~M.}\ \bibnamefont {Rey}}, \bibinfo {author} {\bibfnamefont
  {D.~S.}\ \bibnamefont {Jin}}, \ and\ \bibinfo {author} {\bibfnamefont
  {J.}~\bibnamefont {Ye}},\ }\href {\doibase 10.1038/nature12483} {\bibfield
  {journal} {\bibinfo  {journal} {Nature}\ }\textbf {\bibinfo {volume} {501}},\
  \bibinfo {pages} {521} (\bibinfo {year} {2013})}\BibitemShut {NoStop}%
\bibitem [{\citenamefont {Labuhn}\ \emph {et~al.}(2016)\citenamefont {Labuhn},
  \citenamefont {Barredo}, \citenamefont {Ravets}, \citenamefont
  {de~L{\'e}s{\'e}leuc}, \citenamefont {Macr{\`\i}}, \citenamefont {Lahaye},\
  and\ \citenamefont {Browaeys}}]{Browaeys2016}%
  \BibitemOpen
  \bibfield  {author} {\bibinfo {author} {\bibfnamefont {H.}~\bibnamefont
  {Labuhn}}, \bibinfo {author} {\bibfnamefont {D.}~\bibnamefont {Barredo}},
  \bibinfo {author} {\bibfnamefont {S.}~\bibnamefont {Ravets}}, \bibinfo
  {author} {\bibfnamefont {S.}~\bibnamefont {de~L{\'e}s{\'e}leuc}}, \bibinfo
  {author} {\bibfnamefont {T.}~\bibnamefont {Macr{\`\i}}}, \bibinfo {author}
  {\bibfnamefont {T.}~\bibnamefont {Lahaye}}, \ and\ \bibinfo {author}
  {\bibfnamefont {A.}~\bibnamefont {Browaeys}},\ }\href {\doibase
  10.1038/nature18274} {\bibfield  {journal} {\bibinfo  {journal} {Nature}\
  }\textbf {\bibinfo {volume} {534}},\ \bibinfo {pages} {667} (\bibinfo {year}
  {2016})}\BibitemShut {NoStop}%
\bibitem [{\citenamefont {Peter}\ \emph {et~al.}(2015)\citenamefont {Peter},
  \citenamefont {Yao}, \citenamefont {Lang}, \citenamefont {Huber},
  \citenamefont {Lukin},\ and\ \citenamefont {B\"uchler}}]{Buchler2015}%
  \BibitemOpen
  \bibfield  {author} {\bibinfo {author} {\bibfnamefont {D.}~\bibnamefont
  {Peter}}, \bibinfo {author} {\bibfnamefont {N.~Y.}\ \bibnamefont {Yao}},
  \bibinfo {author} {\bibfnamefont {N.}~\bibnamefont {Lang}}, \bibinfo {author}
  {\bibfnamefont {S.~D.}\ \bibnamefont {Huber}}, \bibinfo {author}
  {\bibfnamefont {M.~D.}\ \bibnamefont {Lukin}}, \ and\ \bibinfo {author}
  {\bibfnamefont {H.~P.}\ \bibnamefont {B\"uchler}},\ }\href {\doibase
  10.1103/PhysRevA.91.053617} {\bibfield  {journal} {\bibinfo  {journal} {Phys.
  Rev. A}\ }\textbf {\bibinfo {volume} {91}},\ \bibinfo {pages} {053617}
  (\bibinfo {year} {2015})}\BibitemShut {NoStop}%
\bibitem [{\citenamefont {Weber}\ \emph {et~al.}(2018)\citenamefont {Weber},
  \citenamefont {de~L{\'{e}}s{\'{e}}leuc}, \citenamefont {Lienhard},
  \citenamefont {Barredo}, \citenamefont {Lahaye}, \citenamefont {Browaeys},\
  and\ \citenamefont {B{\"u}chler}}]{Buchler2018}%
  \BibitemOpen
  \bibfield  {author} {\bibinfo {author} {\bibfnamefont {S.}~\bibnamefont
  {Weber}}, \bibinfo {author} {\bibfnamefont {S.}~\bibnamefont
  {de~L{\'{e}}s{\'{e}}leuc}}, \bibinfo {author} {\bibfnamefont
  {V.}~\bibnamefont {Lienhard}}, \bibinfo {author} {\bibfnamefont
  {D.}~\bibnamefont {Barredo}}, \bibinfo {author} {\bibfnamefont
  {T.}~\bibnamefont {Lahaye}}, \bibinfo {author} {\bibfnamefont
  {A.}~\bibnamefont {Browaeys}}, \ and\ \bibinfo {author} {\bibfnamefont
  {H.~P.}\ \bibnamefont {B{\"u}chler}},\ }\href {\doibase
  10.1088/2058-9565/aaca47} {\bibfield  {journal} {\bibinfo  {journal} {Quantum
  Science and Technology}\ }\textbf {\bibinfo {volume} {3}},\ \bibinfo {pages}
  {044001} (\bibinfo {year} {2018})}\BibitemShut {NoStop}%
\bibitem [{\citenamefont {Schuster}\ \emph {et~al.}(2021)\citenamefont
  {Schuster}, \citenamefont {Flicker}, \citenamefont {Li}, \citenamefont
  {Kotochigova}, \citenamefont {Moore}, \citenamefont {Ye},\ and\ \citenamefont
  {Yao}}]{Schuster2021}%
  \BibitemOpen
  \bibfield  {author} {\bibinfo {author} {\bibfnamefont {T.}~\bibnamefont
  {Schuster}}, \bibinfo {author} {\bibfnamefont {F.}~\bibnamefont {Flicker}},
  \bibinfo {author} {\bibfnamefont {M.}~\bibnamefont {Li}}, \bibinfo {author}
  {\bibfnamefont {S.}~\bibnamefont {Kotochigova}}, \bibinfo {author}
  {\bibfnamefont {J.~E.}\ \bibnamefont {Moore}}, \bibinfo {author}
  {\bibfnamefont {J.}~\bibnamefont {Ye}}, \ and\ \bibinfo {author}
  {\bibfnamefont {N.~Y.}\ \bibnamefont {Yao}},\ }\href {\doibase
  10.1103/PhysRevLett.127.015301} {\bibfield  {journal} {\bibinfo  {journal}
  {Phys. Rev. Lett.}\ }\textbf {\bibinfo {volume} {127}},\ \bibinfo {pages}
  {015301} (\bibinfo {year} {2021})}\BibitemShut {NoStop}%
\bibitem [{\citenamefont {Syzranov}\ \emph {et~al.}(2014)\citenamefont
  {Syzranov}, \citenamefont {Wall}, \citenamefont {Gurarie},\ and\
  \citenamefont {Rey}}]{Rey2014}%
  \BibitemOpen
  \bibfield  {author} {\bibinfo {author} {\bibfnamefont {S.~V.}\ \bibnamefont
  {Syzranov}}, \bibinfo {author} {\bibfnamefont {M.~L.}\ \bibnamefont {Wall}},
  \bibinfo {author} {\bibfnamefont {V.}~\bibnamefont {Gurarie}}, \ and\
  \bibinfo {author} {\bibfnamefont {A.~M.}\ \bibnamefont {Rey}},\ }\href
  {\doibase 10.1038/ncomms6391} {\bibfield  {journal} {\bibinfo  {journal}
  {Nature Communications}\ }\textbf {\bibinfo {volume} {5}},\ \bibinfo {pages}
  {5391} (\bibinfo {year} {2014})}\BibitemShut {NoStop}%
\bibitem [{\citenamefont {Syzranov}\ \emph {et~al.}(2016)\citenamefont
  {Syzranov}, \citenamefont {Wall}, \citenamefont {Zhu}, \citenamefont
  {Gurarie},\ and\ \citenamefont {Rey}}]{Rey2016}%
  \BibitemOpen
  \bibfield  {author} {\bibinfo {author} {\bibfnamefont {S.~V.}\ \bibnamefont
  {Syzranov}}, \bibinfo {author} {\bibfnamefont {M.~L.}\ \bibnamefont {Wall}},
  \bibinfo {author} {\bibfnamefont {B.}~\bibnamefont {Zhu}}, \bibinfo {author}
  {\bibfnamefont {V.}~\bibnamefont {Gurarie}}, \ and\ \bibinfo {author}
  {\bibfnamefont {A.~M.}\ \bibnamefont {Rey}},\ }\href {\doibase
  10.1038/ncomms13543} {\bibfield  {journal} {\bibinfo  {journal} {Nature
  Communications}\ }\textbf {\bibinfo {volume} {7}},\ \bibinfo {pages} {13543}
  (\bibinfo {year} {2016})}\BibitemShut {NoStop}%
\bibitem [{\citenamefont {Salerno}\ \emph {et~al.}(2020)\citenamefont
  {Salerno}, \citenamefont {Palumbo}, \citenamefont {Goldman},\ and\
  \citenamefont {Di~Liberto}}]{Salerno2020}%
  \BibitemOpen
  \bibfield  {author} {\bibinfo {author} {\bibfnamefont {G.}~\bibnamefont
  {Salerno}}, \bibinfo {author} {\bibfnamefont {G.}~\bibnamefont {Palumbo}},
  \bibinfo {author} {\bibfnamefont {N.}~\bibnamefont {Goldman}}, \ and\
  \bibinfo {author} {\bibfnamefont {M.}~\bibnamefont {Di~Liberto}},\ }\href
  {\doibase 10.1103/PhysRevResearch.2.013348} {\bibfield  {journal} {\bibinfo
  {journal} {Phys. Rev. Research}\ }\textbf {\bibinfo {volume} {2}},\ \bibinfo
  {pages} {013348} (\bibinfo {year} {2020})}\BibitemShut {NoStop}%
\bibitem [{\citenamefont {Manmana}\ \emph {et~al.}(2013)\citenamefont
  {Manmana}, \citenamefont {Stoudenmire}, \citenamefont {Hazzard},
  \citenamefont {Rey},\ and\ \citenamefont {Gorshkov}}]{Manmana2013}%
  \BibitemOpen
  \bibfield  {author} {\bibinfo {author} {\bibfnamefont {S.~R.}\ \bibnamefont
  {Manmana}}, \bibinfo {author} {\bibfnamefont {E.~M.}\ \bibnamefont
  {Stoudenmire}}, \bibinfo {author} {\bibfnamefont {K.~R.~A.}\ \bibnamefont
  {Hazzard}}, \bibinfo {author} {\bibfnamefont {A.~M.}\ \bibnamefont {Rey}}, \
  and\ \bibinfo {author} {\bibfnamefont {A.~V.}\ \bibnamefont {Gorshkov}},\
  }\href {\doibase 10.1103/PhysRevB.87.081106} {\bibfield  {journal} {\bibinfo
  {journal} {Phys. Rev. B}\ }\textbf {\bibinfo {volume} {87}},\ \bibinfo
  {pages} {081106} (\bibinfo {year} {2013})}\BibitemShut {NoStop}%
\bibitem [{\citenamefont {Yao}\ \emph {et~al.}(2013)\citenamefont {Yao},
  \citenamefont {Gorshkov}, \citenamefont {Laumann}, \citenamefont {L\"auchli},
  \citenamefont {Ye},\ and\ \citenamefont {Lukin}}]{Yao2013}%
  \BibitemOpen
  \bibfield  {author} {\bibinfo {author} {\bibfnamefont {N.~Y.}\ \bibnamefont
  {Yao}}, \bibinfo {author} {\bibfnamefont {A.~V.}\ \bibnamefont {Gorshkov}},
  \bibinfo {author} {\bibfnamefont {C.~R.}\ \bibnamefont {Laumann}}, \bibinfo
  {author} {\bibfnamefont {A.~M.}\ \bibnamefont {L\"auchli}}, \bibinfo {author}
  {\bibfnamefont {J.}~\bibnamefont {Ye}}, \ and\ \bibinfo {author}
  {\bibfnamefont {M.~D.}\ \bibnamefont {Lukin}},\ }\href {\doibase
  10.1103/PhysRevLett.110.185302} {\bibfield  {journal} {\bibinfo  {journal}
  {Phys. Rev. Lett.}\ }\textbf {\bibinfo {volume} {110}},\ \bibinfo {pages}
  {185302} (\bibinfo {year} {2013})}\BibitemShut {NoStop}%
\bibitem [{\citenamefont {Yao}\ \emph {et~al.}(2018)\citenamefont {Yao},
  \citenamefont {Zaletel}, \citenamefont {Stamper-Kurn},\ and\ \citenamefont
  {Vishwanath}}]{Yao2018}%
  \BibitemOpen
  \bibfield  {author} {\bibinfo {author} {\bibfnamefont {N.~Y.}\ \bibnamefont
  {Yao}}, \bibinfo {author} {\bibfnamefont {M.~P.}\ \bibnamefont {Zaletel}},
  \bibinfo {author} {\bibfnamefont {D.~M.}\ \bibnamefont {Stamper-Kurn}}, \
  and\ \bibinfo {author} {\bibfnamefont {A.}~\bibnamefont {Vishwanath}},\
  }\href {\doibase 10.1038/s41567-017-0030-7} {\bibfield  {journal} {\bibinfo
  {journal} {Nature Physics}\ }\textbf {\bibinfo {volume} {14}},\ \bibinfo
  {pages} {405} (\bibinfo {year} {2018})}\BibitemShut {NoStop}%
\bibitem [{\citenamefont {Lienhard}\ \emph {et~al.}(2020)\citenamefont
  {Lienhard}, \citenamefont {Scholl}, \citenamefont {Weber}, \citenamefont
  {Barredo}, \citenamefont {de~L\'es\'eleuc}, \citenamefont {Bai},
  \citenamefont {Lang}, \citenamefont {Fleischhauer}, \citenamefont
  {B\"uchler}, \citenamefont {Lahaye},\ and\ \citenamefont
  {Browaeys}}]{Browaeys2020}%
  \BibitemOpen
  \bibfield  {author} {\bibinfo {author} {\bibfnamefont {V.}~\bibnamefont
  {Lienhard}}, \bibinfo {author} {\bibfnamefont {P.}~\bibnamefont {Scholl}},
  \bibinfo {author} {\bibfnamefont {S.}~\bibnamefont {Weber}}, \bibinfo
  {author} {\bibfnamefont {D.}~\bibnamefont {Barredo}}, \bibinfo {author}
  {\bibfnamefont {S.}~\bibnamefont {de~L\'es\'eleuc}}, \bibinfo {author}
  {\bibfnamefont {R.}~\bibnamefont {Bai}}, \bibinfo {author} {\bibfnamefont
  {N.}~\bibnamefont {Lang}}, \bibinfo {author} {\bibfnamefont {M.}~\bibnamefont
  {Fleischhauer}}, \bibinfo {author} {\bibfnamefont {H.~P.}\ \bibnamefont
  {B\"uchler}}, \bibinfo {author} {\bibfnamefont {T.}~\bibnamefont {Lahaye}}, \
  and\ \bibinfo {author} {\bibfnamefont {A.}~\bibnamefont {Browaeys}},\ }\href
  {\doibase 10.1103/PhysRevX.10.021031} {\bibfield  {journal} {\bibinfo
  {journal} {Phys. Rev. X}\ }\textbf {\bibinfo {volume} {10}},\ \bibinfo
  {pages} {021031} (\bibinfo {year} {2020})}\BibitemShut {NoStop}%
\bibitem [{\citenamefont {de~L{\'e}s{\'e}leuc}\ \emph
  {et~al.}(2019)\citenamefont {de~L{\'e}s{\'e}leuc}, \citenamefont {Lienhard},
  \citenamefont {Scholl}, \citenamefont {Barredo}, \citenamefont {Weber},
  \citenamefont {Lang}, \citenamefont {B{\"u}chler}, \citenamefont {Lahaye},\
  and\ \citenamefont {Browaeys}}]{Browaeys2019}%
  \BibitemOpen
  \bibfield  {author} {\bibinfo {author} {\bibfnamefont {S.}~\bibnamefont
  {de~L{\'e}s{\'e}leuc}}, \bibinfo {author} {\bibfnamefont {V.}~\bibnamefont
  {Lienhard}}, \bibinfo {author} {\bibfnamefont {P.}~\bibnamefont {Scholl}},
  \bibinfo {author} {\bibfnamefont {D.}~\bibnamefont {Barredo}}, \bibinfo
  {author} {\bibfnamefont {S.}~\bibnamefont {Weber}}, \bibinfo {author}
  {\bibfnamefont {N.}~\bibnamefont {Lang}}, \bibinfo {author} {\bibfnamefont
  {H.~P.}\ \bibnamefont {B{\"u}chler}}, \bibinfo {author} {\bibfnamefont
  {T.}~\bibnamefont {Lahaye}}, \ and\ \bibinfo {author} {\bibfnamefont
  {A.}~\bibnamefont {Browaeys}},\ }\href {\doibase 10.1126/science.aav9105}
  {\bibfield  {journal} {\bibinfo  {journal} {Science}\ }\textbf {\bibinfo
  {volume} {365}},\ \bibinfo {pages} {775} (\bibinfo {year}
  {2019})}\BibitemShut {NoStop}%
\bibitem [{\citenamefont {Li}\ and\ \citenamefont {Liu}(2016)}]{Liu2016}%
  \BibitemOpen
  \bibfield  {author} {\bibinfo {author} {\bibfnamefont {X.}~\bibnamefont
  {Li}}\ and\ \bibinfo {author} {\bibfnamefont {W.~V.}\ \bibnamefont {Liu}},\
  }\href {\doibase 10.1088/0034-4885/79/11/116401} {\bibfield  {journal}
  {\bibinfo  {journal} {Reports on Progress in Physics}\ }\textbf {\bibinfo
  {volume} {79}},\ \bibinfo {pages} {116401} (\bibinfo {year}
  {2016})}\BibitemShut {NoStop}%
\bibitem [{\citenamefont {Isacsson}\ and\ \citenamefont
  {Girvin}(2005)}]{Girvin2005}%
  \BibitemOpen
  \bibfield  {author} {\bibinfo {author} {\bibfnamefont {A.}~\bibnamefont
  {Isacsson}}\ and\ \bibinfo {author} {\bibfnamefont {S.~M.}\ \bibnamefont
  {Girvin}},\ }\href {\doibase 10.1103/PhysRevA.72.053604} {\bibfield
  {journal} {\bibinfo  {journal} {Phys. Rev. A}\ }\textbf {\bibinfo {volume}
  {72}},\ \bibinfo {pages} {053604} (\bibinfo {year} {2005})}\BibitemShut
  {NoStop}%
\bibitem [{\citenamefont {M\"uller}\ \emph {et~al.}(2007)\citenamefont
  {M\"uller}, \citenamefont {F\"olling}, \citenamefont {Widera},\ and\
  \citenamefont {Bloch}}]{Muller2007}%
  \BibitemOpen
  \bibfield  {author} {\bibinfo {author} {\bibfnamefont {T.}~\bibnamefont
  {M\"uller}}, \bibinfo {author} {\bibfnamefont {S.}~\bibnamefont {F\"olling}},
  \bibinfo {author} {\bibfnamefont {A.}~\bibnamefont {Widera}}, \ and\ \bibinfo
  {author} {\bibfnamefont {I.}~\bibnamefont {Bloch}},\ }\href {\doibase
  10.1103/PhysRevLett.99.200405} {\bibfield  {journal} {\bibinfo  {journal}
  {Phys. Rev. Lett.}\ }\textbf {\bibinfo {volume} {99}},\ \bibinfo {pages}
  {200405} (\bibinfo {year} {2007})}\BibitemShut {NoStop}%
\bibitem [{\citenamefont {Wirth}\ \emph {et~al.}(2011)\citenamefont {Wirth},
  \citenamefont {{\"O}lschl{\"a}ger},\ and\ \citenamefont
  {Hemmerich}}]{Hemmerich2011}%
  \BibitemOpen
  \bibfield  {author} {\bibinfo {author} {\bibfnamefont {G.}~\bibnamefont
  {Wirth}}, \bibinfo {author} {\bibfnamefont {M.}~\bibnamefont
  {{\"O}lschl{\"a}ger}}, \ and\ \bibinfo {author} {\bibfnamefont
  {A.}~\bibnamefont {Hemmerich}},\ }\href {\doibase 10.1038/nphys1857}
  {\bibfield  {journal} {\bibinfo  {journal} {Nature Physics}\ }\textbf
  {\bibinfo {volume} {7}},\ \bibinfo {pages} {147} (\bibinfo {year}
  {2011})}\BibitemShut {NoStop}%
\bibitem [{\citenamefont {Prodan}\ and\ \citenamefont
  {Prodan}(2009)}]{Prodan2009}%
  \BibitemOpen
  \bibfield  {author} {\bibinfo {author} {\bibfnamefont {E.}~\bibnamefont
  {Prodan}}\ and\ \bibinfo {author} {\bibfnamefont {C.}~\bibnamefont
  {Prodan}},\ }\href {\doibase 10.1103/PhysRevLett.103.248101} {\bibfield
  {journal} {\bibinfo  {journal} {Phys. Rev. Lett.}\ }\textbf {\bibinfo
  {volume} {103}},\ \bibinfo {pages} {248101} (\bibinfo {year}
  {2009})}\BibitemShut {NoStop}%
\bibitem [{\citenamefont {Bermudez}\ \emph {et~al.}(2011)\citenamefont
  {Bermudez}, \citenamefont {Schaetz},\ and\ \citenamefont
  {Porras}}]{Bermudez2011}%
  \BibitemOpen
  \bibfield  {author} {\bibinfo {author} {\bibfnamefont {A.}~\bibnamefont
  {Bermudez}}, \bibinfo {author} {\bibfnamefont {T.}~\bibnamefont {Schaetz}}, \
  and\ \bibinfo {author} {\bibfnamefont {D.}~\bibnamefont {Porras}},\ }\href
  {\doibase 10.1103/PhysRevLett.107.150501} {\bibfield  {journal} {\bibinfo
  {journal} {Phys. Rev. Lett.}\ }\textbf {\bibinfo {volume} {107}},\ \bibinfo
  {pages} {150501} (\bibinfo {year} {2011})}\BibitemShut {NoStop}%
\bibitem [{\citenamefont {Kane}\ and\ \citenamefont
  {Lubensky}(2014)}]{Kane2014}%
  \BibitemOpen
  \bibfield  {author} {\bibinfo {author} {\bibfnamefont {C.~L.}\ \bibnamefont
  {Kane}}\ and\ \bibinfo {author} {\bibfnamefont {T.~C.}\ \bibnamefont
  {Lubensky}},\ }\href {\doibase 10.1038/nphys2835} {\bibfield  {journal}
  {\bibinfo  {journal} {Nature Physics}\ }\textbf {\bibinfo {volume} {10}},\
  \bibinfo {pages} {39} (\bibinfo {year} {2014})}\BibitemShut {NoStop}%
\bibitem [{\citenamefont {S{\"u}sstrunk}\ and\ \citenamefont
  {Huber}(2015)}]{Huber2015}%
  \BibitemOpen
  \bibfield  {author} {\bibinfo {author} {\bibfnamefont {R.}~\bibnamefont
  {S{\"u}sstrunk}}\ and\ \bibinfo {author} {\bibfnamefont {S.~D.}\ \bibnamefont
  {Huber}},\ }\href {\doibase 10.1126/science.aab0239} {\bibfield  {journal}
  {\bibinfo  {journal} {Science}\ }\textbf {\bibinfo {volume} {349}},\ \bibinfo
  {pages} {47} (\bibinfo {year} {2015})}\BibitemShut {NoStop}%
\bibitem [{\citenamefont {Stenull}\ \emph {et~al.}(2016)\citenamefont
  {Stenull}, \citenamefont {Kane},\ and\ \citenamefont
  {Lubensky}}]{Lubensky2016}%
  \BibitemOpen
  \bibfield  {author} {\bibinfo {author} {\bibfnamefont {O.}~\bibnamefont
  {Stenull}}, \bibinfo {author} {\bibfnamefont {C.~L.}\ \bibnamefont {Kane}}, \
  and\ \bibinfo {author} {\bibfnamefont {T.~C.}\ \bibnamefont {Lubensky}},\
  }\href {\doibase 10.1103/PhysRevLett.117.068001} {\bibfield  {journal}
  {\bibinfo  {journal} {Phys. Rev. Lett.}\ }\textbf {\bibinfo {volume} {117}},\
  \bibinfo {pages} {068001} (\bibinfo {year} {2016})}\BibitemShut {NoStop}%
\bibitem [{\citenamefont {Salerno}\ \emph {et~al.}(2016)\citenamefont
  {Salerno}, \citenamefont {Ozawa}, \citenamefont {Price},\ and\ \citenamefont
  {Carusotto}}]{Salerno2016}%
  \BibitemOpen
  \bibfield  {author} {\bibinfo {author} {\bibfnamefont {G.}~\bibnamefont
  {Salerno}}, \bibinfo {author} {\bibfnamefont {T.}~\bibnamefont {Ozawa}},
  \bibinfo {author} {\bibfnamefont {H.~M.}\ \bibnamefont {Price}}, \ and\
  \bibinfo {author} {\bibfnamefont {I.}~\bibnamefont {Carusotto}},\ }\href
  {\doibase 10.1103/PhysRevB.93.085105} {\bibfield  {journal} {\bibinfo
  {journal} {Phys. Rev. B}\ }\textbf {\bibinfo {volume} {93}},\ \bibinfo
  {pages} {085105} (\bibinfo {year} {2016})}\BibitemShut {NoStop}%
\bibitem [{\citenamefont {Li}\ \emph {et~al.}(2020)\citenamefont {Li},
  \citenamefont {Wang}, \citenamefont {Liu}, \citenamefont {Li}, \citenamefont
  {Zhang},\ and\ \citenamefont {Chen}}]{Li2020}%
  \BibitemOpen
  \bibfield  {author} {\bibinfo {author} {\bibfnamefont {J.}~\bibnamefont
  {Li}}, \bibinfo {author} {\bibfnamefont {L.}~\bibnamefont {Wang}}, \bibinfo
  {author} {\bibfnamefont {J.}~\bibnamefont {Liu}}, \bibinfo {author}
  {\bibfnamefont {R.}~\bibnamefont {Li}}, \bibinfo {author} {\bibfnamefont
  {Z.}~\bibnamefont {Zhang}}, \ and\ \bibinfo {author} {\bibfnamefont {X.-Q.}\
  \bibnamefont {Chen}},\ }\href {\doibase 10.1103/PhysRevB.101.081403}
  {\bibfield  {journal} {\bibinfo  {journal} {Phys. Rev. B}\ }\textbf {\bibinfo
  {volume} {101}},\ \bibinfo {pages} {081403} (\bibinfo {year}
  {2020})}\BibitemShut {NoStop}%
\bibitem [{\citenamefont {Wei}\ \emph {et~al.}(2021)\citenamefont {Wei},
  \citenamefont {Zhang}, \citenamefont {Deng}, \citenamefont {Lu},
  \citenamefont {Huang}, \citenamefont {Yan}, \citenamefont {Chen},
  \citenamefont {Liu},\ and\ \citenamefont {Jia}}]{Wei2021}%
  \BibitemOpen
  \bibfield  {author} {\bibinfo {author} {\bibfnamefont {Q.}~\bibnamefont
  {Wei}}, \bibinfo {author} {\bibfnamefont {X.}~\bibnamefont {Zhang}}, \bibinfo
  {author} {\bibfnamefont {W.}~\bibnamefont {Deng}}, \bibinfo {author}
  {\bibfnamefont {J.}~\bibnamefont {Lu}}, \bibinfo {author} {\bibfnamefont
  {X.}~\bibnamefont {Huang}}, \bibinfo {author} {\bibfnamefont
  {M.}~\bibnamefont {Yan}}, \bibinfo {author} {\bibfnamefont {G.}~\bibnamefont
  {Chen}}, \bibinfo {author} {\bibfnamefont {Z.}~\bibnamefont {Liu}}, \ and\
  \bibinfo {author} {\bibfnamefont {S.}~\bibnamefont {Jia}},\ }\href {\doibase
  10.1038/s41563-021-00933-4} {\bibfield  {journal} {\bibinfo  {journal}
  {Nature Materials}\ }\textbf {\bibinfo {volume} {20}},\ \bibinfo {pages}
  {812} (\bibinfo {year} {2021})}\BibitemShut {NoStop}%
\bibitem [{\citenamefont {Luo}\ \emph {et~al.}(2021)\citenamefont {Luo},
  \citenamefont {Wang}, \citenamefont {Lin}, \citenamefont {Jiang},
  \citenamefont {Wu}, \citenamefont {Li},\ and\ \citenamefont
  {Jiang}}]{Luo2021}%
  \BibitemOpen
  \bibfield  {author} {\bibinfo {author} {\bibfnamefont {L.}~\bibnamefont
  {Luo}}, \bibinfo {author} {\bibfnamefont {H.-X.}\ \bibnamefont {Wang}},
  \bibinfo {author} {\bibfnamefont {Z.-K.}\ \bibnamefont {Lin}}, \bibinfo
  {author} {\bibfnamefont {B.}~\bibnamefont {Jiang}}, \bibinfo {author}
  {\bibfnamefont {Y.}~\bibnamefont {Wu}}, \bibinfo {author} {\bibfnamefont
  {F.}~\bibnamefont {Li}}, \ and\ \bibinfo {author} {\bibfnamefont {J.-H.}\
  \bibnamefont {Jiang}},\ }\href {\doibase 10.1038/s41563-021-00985-6}
  {\bibfield  {journal} {\bibinfo  {journal} {Nature Materials}\ }\textbf
  {\bibinfo {volume} {20}},\ \bibinfo {pages} {794} (\bibinfo {year}
  {2021})}\BibitemShut {NoStop}%
\bibitem [{\citenamefont {Lange}\ \emph {et~al.}(2022)\citenamefont {Lange},
  \citenamefont {Bouhon}, \citenamefont {Monserrat},\ and\ \citenamefont
  {Slager}}]{Slager2022}%
  \BibitemOpen
  \bibfield  {author} {\bibinfo {author} {\bibfnamefont {G.~F.}\ \bibnamefont
  {Lange}}, \bibinfo {author} {\bibfnamefont {A.}~\bibnamefont {Bouhon}},
  \bibinfo {author} {\bibfnamefont {B.}~\bibnamefont {Monserrat}}, \ and\
  \bibinfo {author} {\bibfnamefont {R.-J.}\ \bibnamefont {Slager}},\ }\href
  {\doibase 10.1103/PhysRevB.105.064301} {\bibfield  {journal} {\bibinfo
  {journal} {Phys. Rev. B}\ }\textbf {\bibinfo {volume} {105}},\ \bibinfo
  {pages} {064301} (\bibinfo {year} {2022})}\BibitemShut {NoStop}%
\bibitem [{\citenamefont {Bruus}\ and\ \citenamefont
  {Flensberg}(2004)}]{Bruus}%
  \BibitemOpen
  \bibfield  {author} {\bibinfo {author} {\bibfnamefont {H.}~\bibnamefont
  {Bruus}}\ and\ \bibinfo {author} {\bibfnamefont {K.}~\bibnamefont
  {Flensberg}},\ }\href@noop {} {\emph {\bibinfo {title} {Many-Body Quantum
  Theory in Condensed Matter Physics: An Introduction}}}\ (\bibinfo
  {publisher} {Oxford University Press},\ \bibinfo {year} {2004})\BibitemShut
  {NoStop}%
\bibitem [{\citenamefont {Schnyder}(2018)}]{Schnyder2018}%
  \BibitemOpen
  \bibfield  {author} {\bibinfo {author} {\bibfnamefont {A.~P.}\ \bibnamefont
  {Schnyder}},\ }\href
  {https://www.fkf.mpg.de/6431357/topo_lecture_notes_schnyder_TMS18.pdf}
  {\bibfield  {journal} {\bibinfo  {journal} {Topological Matter School}\ }
  (\bibinfo {year} {2018})}\BibitemShut {NoStop}%
\bibitem [{\citenamefont {Feng}\ \emph {et~al.}(2021)\citenamefont {Feng},
  \citenamefont {Zhu}, \citenamefont {Wu},\ and\ \citenamefont
  {Yang}}]{Feng2021}%
  \BibitemOpen
  \bibfield  {author} {\bibinfo {author} {\bibfnamefont {X.}~\bibnamefont
  {Feng}}, \bibinfo {author} {\bibfnamefont {J.}~\bibnamefont {Zhu}}, \bibinfo
  {author} {\bibfnamefont {W.}~\bibnamefont {Wu}}, \ and\ \bibinfo {author}
  {\bibfnamefont {S.~A.}\ \bibnamefont {Yang}},\ }\href {\doibase
  10.1088/1674-1056/ac1f0c} {\bibfield  {journal} {\bibinfo  {journal} {Chinese
  Physics B}\ }\textbf {\bibinfo {volume} {30}},\ \bibinfo {pages} {107304}
  (\bibinfo {year} {2021})}\BibitemShut {NoStop}%
\bibitem [{\citenamefont {Wu}\ and\ \citenamefont {Das~Sarma}(2008)}]{Wu2008}%
  \BibitemOpen
  \bibfield  {author} {\bibinfo {author} {\bibfnamefont {C.}~\bibnamefont
  {Wu}}\ and\ \bibinfo {author} {\bibfnamefont {S.}~\bibnamefont {Das~Sarma}},\
  }\href {\doibase 10.1103/PhysRevB.77.235107} {\bibfield  {journal} {\bibinfo
  {journal} {Phys. Rev. B}\ }\textbf {\bibinfo {volume} {77}},\ \bibinfo
  {pages} {235107} (\bibinfo {year} {2008})}\BibitemShut {NoStop}%
\bibitem [{\citenamefont {Jacqmin}\ \emph {et~al.}(2014)\citenamefont
  {Jacqmin}, \citenamefont {Carusotto}, \citenamefont {Sagnes}, \citenamefont
  {Abbarchi}, \citenamefont {Solnyshkov}, \citenamefont {Malpuech},
  \citenamefont {Galopin}, \citenamefont {Lema\^{\i}tre}, \citenamefont
  {Bloch},\ and\ \citenamefont {Amo}}]{Amo2014}%
  \BibitemOpen
  \bibfield  {author} {\bibinfo {author} {\bibfnamefont {T.}~\bibnamefont
  {Jacqmin}}, \bibinfo {author} {\bibfnamefont {I.}~\bibnamefont {Carusotto}},
  \bibinfo {author} {\bibfnamefont {I.}~\bibnamefont {Sagnes}}, \bibinfo
  {author} {\bibfnamefont {M.}~\bibnamefont {Abbarchi}}, \bibinfo {author}
  {\bibfnamefont {D.~D.}\ \bibnamefont {Solnyshkov}}, \bibinfo {author}
  {\bibfnamefont {G.}~\bibnamefont {Malpuech}}, \bibinfo {author}
  {\bibfnamefont {E.}~\bibnamefont {Galopin}}, \bibinfo {author} {\bibfnamefont
  {A.}~\bibnamefont {Lema\^{\i}tre}}, \bibinfo {author} {\bibfnamefont
  {J.}~\bibnamefont {Bloch}}, \ and\ \bibinfo {author} {\bibfnamefont
  {A.}~\bibnamefont {Amo}},\ }\href {\doibase 10.1103/PhysRevLett.112.116402}
  {\bibfield  {journal} {\bibinfo  {journal} {Phys. Rev. Lett.}\ }\textbf
  {\bibinfo {volume} {112}},\ \bibinfo {pages} {116402} (\bibinfo {year}
  {2014})}\BibitemShut {NoStop}%
\bibitem [{\citenamefont {Gardenier}\ \emph {et~al.}(2020)\citenamefont
  {Gardenier}, \citenamefont {van~den Broeke}, \citenamefont {Moes},
  \citenamefont {Swart}, \citenamefont {Delerue}, \citenamefont {Slot},
  \citenamefont {Smith},\ and\ \citenamefont {Vanmaekelbergh}}]{Gardenier2020}%
  \BibitemOpen
  \bibfield  {author} {\bibinfo {author} {\bibfnamefont {T.~S.}\ \bibnamefont
  {Gardenier}}, \bibinfo {author} {\bibfnamefont {J.~J.}\ \bibnamefont {van~den
  Broeke}}, \bibinfo {author} {\bibfnamefont {J.~R.}\ \bibnamefont {Moes}},
  \bibinfo {author} {\bibfnamefont {I.}~\bibnamefont {Swart}}, \bibinfo
  {author} {\bibfnamefont {C.}~\bibnamefont {Delerue}}, \bibinfo {author}
  {\bibfnamefont {M.~R.}\ \bibnamefont {Slot}}, \bibinfo {author}
  {\bibfnamefont {C.~M.}\ \bibnamefont {Smith}}, \ and\ \bibinfo {author}
  {\bibfnamefont {D.}~\bibnamefont {Vanmaekelbergh}},\ }\bibfield  {booktitle}
  {\emph {\bibinfo {booktitle} {ACS Nano}},\ }\href {\doibase
  10.1021/acsnano.0c05747} {\bibfield  {journal} {\bibinfo  {journal} {ACS
  Nano}\ }\textbf {\bibinfo {volume} {14}},\ \bibinfo {pages} {13638} (\bibinfo
  {year} {2020})}\BibitemShut {NoStop}%
\bibitem [{\citenamefont {Castro~Neto}\ \emph {et~al.}(2009)\citenamefont
  {Castro~Neto}, \citenamefont {Guinea}, \citenamefont {Peres}, \citenamefont
  {Novoselov},\ and\ \citenamefont {Geim}}]{CastroNeto2009}%
  \BibitemOpen
  \bibfield  {author} {\bibinfo {author} {\bibfnamefont {A.~H.}\ \bibnamefont
  {Castro~Neto}}, \bibinfo {author} {\bibfnamefont {F.}~\bibnamefont {Guinea}},
  \bibinfo {author} {\bibfnamefont {N.~M.~R.}\ \bibnamefont {Peres}}, \bibinfo
  {author} {\bibfnamefont {K.~S.}\ \bibnamefont {Novoselov}}, \ and\ \bibinfo
  {author} {\bibfnamefont {A.~K.}\ \bibnamefont {Geim}},\ }\href {\doibase
  10.1103/RevModPhys.81.109} {\bibfield  {journal} {\bibinfo  {journal} {Rev.
  Mod. Phys.}\ }\textbf {\bibinfo {volume} {81}},\ \bibinfo {pages} {109}
  (\bibinfo {year} {2009})}\BibitemShut {NoStop}%
\bibitem [{\citenamefont {Park}\ \emph {et~al.}(2021)\citenamefont {Park},
  \citenamefont {Hwang}, \citenamefont {Choi},\ and\ \citenamefont
  {Yang}}]{Park2021}%
  \BibitemOpen
  \bibfield  {author} {\bibinfo {author} {\bibfnamefont {S.}~\bibnamefont
  {Park}}, \bibinfo {author} {\bibfnamefont {Y.}~\bibnamefont {Hwang}},
  \bibinfo {author} {\bibfnamefont {H.~C.}\ \bibnamefont {Choi}}, \ and\
  \bibinfo {author} {\bibfnamefont {B.-J.}\ \bibnamefont {Yang}},\ }\href
  {\doibase 10.1038/s41467-021-27158-y} {\bibfield  {journal} {\bibinfo
  {journal} {Nature Communications}\ }\textbf {\bibinfo {volume} {12}},\
  \bibinfo {pages} {6781} (\bibinfo {year} {2021})}\BibitemShut {NoStop}%
\bibitem [{\citenamefont {Bernevig}\ and\ \citenamefont
  {Hughes}(2013)}]{Bernevig2013}%
  \BibitemOpen
  \bibfield  {author} {\bibinfo {author} {\bibfnamefont {B.~A.}\ \bibnamefont
  {Bernevig}}\ and\ \bibinfo {author} {\bibfnamefont {T.~L.}\ \bibnamefont
  {Hughes}},\ }\href@noop {} {\emph {\bibinfo {title} {Topological Insulators
  and Topological Superconductors}}}\ (\bibinfo  {publisher} {Princeton
  University Press},\ \bibinfo {year} {2013})\BibitemShut {NoStop}%
\bibitem [{\citenamefont {Sun}\ \emph {et~al.}(2009)\citenamefont {Sun},
  \citenamefont {Yao}, \citenamefont {Fradkin},\ and\ \citenamefont
  {Kivelson}}]{Sun2009}%
  \BibitemOpen
  \bibfield  {author} {\bibinfo {author} {\bibfnamefont {K.}~\bibnamefont
  {Sun}}, \bibinfo {author} {\bibfnamefont {H.}~\bibnamefont {Yao}}, \bibinfo
  {author} {\bibfnamefont {E.}~\bibnamefont {Fradkin}}, \ and\ \bibinfo
  {author} {\bibfnamefont {S.~A.}\ \bibnamefont {Kivelson}},\ }\href {\doibase
  10.1103/PhysRevLett.103.046811} {\bibfield  {journal} {\bibinfo  {journal}
  {Phys. Rev. Lett.}\ }\textbf {\bibinfo {volume} {103}},\ \bibinfo {pages}
  {046811} (\bibinfo {year} {2009})}\BibitemShut {NoStop}%
\bibitem [{\citenamefont {Montambaux}\ \emph {et~al.}(2018)\citenamefont
  {Montambaux}, \citenamefont {Lim}, \citenamefont {Fuchs},\ and\ \citenamefont
  {Pi\'echon}}]{Montambaux2018}%
  \BibitemOpen
  \bibfield  {author} {\bibinfo {author} {\bibfnamefont {G.}~\bibnamefont
  {Montambaux}}, \bibinfo {author} {\bibfnamefont {L.-K.}\ \bibnamefont {Lim}},
  \bibinfo {author} {\bibfnamefont {J.-N.}\ \bibnamefont {Fuchs}}, \ and\
  \bibinfo {author} {\bibfnamefont {F.}~\bibnamefont {Pi\'echon}},\ }\href
  {\doibase 10.1103/PhysRevLett.121.256402} {\bibfield  {journal} {\bibinfo
  {journal} {Phys. Rev. Lett.}\ }\textbf {\bibinfo {volume} {121}},\ \bibinfo
  {pages} {256402} (\bibinfo {year} {2018})}\BibitemShut {NoStop}%
\bibitem [{\citenamefont {He}\ \emph {et~al.}(2012)\citenamefont {He},
  \citenamefont {Moore},\ and\ \citenamefont {Varma}}]{Varma2012}%
  \BibitemOpen
  \bibfield  {author} {\bibinfo {author} {\bibfnamefont {Y.}~\bibnamefont
  {He}}, \bibinfo {author} {\bibfnamefont {J.}~\bibnamefont {Moore}}, \ and\
  \bibinfo {author} {\bibfnamefont {C.~M.}\ \bibnamefont {Varma}},\ }\href
  {\doibase 10.1103/PhysRevB.85.155106} {\bibfield  {journal} {\bibinfo
  {journal} {Phys. Rev. B}\ }\textbf {\bibinfo {volume} {85}},\ \bibinfo
  {pages} {155106} (\bibinfo {year} {2012})}\BibitemShut {NoStop}%
\bibitem [{\citenamefont {Montambaux}\ \emph {et~al.}(2009)\citenamefont
  {Montambaux}, \citenamefont {Pi\'echon}, \citenamefont {Fuchs},\ and\
  \citenamefont {Goerbig}}]{Montambaux2009}%
  \BibitemOpen
  \bibfield  {author} {\bibinfo {author} {\bibfnamefont {G.}~\bibnamefont
  {Montambaux}}, \bibinfo {author} {\bibfnamefont {F.}~\bibnamefont
  {Pi\'echon}}, \bibinfo {author} {\bibfnamefont {J.-N.}\ \bibnamefont
  {Fuchs}}, \ and\ \bibinfo {author} {\bibfnamefont {M.~O.}\ \bibnamefont
  {Goerbig}},\ }\href {\doibase 10.1103/PhysRevB.80.153412} {\bibfield
  {journal} {\bibinfo  {journal} {Phys. Rev. B}\ }\textbf {\bibinfo {volume}
  {80}},\ \bibinfo {pages} {153412} (\bibinfo {year} {2009})}\BibitemShut
  {NoStop}%
\bibitem [{\citenamefont {Tarruell}\ \emph {et~al.}(2012)\citenamefont
  {Tarruell}, \citenamefont {Greif}, \citenamefont {Uehlinger}, \citenamefont
  {Jotzu},\ and\ \citenamefont {Esslinger}}]{Tarruell2012}%
  \BibitemOpen
  \bibfield  {author} {\bibinfo {author} {\bibfnamefont {L.}~\bibnamefont
  {Tarruell}}, \bibinfo {author} {\bibfnamefont {D.}~\bibnamefont {Greif}},
  \bibinfo {author} {\bibfnamefont {T.}~\bibnamefont {Uehlinger}}, \bibinfo
  {author} {\bibfnamefont {G.}~\bibnamefont {Jotzu}}, \ and\ \bibinfo {author}
  {\bibfnamefont {T.}~\bibnamefont {Esslinger}},\ }\href {\doibase
  10.1038/nature10871} {\bibfield  {journal} {\bibinfo  {journal} {Nature}\
  }\textbf {\bibinfo {volume} {483}},\ \bibinfo {pages} {302} (\bibinfo {year}
  {2012})}\BibitemShut {NoStop}%
\bibitem [{\citenamefont {Bellec}\ \emph {et~al.}(2013)\citenamefont {Bellec},
  \citenamefont {Kuhl}, \citenamefont {Montambaux},\ and\ \citenamefont
  {Mortessagne}}]{Bellec2013}%
  \BibitemOpen
  \bibfield  {author} {\bibinfo {author} {\bibfnamefont {M.}~\bibnamefont
  {Bellec}}, \bibinfo {author} {\bibfnamefont {U.}~\bibnamefont {Kuhl}},
  \bibinfo {author} {\bibfnamefont {G.}~\bibnamefont {Montambaux}}, \ and\
  \bibinfo {author} {\bibfnamefont {F.}~\bibnamefont {Mortessagne}},\ }\href
  {\doibase 10.1103/PhysRevLett.110.033902} {\bibfield  {journal} {\bibinfo
  {journal} {Phys. Rev. Lett.}\ }\textbf {\bibinfo {volume} {110}},\ \bibinfo
  {pages} {033902} (\bibinfo {year} {2013})}\BibitemShut {NoStop}%
\bibitem [{\citenamefont {Rechtsman}\ \emph
  {et~al.}(2013{\natexlab{a}})\citenamefont {Rechtsman}, \citenamefont
  {Plotnik}, \citenamefont {Zeuner}, \citenamefont {Song}, \citenamefont
  {Chen}, \citenamefont {Szameit},\ and\ \citenamefont
  {Segev}}]{Rechtsman2013}%
  \BibitemOpen
  \bibfield  {author} {\bibinfo {author} {\bibfnamefont {M.~C.}\ \bibnamefont
  {Rechtsman}}, \bibinfo {author} {\bibfnamefont {Y.}~\bibnamefont {Plotnik}},
  \bibinfo {author} {\bibfnamefont {J.~M.}\ \bibnamefont {Zeuner}}, \bibinfo
  {author} {\bibfnamefont {D.}~\bibnamefont {Song}}, \bibinfo {author}
  {\bibfnamefont {Z.}~\bibnamefont {Chen}}, \bibinfo {author} {\bibfnamefont
  {A.}~\bibnamefont {Szameit}}, \ and\ \bibinfo {author} {\bibfnamefont
  {M.}~\bibnamefont {Segev}},\ }\href {\doibase 10.1103/PhysRevLett.111.103901}
  {\bibfield  {journal} {\bibinfo  {journal} {Phys. Rev. Lett.}\ }\textbf
  {\bibinfo {volume} {111}},\ \bibinfo {pages} {103901} (\bibinfo {year}
  {2013}{\natexlab{a}})}\BibitemShut {NoStop}%
\bibitem [{\citenamefont {Mili\ifmmode \acute{c}\else
  \'{c}\fi{}evi\ifmmode~\acute{c}\else \'{c}\fi{}}\ \emph
  {et~al.}(2019)\citenamefont {Mili\ifmmode \acute{c}\else
  \'{c}\fi{}evi\ifmmode~\acute{c}\else \'{c}\fi{}}, \citenamefont {Montambaux},
  \citenamefont {Ozawa}, \citenamefont {Jamadi}, \citenamefont {Real},
  \citenamefont {Sagnes}, \citenamefont {Lema\^{\i}tre}, \citenamefont
  {Le~Gratiet}, \citenamefont {Harouri}, \citenamefont {Bloch},\ and\
  \citenamefont {Amo}}]{Amo2019}%
  \BibitemOpen
  \bibfield  {author} {\bibinfo {author} {\bibfnamefont {M.}~\bibnamefont
  {Mili\ifmmode \acute{c}\else \'{c}\fi{}evi\ifmmode~\acute{c}\else
  \'{c}\fi{}}}, \bibinfo {author} {\bibfnamefont {G.}~\bibnamefont
  {Montambaux}}, \bibinfo {author} {\bibfnamefont {T.}~\bibnamefont {Ozawa}},
  \bibinfo {author} {\bibfnamefont {O.}~\bibnamefont {Jamadi}}, \bibinfo
  {author} {\bibfnamefont {B.}~\bibnamefont {Real}}, \bibinfo {author}
  {\bibfnamefont {I.}~\bibnamefont {Sagnes}}, \bibinfo {author} {\bibfnamefont
  {A.}~\bibnamefont {Lema\^{\i}tre}}, \bibinfo {author} {\bibfnamefont
  {L.}~\bibnamefont {Le~Gratiet}}, \bibinfo {author} {\bibfnamefont
  {A.}~\bibnamefont {Harouri}}, \bibinfo {author} {\bibfnamefont
  {J.}~\bibnamefont {Bloch}}, \ and\ \bibinfo {author} {\bibfnamefont
  {A.}~\bibnamefont {Amo}},\ }\href {\doibase 10.1103/PhysRevX.9.031010}
  {\bibfield  {journal} {\bibinfo  {journal} {Phys. Rev. X}\ }\textbf {\bibinfo
  {volume} {9}},\ \bibinfo {pages} {031010} (\bibinfo {year}
  {2019})}\BibitemShut {NoStop}%
\bibitem [{\citenamefont {Kim}\ \emph {et~al.}(2015)\citenamefont {Kim},
  \citenamefont {Baik}, \citenamefont {Ryu}, \citenamefont {Sohn},
  \citenamefont {Park}, \citenamefont {Park}, \citenamefont {Denlinger},
  \citenamefont {Yi}, \citenamefont {Choi},\ and\ \citenamefont
  {Kim}}]{Su2015}%
  \BibitemOpen
  \bibfield  {author} {\bibinfo {author} {\bibfnamefont {J.}~\bibnamefont
  {Kim}}, \bibinfo {author} {\bibfnamefont {S.~S.}\ \bibnamefont {Baik}},
  \bibinfo {author} {\bibfnamefont {S.~H.}\ \bibnamefont {Ryu}}, \bibinfo
  {author} {\bibfnamefont {Y.}~\bibnamefont {Sohn}}, \bibinfo {author}
  {\bibfnamefont {S.}~\bibnamefont {Park}}, \bibinfo {author} {\bibfnamefont
  {B.-G.}\ \bibnamefont {Park}}, \bibinfo {author} {\bibfnamefont
  {J.}~\bibnamefont {Denlinger}}, \bibinfo {author} {\bibfnamefont
  {Y.}~\bibnamefont {Yi}}, \bibinfo {author} {\bibfnamefont {H.~J.}\
  \bibnamefont {Choi}}, \ and\ \bibinfo {author} {\bibfnamefont {K.~S.}\
  \bibnamefont {Kim}},\ }\href {\doibase 10.1126/science.aaa6486} {\bibfield
  {journal} {\bibinfo  {journal} {Science}\ }\textbf {\bibinfo {volume}
  {349}},\ \bibinfo {pages} {723} (\bibinfo {year} {2015})}\BibitemShut
  {NoStop}%
\bibitem [{\citenamefont {Kitagawa}\ \emph {et~al.}(2010)\citenamefont
  {Kitagawa}, \citenamefont {Berg}, \citenamefont {Rudner},\ and\ \citenamefont
  {Demler}}]{Kitagawa2010}%
  \BibitemOpen
  \bibfield  {author} {\bibinfo {author} {\bibfnamefont {T.}~\bibnamefont
  {Kitagawa}}, \bibinfo {author} {\bibfnamefont {E.}~\bibnamefont {Berg}},
  \bibinfo {author} {\bibfnamefont {M.}~\bibnamefont {Rudner}}, \ and\ \bibinfo
  {author} {\bibfnamefont {E.}~\bibnamefont {Demler}},\ }\href {\doibase
  10.1103/PhysRevB.82.235114} {\bibfield  {journal} {\bibinfo  {journal} {Phys.
  Rev. B}\ }\textbf {\bibinfo {volume} {82}},\ \bibinfo {pages} {235114}
  (\bibinfo {year} {2010})}\BibitemShut {NoStop}%
\bibitem [{\citenamefont {Mukherjee}\ \emph {et~al.}(2017)\citenamefont
  {Mukherjee}, \citenamefont {Spracklen}, \citenamefont {Valiente},
  \citenamefont {Andersson}, \citenamefont {{\"O}hberg}, \citenamefont
  {Goldman},\ and\ \citenamefont {Thomson}}]{Mukherjee2017}%
  \BibitemOpen
  \bibfield  {author} {\bibinfo {author} {\bibfnamefont {S.}~\bibnamefont
  {Mukherjee}}, \bibinfo {author} {\bibfnamefont {A.}~\bibnamefont
  {Spracklen}}, \bibinfo {author} {\bibfnamefont {M.}~\bibnamefont {Valiente}},
  \bibinfo {author} {\bibfnamefont {E.}~\bibnamefont {Andersson}}, \bibinfo
  {author} {\bibfnamefont {P.}~\bibnamefont {{\"O}hberg}}, \bibinfo {author}
  {\bibfnamefont {N.}~\bibnamefont {Goldman}}, \ and\ \bibinfo {author}
  {\bibfnamefont {R.~R.}\ \bibnamefont {Thomson}},\ }\href {\doibase
  10.1038/ncomms13918} {\bibfield  {journal} {\bibinfo  {journal} {Nature
  Communications}\ }\textbf {\bibinfo {volume} {8}},\ \bibinfo {pages} {13918}
  (\bibinfo {year} {2017})}\BibitemShut {NoStop}%
\bibitem [{\citenamefont {Maczewsky}\ \emph {et~al.}(2017)\citenamefont
  {Maczewsky}, \citenamefont {Zeuner}, \citenamefont {Nolte},\ and\
  \citenamefont {Szameit}}]{Szameit2017}%
  \BibitemOpen
  \bibfield  {author} {\bibinfo {author} {\bibfnamefont {L.~J.}\ \bibnamefont
  {Maczewsky}}, \bibinfo {author} {\bibfnamefont {J.~M.}\ \bibnamefont
  {Zeuner}}, \bibinfo {author} {\bibfnamefont {S.}~\bibnamefont {Nolte}}, \
  and\ \bibinfo {author} {\bibfnamefont {A.}~\bibnamefont {Szameit}},\ }\href
  {\doibase 10.1038/ncomms13756} {\bibfield  {journal} {\bibinfo  {journal}
  {Nature Communications}\ }\textbf {\bibinfo {volume} {8}},\ \bibinfo {pages}
  {13756} (\bibinfo {year} {2017})}\BibitemShut {NoStop}%
\bibitem [{\citenamefont {Wintersperger}\ \emph {et~al.}(2020)\citenamefont
  {Wintersperger}, \citenamefont {Braun}, \citenamefont {{\"U}nal},
  \citenamefont {Eckardt}, \citenamefont {Di~Liberto}, \citenamefont {Goldman},
  \citenamefont {Bloch},\ and\ \citenamefont
  {Aidelsburger}}]{Wintersperger2020}%
  \BibitemOpen
  \bibfield  {author} {\bibinfo {author} {\bibfnamefont {K.}~\bibnamefont
  {Wintersperger}}, \bibinfo {author} {\bibfnamefont {C.}~\bibnamefont
  {Braun}}, \bibinfo {author} {\bibfnamefont {F.~N.}\ \bibnamefont {{\"U}nal}},
  \bibinfo {author} {\bibfnamefont {A.}~\bibnamefont {Eckardt}}, \bibinfo
  {author} {\bibfnamefont {M.}~\bibnamefont {Di~Liberto}}, \bibinfo {author}
  {\bibfnamefont {N.}~\bibnamefont {Goldman}}, \bibinfo {author} {\bibfnamefont
  {I.}~\bibnamefont {Bloch}}, \ and\ \bibinfo {author} {\bibfnamefont
  {M.}~\bibnamefont {Aidelsburger}},\ }\href {\doibase
  10.1038/s41567-020-0949-y} {\bibfield  {journal} {\bibinfo  {journal} {Nature
  Physics}\ }\textbf {\bibinfo {volume} {16}},\ \bibinfo {pages} {1058}
  (\bibinfo {year} {2020})}\BibitemShut {NoStop}%
\bibitem [{\citenamefont {Quelle}\ \emph {et~al.}(2017)\citenamefont {Quelle},
  \citenamefont {Weitenberg}, \citenamefont {Sengstock},\ and\ \citenamefont
  {Morais~Smith}}]{Quelle2017}%
  \BibitemOpen
  \bibfield  {author} {\bibinfo {author} {\bibfnamefont {A.}~\bibnamefont
  {Quelle}}, \bibinfo {author} {\bibfnamefont {C.}~\bibnamefont {Weitenberg}},
  \bibinfo {author} {\bibfnamefont {K.}~\bibnamefont {Sengstock}}, \ and\
  \bibinfo {author} {\bibfnamefont {C.}~\bibnamefont {Morais~Smith}},\ }\href
  {\doibase 10.1088/1367-2630/aa8646} {\bibfield  {journal} {\bibinfo
  {journal} {New Journal of Physics}\ }\textbf {\bibinfo {volume} {19}},\
  \bibinfo {pages} {113010} (\bibinfo {year} {2017})}\BibitemShut {NoStop}%
\bibitem [{\citenamefont {Giovanazzi}\ \emph {et~al.}(2002)\citenamefont
  {Giovanazzi}, \citenamefont {G\"orlitz},\ and\ \citenamefont
  {Pfau}}]{Pfau2002}%
  \BibitemOpen
  \bibfield  {author} {\bibinfo {author} {\bibfnamefont {S.}~\bibnamefont
  {Giovanazzi}}, \bibinfo {author} {\bibfnamefont {A.}~\bibnamefont
  {G\"orlitz}}, \ and\ \bibinfo {author} {\bibfnamefont {T.}~\bibnamefont
  {Pfau}},\ }\href {\doibase 10.1103/PhysRevLett.89.130401} {\bibfield
  {journal} {\bibinfo  {journal} {Phys. Rev. Lett.}\ }\textbf {\bibinfo
  {volume} {89}},\ \bibinfo {pages} {130401} (\bibinfo {year}
  {2002})}\BibitemShut {NoStop}%
\bibitem [{\citenamefont {Tang}\ \emph {et~al.}(2018)\citenamefont {Tang},
  \citenamefont {Kao}, \citenamefont {Li},\ and\ \citenamefont
  {Lev}}]{Lev2018}%
  \BibitemOpen
  \bibfield  {author} {\bibinfo {author} {\bibfnamefont {Y.}~\bibnamefont
  {Tang}}, \bibinfo {author} {\bibfnamefont {W.}~\bibnamefont {Kao}}, \bibinfo
  {author} {\bibfnamefont {K.-Y.}\ \bibnamefont {Li}}, \ and\ \bibinfo {author}
  {\bibfnamefont {B.~L.}\ \bibnamefont {Lev}},\ }\href {\doibase
  10.1103/PhysRevLett.120.230401} {\bibfield  {journal} {\bibinfo  {journal}
  {Phys. Rev. Lett.}\ }\textbf {\bibinfo {volume} {120}},\ \bibinfo {pages}
  {230401} (\bibinfo {year} {2018})}\BibitemShut {NoStop}%
\bibitem [{\citenamefont {Celi}\ \emph {et~al.}(2014)\citenamefont {Celi},
  \citenamefont {Massignan}, \citenamefont {Ruseckas}, \citenamefont {Goldman},
  \citenamefont {Spielman}, \citenamefont {Juzeli\ifmmode~\bar{u}\else
  \={u}\fi{}nas},\ and\ \citenamefont {Lewenstein}}]{Celi2014}%
  \BibitemOpen
  \bibfield  {author} {\bibinfo {author} {\bibfnamefont {A.}~\bibnamefont
  {Celi}}, \bibinfo {author} {\bibfnamefont {P.}~\bibnamefont {Massignan}},
  \bibinfo {author} {\bibfnamefont {J.}~\bibnamefont {Ruseckas}}, \bibinfo
  {author} {\bibfnamefont {N.}~\bibnamefont {Goldman}}, \bibinfo {author}
  {\bibfnamefont {I.~B.}\ \bibnamefont {Spielman}}, \bibinfo {author}
  {\bibfnamefont {G.}~\bibnamefont {Juzeli\ifmmode~\bar{u}\else
  \={u}\fi{}nas}}, \ and\ \bibinfo {author} {\bibfnamefont {M.}~\bibnamefont
  {Lewenstein}},\ }\href {\doibase 10.1103/PhysRevLett.112.043001} {\bibfield
  {journal} {\bibinfo  {journal} {Phys. Rev. Lett.}\ }\textbf {\bibinfo
  {volume} {112}},\ \bibinfo {pages} {043001} (\bibinfo {year}
  {2014})}\BibitemShut {NoStop}%
\bibitem [{\citenamefont {Ozawa}\ and\ \citenamefont
  {Price}(2019)}]{Price2019}%
  \BibitemOpen
  \bibfield  {author} {\bibinfo {author} {\bibfnamefont {T.}~\bibnamefont
  {Ozawa}}\ and\ \bibinfo {author} {\bibfnamefont {H.~M.}\ \bibnamefont
  {Price}},\ }\href {\doibase 10.1038/s42254-019-0045-3} {\bibfield  {journal}
  {\bibinfo  {journal} {Nature Reviews Physics}\ }\textbf {\bibinfo {volume}
  {1}},\ \bibinfo {pages} {349} (\bibinfo {year} {2019})}\BibitemShut {NoStop}%
\bibitem [{\citenamefont {Bradlyn}\ \emph {et~al.}(2016)\citenamefont
  {Bradlyn}, \citenamefont {Cano}, \citenamefont {Wang}, \citenamefont
  {Vergniory}, \citenamefont {Felser}, \citenamefont {Cava},\ and\
  \citenamefont {Bernevig}}]{Bradlyn2016}%
  \BibitemOpen
  \bibfield  {author} {\bibinfo {author} {\bibfnamefont {B.}~\bibnamefont
  {Bradlyn}}, \bibinfo {author} {\bibfnamefont {J.}~\bibnamefont {Cano}},
  \bibinfo {author} {\bibfnamefont {Z.}~\bibnamefont {Wang}}, \bibinfo {author}
  {\bibfnamefont {M.~G.}\ \bibnamefont {Vergniory}}, \bibinfo {author}
  {\bibfnamefont {C.}~\bibnamefont {Felser}}, \bibinfo {author} {\bibfnamefont
  {R.~J.}\ \bibnamefont {Cava}}, \ and\ \bibinfo {author} {\bibfnamefont
  {B.~A.}\ \bibnamefont {Bernevig}},\ }\href
  {https://science.sciencemag.org/content/353/6299/aaf5037} {\bibfield
  {journal} {\bibinfo  {journal} {Science}\ }\textbf {\bibinfo {volume}
  {353}},\ \bibinfo {pages} {aaf5037} (\bibinfo {year} {2016})}\BibitemShut
  {NoStop}%
\bibitem [{\citenamefont {Green}\ \emph {et~al.}(2010)\citenamefont {Green},
  \citenamefont {Santos},\ and\ \citenamefont {Chamon}}]{Chamon2010}%
  \BibitemOpen
  \bibfield  {author} {\bibinfo {author} {\bibfnamefont {D.}~\bibnamefont
  {Green}}, \bibinfo {author} {\bibfnamefont {L.}~\bibnamefont {Santos}}, \
  and\ \bibinfo {author} {\bibfnamefont {C.}~\bibnamefont {Chamon}},\ }\href
  {\doibase 10.1103/PhysRevB.82.075104} {\bibfield  {journal} {\bibinfo
  {journal} {Phys. Rev. B}\ }\textbf {\bibinfo {volume} {82}},\ \bibinfo
  {pages} {075104} (\bibinfo {year} {2010})}\BibitemShut {NoStop}%
\bibitem [{\citenamefont {Fulga}\ and\ \citenamefont
  {Stern}(2017)}]{Fulga2017}%
  \BibitemOpen
  \bibfield  {author} {\bibinfo {author} {\bibfnamefont {I.~C.}\ \bibnamefont
  {Fulga}}\ and\ \bibinfo {author} {\bibfnamefont {A.}~\bibnamefont {Stern}},\
  }\href {\doibase 10.1103/PhysRevB.95.241116} {\bibfield  {journal} {\bibinfo
  {journal} {Phys. Rev. B}\ }\textbf {\bibinfo {volume} {95}},\ \bibinfo
  {pages} {241116} (\bibinfo {year} {2017})}\BibitemShut {NoStop}%
\bibitem [{\citenamefont {Zhu}\ \emph {et~al.}(2017)\citenamefont {Zhu},
  \citenamefont {Zhang}, \citenamefont {Yan}, \citenamefont {Xing},\ and\
  \citenamefont {Zhu}}]{Zhu2017}%
  \BibitemOpen
  \bibfield  {author} {\bibinfo {author} {\bibfnamefont {Y.-Q.}\ \bibnamefont
  {Zhu}}, \bibinfo {author} {\bibfnamefont {D.-W.}\ \bibnamefont {Zhang}},
  \bibinfo {author} {\bibfnamefont {H.}~\bibnamefont {Yan}}, \bibinfo {author}
  {\bibfnamefont {D.-Y.}\ \bibnamefont {Xing}}, \ and\ \bibinfo {author}
  {\bibfnamefont {S.-L.}\ \bibnamefont {Zhu}},\ }\href {\doibase
  10.1103/PhysRevA.96.033634} {\bibfield  {journal} {\bibinfo  {journal} {Phys.
  Rev. A}\ }\textbf {\bibinfo {volume} {96}},\ \bibinfo {pages} {033634}
  (\bibinfo {year} {2017})}\BibitemShut {NoStop}%
\bibitem [{\citenamefont {Tan}\ \emph {et~al.}(2018)\citenamefont {Tan},
  \citenamefont {Zhang}, \citenamefont {Liu}, \citenamefont {Xue},
  \citenamefont {Yu}, \citenamefont {Zhu}, \citenamefont {Yan}, \citenamefont
  {Zhu},\ and\ \citenamefont {Yu}}]{Yang2018}%
  \BibitemOpen
  \bibfield  {author} {\bibinfo {author} {\bibfnamefont {X.}~\bibnamefont
  {Tan}}, \bibinfo {author} {\bibfnamefont {D.-W.}\ \bibnamefont {Zhang}},
  \bibinfo {author} {\bibfnamefont {Q.}~\bibnamefont {Liu}}, \bibinfo {author}
  {\bibfnamefont {G.}~\bibnamefont {Xue}}, \bibinfo {author} {\bibfnamefont
  {H.-F.}\ \bibnamefont {Yu}}, \bibinfo {author} {\bibfnamefont {Y.-Q.}\
  \bibnamefont {Zhu}}, \bibinfo {author} {\bibfnamefont {H.}~\bibnamefont
  {Yan}}, \bibinfo {author} {\bibfnamefont {S.-L.}\ \bibnamefont {Zhu}}, \ and\
  \bibinfo {author} {\bibfnamefont {Y.}~\bibnamefont {Yu}},\ }\href {\doibase
  10.1103/PhysRevLett.120.130503} {\bibfield  {journal} {\bibinfo  {journal}
  {Phys. Rev. Lett.}\ }\textbf {\bibinfo {volume} {120}},\ \bibinfo {pages}
  {130503} (\bibinfo {year} {2018})}\BibitemShut {NoStop}%
\bibitem [{\citenamefont {Hu}\ and\ \citenamefont {Zhang}(2018)}]{Zhang2018}%
  \BibitemOpen
  \bibfield  {author} {\bibinfo {author} {\bibfnamefont {H.}~\bibnamefont
  {Hu}}\ and\ \bibinfo {author} {\bibfnamefont {C.}~\bibnamefont {Zhang}},\
  }\href {\doibase 10.1103/PhysRevA.98.013627} {\bibfield  {journal} {\bibinfo
  {journal} {Phys. Rev. A}\ }\textbf {\bibinfo {volume} {98}},\ \bibinfo
  {pages} {013627} (\bibinfo {year} {2018})}\BibitemShut {NoStop}%
\bibitem [{\citenamefont {Armitage}\ \emph {et~al.}(2018)\citenamefont
  {Armitage}, \citenamefont {Mele},\ and\ \citenamefont
  {Vishwanath}}]{Armitage2018}%
  \BibitemOpen
  \bibfield  {author} {\bibinfo {author} {\bibfnamefont {N.}~\bibnamefont
  {Armitage}}, \bibinfo {author} {\bibfnamefont {E.}~\bibnamefont {Mele}}, \
  and\ \bibinfo {author} {\bibfnamefont {A.}~\bibnamefont {Vishwanath}},\
  }\href {https://doi.org/10.1103/RevModPhys.90.015001} {\bibfield  {journal}
  {\bibinfo  {journal} {Reviews of Modern Physics}\ }\textbf {\bibinfo {volume}
  {90}},\ \bibinfo {pages} {015001} (\bibinfo {year} {2018})}\BibitemShut
  {NoStop}%
\bibitem [{\citenamefont {Xiao}\ \emph {et~al.}(2010)\citenamefont {Xiao},
  \citenamefont {Chang},\ and\ \citenamefont {Niu}}]{Niu2010}%
  \BibitemOpen
  \bibfield  {author} {\bibinfo {author} {\bibfnamefont {D.}~\bibnamefont
  {Xiao}}, \bibinfo {author} {\bibfnamefont {M.-C.}\ \bibnamefont {Chang}}, \
  and\ \bibinfo {author} {\bibfnamefont {Q.}~\bibnamefont {Niu}},\ }\href
  {\doibase 10.1103/RevModPhys.82.1959} {\bibfield  {journal} {\bibinfo
  {journal} {Rev. Mod. Phys.}\ }\textbf {\bibinfo {volume} {82}},\ \bibinfo
  {pages} {1959} (\bibinfo {year} {2010})}\BibitemShut {NoStop}%
\bibitem [{\citenamefont {Wang}\ \emph {et~al.}(2021)\citenamefont {Wang},
  \citenamefont {Cheng}, \citenamefont {Wang}, \citenamefont {Zhang},
  \citenamefont {Lu}, \citenamefont {Yi}, \citenamefont {Niu}, \citenamefont
  {Deng}, \citenamefont {Liu}, \citenamefont {Chen},\ and\ \citenamefont
  {Pan}}]{Pan2021}%
  \BibitemOpen
  \bibfield  {author} {\bibinfo {author} {\bibfnamefont {Z.-Y.}\ \bibnamefont
  {Wang}}, \bibinfo {author} {\bibfnamefont {X.-C.}\ \bibnamefont {Cheng}},
  \bibinfo {author} {\bibfnamefont {B.-Z.}\ \bibnamefont {Wang}}, \bibinfo
  {author} {\bibfnamefont {J.-Y.}\ \bibnamefont {Zhang}}, \bibinfo {author}
  {\bibfnamefont {Y.-H.}\ \bibnamefont {Lu}}, \bibinfo {author} {\bibfnamefont
  {C.-R.}\ \bibnamefont {Yi}}, \bibinfo {author} {\bibfnamefont
  {S.}~\bibnamefont {Niu}}, \bibinfo {author} {\bibfnamefont {Y.}~\bibnamefont
  {Deng}}, \bibinfo {author} {\bibfnamefont {X.-J.}\ \bibnamefont {Liu}},
  \bibinfo {author} {\bibfnamefont {S.}~\bibnamefont {Chen}}, \ and\ \bibinfo
  {author} {\bibfnamefont {J.-W.}\ \bibnamefont {Pan}},\ }\href {\doibase
  10.1126/science.abc0105} {\bibfield  {journal} {\bibinfo  {journal}
  {Science}\ }\textbf {\bibinfo {volume} {372}},\ \bibinfo {pages} {271}
  (\bibinfo {year} {2021})}\BibitemShut {NoStop}%
\bibitem [{\citenamefont {Sugawa}\ \emph {et~al.}(2018)\citenamefont {Sugawa},
  \citenamefont {Salces-Carcoba}, \citenamefont {Perry}, \citenamefont {Yue},\
  and\ \citenamefont {Spielman}}]{Spielman2018}%
  \BibitemOpen
  \bibfield  {author} {\bibinfo {author} {\bibfnamefont {S.}~\bibnamefont
  {Sugawa}}, \bibinfo {author} {\bibfnamefont {F.}~\bibnamefont
  {Salces-Carcoba}}, \bibinfo {author} {\bibfnamefont {A.~R.}\ \bibnamefont
  {Perry}}, \bibinfo {author} {\bibfnamefont {Y.}~\bibnamefont {Yue}}, \ and\
  \bibinfo {author} {\bibfnamefont {I.~B.}\ \bibnamefont {Spielman}},\ }\href
  {\doibase 10.1126/science.aam9031} {\bibfield  {journal} {\bibinfo  {journal}
  {Science}\ }\textbf {\bibinfo {volume} {360}},\ \bibinfo {pages} {1429}
  (\bibinfo {year} {2018})}\BibitemShut {NoStop}%
\bibitem [{\citenamefont {Guinea}\ \emph {et~al.}(2010)\citenamefont {Guinea},
  \citenamefont {Katsnelson},\ and\ \citenamefont {Geim}}]{Guinea2010}%
  \BibitemOpen
  \bibfield  {author} {\bibinfo {author} {\bibfnamefont {F.}~\bibnamefont
  {Guinea}}, \bibinfo {author} {\bibfnamefont {M.~I.}\ \bibnamefont
  {Katsnelson}}, \ and\ \bibinfo {author} {\bibfnamefont {A.~K.}\ \bibnamefont
  {Geim}},\ }\href {\doibase 10.1038/nphys1420} {\bibfield  {journal} {\bibinfo
   {journal} {Nature Physics}\ }\textbf {\bibinfo {volume} {6}},\ \bibinfo
  {pages} {30} (\bibinfo {year} {2010})}\BibitemShut {NoStop}%
\bibitem [{\citenamefont {Rechtsman}\ \emph
  {et~al.}(2013{\natexlab{b}})\citenamefont {Rechtsman}, \citenamefont
  {Zeuner}, \citenamefont {T{\"u}nnermann}, \citenamefont {Nolte},
  \citenamefont {Segev},\ and\ \citenamefont {Szameit}}]{Rechtsman2013a}%
  \BibitemOpen
  \bibfield  {author} {\bibinfo {author} {\bibfnamefont {M.~C.}\ \bibnamefont
  {Rechtsman}}, \bibinfo {author} {\bibfnamefont {J.~M.}\ \bibnamefont
  {Zeuner}}, \bibinfo {author} {\bibfnamefont {A.}~\bibnamefont
  {T{\"u}nnermann}}, \bibinfo {author} {\bibfnamefont {S.}~\bibnamefont
  {Nolte}}, \bibinfo {author} {\bibfnamefont {M.}~\bibnamefont {Segev}}, \ and\
  \bibinfo {author} {\bibfnamefont {A.}~\bibnamefont {Szameit}},\ }\href
  {\doibase 10.1038/nphoton.2012.302} {\bibfield  {journal} {\bibinfo
  {journal} {Nature Photonics}\ }\textbf {\bibinfo {volume} {7}},\ \bibinfo
  {pages} {153} (\bibinfo {year} {2013}{\natexlab{b}})}\BibitemShut {NoStop}%
\bibitem [{\citenamefont {Jamadi}\ \emph {et~al.}(2020)\citenamefont {Jamadi},
  \citenamefont {Rozas}, \citenamefont {Salerno}, \citenamefont
  {Mili{\'c}evi{\'c}}, \citenamefont {Ozawa}, \citenamefont {Sagnes},
  \citenamefont {Lema{\^\i}tre}, \citenamefont {Le~Gratiet}, \citenamefont
  {Harouri}, \citenamefont {Carusotto}, \citenamefont {Bloch},\ and\
  \citenamefont {Amo}}]{Jamadi2020}%
  \BibitemOpen
  \bibfield  {author} {\bibinfo {author} {\bibfnamefont {O.}~\bibnamefont
  {Jamadi}}, \bibinfo {author} {\bibfnamefont {E.}~\bibnamefont {Rozas}},
  \bibinfo {author} {\bibfnamefont {G.}~\bibnamefont {Salerno}}, \bibinfo
  {author} {\bibfnamefont {M.}~\bibnamefont {Mili{\'c}evi{\'c}}}, \bibinfo
  {author} {\bibfnamefont {T.}~\bibnamefont {Ozawa}}, \bibinfo {author}
  {\bibfnamefont {I.}~\bibnamefont {Sagnes}}, \bibinfo {author} {\bibfnamefont
  {A.}~\bibnamefont {Lema{\^\i}tre}}, \bibinfo {author} {\bibfnamefont
  {L.}~\bibnamefont {Le~Gratiet}}, \bibinfo {author} {\bibfnamefont
  {A.}~\bibnamefont {Harouri}}, \bibinfo {author} {\bibfnamefont
  {I.}~\bibnamefont {Carusotto}}, \bibinfo {author} {\bibfnamefont
  {J.}~\bibnamefont {Bloch}}, \ and\ \bibinfo {author} {\bibfnamefont
  {A.}~\bibnamefont {Amo}},\ }\href {\doibase 10.1038/s41377-020-00377-6}
  {\bibfield  {journal} {\bibinfo  {journal} {Light: Science \& Applications}\
  }\textbf {\bibinfo {volume} {9}},\ \bibinfo {pages} {144} (\bibinfo {year}
  {2020})}\BibitemShut {NoStop}%
\bibitem [{\citenamefont {Tian}\ \emph {et~al.}(2015)\citenamefont {Tian},
  \citenamefont {Endres},\ and\ \citenamefont {Pekker}}]{Pekker2015}%
  \BibitemOpen
  \bibfield  {author} {\bibinfo {author} {\bibfnamefont {B.}~\bibnamefont
  {Tian}}, \bibinfo {author} {\bibfnamefont {M.}~\bibnamefont {Endres}}, \ and\
  \bibinfo {author} {\bibfnamefont {D.}~\bibnamefont {Pekker}},\ }\href
  {\doibase 10.1103/PhysRevLett.115.236803} {\bibfield  {journal} {\bibinfo
  {journal} {Phys. Rev. Lett.}\ }\textbf {\bibinfo {volume} {115}},\ \bibinfo
  {pages} {236803} (\bibinfo {year} {2015})}\BibitemShut {NoStop}%
\bibitem [{\citenamefont {Jamotte}\ \emph {et~al.}(2022)\citenamefont
  {Jamotte}, \citenamefont {Goldman},\ and\ \citenamefont
  {Di~Liberto}}]{Jamotte2022}%
  \BibitemOpen
  \bibfield  {author} {\bibinfo {author} {\bibfnamefont {M.}~\bibnamefont
  {Jamotte}}, \bibinfo {author} {\bibfnamefont {N.}~\bibnamefont {Goldman}}, \
  and\ \bibinfo {author} {\bibfnamefont {M.}~\bibnamefont {Di~Liberto}},\
  }\href {\doibase 10.1038/s42005-022-00802-9} {\bibfield  {journal} {\bibinfo
  {journal} {Communications Physics}\ }\textbf {\bibinfo {volume} {5}},\
  \bibinfo {pages} {30} (\bibinfo {year} {2022})}\BibitemShut {NoStop}%
\bibitem [{\citenamefont {Salerno}\ \emph {et~al.}(2015)\citenamefont
  {Salerno}, \citenamefont {Ozawa}, \citenamefont {Price},\ and\ \citenamefont
  {Carusotto}}]{Salerno2015}%
  \BibitemOpen
  \bibfield  {author} {\bibinfo {author} {\bibfnamefont {G.}~\bibnamefont
  {Salerno}}, \bibinfo {author} {\bibfnamefont {T.}~\bibnamefont {Ozawa}},
  \bibinfo {author} {\bibfnamefont {H.~M.}\ \bibnamefont {Price}}, \ and\
  \bibinfo {author} {\bibfnamefont {I.}~\bibnamefont {Carusotto}},\ }\href
  {\doibase 10.1088/2053-1583/2/3/034015} {\bibfield  {journal} {\bibinfo
  {journal} {2D Materials}\ }\textbf {\bibinfo {volume} {2}},\ \bibinfo {pages}
  {034015} (\bibinfo {year} {2015})}\BibitemShut {NoStop}%
\bibitem [{\citenamefont {Goerbig}(2011)}]{Goerbig2011}%
  \BibitemOpen
  \bibfield  {author} {\bibinfo {author} {\bibfnamefont {M.~O.}\ \bibnamefont
  {Goerbig}},\ }\href {\doibase 10.1103/RevModPhys.83.1193} {\bibfield
  {journal} {\bibinfo  {journal} {Rev. Mod. Phys.}\ }\textbf {\bibinfo {volume}
  {83}},\ \bibinfo {pages} {1193} (\bibinfo {year} {2011})}\BibitemShut
  {NoStop}%
\bibitem [{\citenamefont {Duca}\ \emph {et~al.}(2015)\citenamefont {Duca},
  \citenamefont {Li}, \citenamefont {Reitter}, \citenamefont {Bloch},
  \citenamefont {Schleier-Smith},\ and\ \citenamefont {Schneider}}]{Duca2015}%
  \BibitemOpen
  \bibfield  {author} {\bibinfo {author} {\bibfnamefont {L.}~\bibnamefont
  {Duca}}, \bibinfo {author} {\bibfnamefont {T.}~\bibnamefont {Li}}, \bibinfo
  {author} {\bibfnamefont {M.}~\bibnamefont {Reitter}}, \bibinfo {author}
  {\bibfnamefont {I.}~\bibnamefont {Bloch}}, \bibinfo {author} {\bibfnamefont
  {M.}~\bibnamefont {Schleier-Smith}}, \ and\ \bibinfo {author} {\bibfnamefont
  {U.}~\bibnamefont {Schneider}},\ }\href {\doibase 10.1126/science.1259052}
  {\bibfield  {journal} {\bibinfo  {journal} {Science}\ }\textbf {\bibinfo
  {volume} {347}},\ \bibinfo {pages} {288} (\bibinfo {year}
  {2015})}\BibitemShut {NoStop}%
\bibitem [{\citenamefont {Ernst}\ \emph {et~al.}(2010)\citenamefont {Ernst},
  \citenamefont {G{\"o}tze}, \citenamefont {Krauser}, \citenamefont {Pyka},
  \citenamefont {L{\"u}hmann}, \citenamefont {Pfannkuche},\ and\ \citenamefont
  {Sengstock}}]{Sengstock2010}%
  \BibitemOpen
  \bibfield  {author} {\bibinfo {author} {\bibfnamefont {P.~T.}\ \bibnamefont
  {Ernst}}, \bibinfo {author} {\bibfnamefont {S.}~\bibnamefont {G{\"o}tze}},
  \bibinfo {author} {\bibfnamefont {J.~S.}\ \bibnamefont {Krauser}}, \bibinfo
  {author} {\bibfnamefont {K.}~\bibnamefont {Pyka}}, \bibinfo {author}
  {\bibfnamefont {D.-S.}\ \bibnamefont {L{\"u}hmann}}, \bibinfo {author}
  {\bibfnamefont {D.}~\bibnamefont {Pfannkuche}}, \ and\ \bibinfo {author}
  {\bibfnamefont {K.}~\bibnamefont {Sengstock}},\ }\href {\doibase
  10.1038/nphys1476} {\bibfield  {journal} {\bibinfo  {journal} {Nature
  Physics}\ }\textbf {\bibinfo {volume} {6}},\ \bibinfo {pages} {56} (\bibinfo
  {year} {2010})}\BibitemShut {NoStop}%
\bibitem [{\citenamefont {Frisch}(2014)}]{Frisch2014}%
  \BibitemOpen
  \bibfield  {author} {\bibinfo {author} {\bibfnamefont {A.}~\bibnamefont
  {Frisch}},\ }\emph {\bibinfo {title} {Dipolar Quantum Gases of Erbium}},\
  \href@noop {} {Ph.D. thesis} (\bibinfo {year} {2014})\BibitemShut {NoStop}%
\bibitem [{\citenamefont {Aymar}\ and\ \citenamefont
  {Dulieu}(2005)}]{Aymar2005}%
  \BibitemOpen
  \bibfield  {author} {\bibinfo {author} {\bibfnamefont {M.}~\bibnamefont
  {Aymar}}\ and\ \bibinfo {author} {\bibfnamefont {O.}~\bibnamefont {Dulieu}},\
  }\bibfield  {booktitle} {\emph {\bibinfo {booktitle} {The Journal of Chemical
  Physics}},\ }\href {\doibase 10.1063/1.1903944} {\bibfield  {journal}
  {\bibinfo  {journal} {The Journal of Chemical Physics}\ }\textbf {\bibinfo
  {volume} {122}},\ \bibinfo {pages} {204302} (\bibinfo {year}
  {2005})}\BibitemShut {NoStop}%
\bibitem [{\citenamefont {Bloch}\ \emph {et~al.}(2008)\citenamefont {Bloch},
  \citenamefont {Dalibard},\ and\ \citenamefont {Zwerger}}]{Bloch2008}%
  \BibitemOpen
  \bibfield  {author} {\bibinfo {author} {\bibfnamefont {I.}~\bibnamefont
  {Bloch}}, \bibinfo {author} {\bibfnamefont {J.}~\bibnamefont {Dalibard}}, \
  and\ \bibinfo {author} {\bibfnamefont {W.}~\bibnamefont {Zwerger}},\ }\href
  {\doibase 10.1103/RevModPhys.80.885} {\bibfield  {journal} {\bibinfo
  {journal} {Rev. Mod. Phys.}\ }\textbf {\bibinfo {volume} {80}},\ \bibinfo
  {pages} {885} (\bibinfo {year} {2008})}\BibitemShut {NoStop}%
\bibitem [{\citenamefont {Lim}\ \emph {et~al.}(2020)\citenamefont {Lim},
  \citenamefont {Fuchs}, \citenamefont {Pi\'echon},\ and\ \citenamefont
  {Montambaux}}]{Lim2020}%
  \BibitemOpen
  \bibfield  {author} {\bibinfo {author} {\bibfnamefont {L.-K.}\ \bibnamefont
  {Lim}}, \bibinfo {author} {\bibfnamefont {J.-N.}\ \bibnamefont {Fuchs}},
  \bibinfo {author} {\bibfnamefont {F.}~\bibnamefont {Pi\'echon}}, \ and\
  \bibinfo {author} {\bibfnamefont {G.}~\bibnamefont {Montambaux}},\ }\href
  {\doibase 10.1103/PhysRevB.101.045131} {\bibfield  {journal} {\bibinfo
  {journal} {Phys. Rev. B}\ }\textbf {\bibinfo {volume} {101}},\ \bibinfo
  {pages} {045131} (\bibinfo {year} {2020})}\BibitemShut {NoStop}%
\bibitem [{\citenamefont {Goldman}\ and\ \citenamefont
  {Dalibard}(2014)}]{Goldman2014b}%
  \BibitemOpen
  \bibfield  {author} {\bibinfo {author} {\bibfnamefont {N.}~\bibnamefont
  {Goldman}}\ and\ \bibinfo {author} {\bibfnamefont {J.}~\bibnamefont
  {Dalibard}},\ }\href {\doibase 10.1103/PhysRevX.4.031027} {\bibfield
  {journal} {\bibinfo  {journal} {Phys. Rev. X}\ }\textbf {\bibinfo {volume}
  {4}},\ \bibinfo {pages} {031027} (\bibinfo {year} {2014})}\BibitemShut
  {NoStop}%
\end{thebibliography}%

\end{document}